\documentclass[preprint,prd,superscriptaddress,nofootinbib]{revtex4-2}

\usepackage[english]{babel}
\usepackage{amssymb}
\usepackage{array}
\usepackage{tikz}
    \usepackage{multirow}

\usepackage{comment}

\newcommand{\IM}{\mathrm{Im}}

\newcommand{\circledstar}{\tikz[baseline=(X.base)] 
  \node[draw,circle,inner sep=0.5pt](X){$\ast$};}

\usepackage[letterpaper,top=2cm,bottom=2cm,left=3cm,right=3cm,marginparwidth=1.75cm]{geometry}

\usepackage{amsmath}
\usepackage{graphicx}
\usepackage[colorlinks=true, allcolors=blue]{hyperref}

\begin{document}

\title{Nevanlinna–Pick interpolation from uncertain data}
\author{Sarah Fields}\affiliation{\CU}
\author{Norman Christ}\affiliation{\CU}
\newcommand*{\CU}{Physics Department, Columbia University, New York City, New York 10027, USA}

\date{Oct 14, 2025}

\begin{abstract}
The calculation of inclusive processes that involve the production of many particles is a challenge for lattice QCD, a Euclidean-space method that is far removed from real-time, multiparticle production. A new approach to this problem based on Nevanlinna-Pick interpolation has been proposed by Bergamaschi, {\it et al}. Here we extend their method by exploring the propagation of the statistical and systematic errors that accompany a lattice QCD calculation through this interpolation process. A simplified example of a multiparticle spectral function is studied with a focus on the possible applications of these methods to the calculation of inclusive heavy-particle decays.   
\end{abstract}

\maketitle

\newpage

\section{Introduction}
\label{sec:intro}

The study of inclusive processes with high precision is key to interpreting experimental data and testing Standard Model predictions. The difficulties in calculating inclusive processes from lattice QCD arise because lattice field theory results are obtained in Euclidean space, not physical Minkowski space, which is relevant for physical scattering and decay processes. While many physical quantities such as low-energy QCD energy eigenvalues and the matrix elements of the corresponding eigenstates can be computed directly in Euclidean space, hadronic decays into many-particle final states depend on real-time particle propagation and interaction and are related to Euclidean space amplitudes by analytic continuation.  Performing such an analytic continuation in practice is an ill-posed problem because of the finite number of Euclidean space results and the errors associated with those results. This difficulty is commonly known as the inverse problem.

Many methods have been applied to tackle the inverse problem in lattice QCD, including Bayesian inference, linear regularization, rational approximations, and machine learning. Reviews and recent examples of possible approaches to this inverse problem can be found in Refs.~\cite{Rothkopf:2022fyo,Rothkopf:2022ctl,Aarts:2025gyp,DelDebbio:2024sfa,Hansen:2017mnd,PhysRevD.100.034521,PhysRevD.99.094508,Bailas:2020qmv,Gambino:2022dvu,Barone:2023tbl,Bruno:2024fqc,Patella:2024cto,DelDebbio:2024lwm,Bergamaschi:2023xzx,Zhang:2024ixe,Horak:2021syv,Karpie:2019eiq}.

In this work, we discuss the implementation of the Nevanlinna–Pick interpolation method first applied to lattice QCD in Ref.~\cite{Bergamaschi:2023xzx}, where renewed interest in the application of this method to physics was initiated by Ref.~\cite{PhysRevLett.126.056402}. In this approach, one exploits the remarkable properties of the analytic function, here represented by the analytic function $G(z)$, whose imaginary part on the real axis determines the spectral function of interest and whose values on the imaginary axis can be computed using lattice QCD. For $N$ values of $G(z)$ determined on the imaginary axis, Nevanlinna-Pick interpolation can be used to bound the possible values allowed for $G(z)$ for $z$ elsewhere in the upper half of the complex plane. As was recognized in Ref.~\cite{Bergamaschi:2023xzx}, while this approach does not exactly solve the inverse problem, it does provide absolute bounds which constrain the resulting interpolated values of $G(z)$. This raises the possibility that predictions with precisely known interpolation errors might be obtained from Euclidean lattice data. Such an approach has the potential to extend the first-principles character of many lattice QCD calculations to this inclusive application. Additional applications of this method to lattice QCD can be found in Refs.~\cite{Salg:2025now,Iskakov:2023zpc,Huang:2023gpb}.
While this paper is focused on the Nevanlinna-Pick interpolation proposed by Bergamaschi, {\it et al.}~\cite{Bergamaschi:2023xzx}, a similar first-principles method that can be directly applied to the time-dependent Green's functions of lattice QCD was recently introduced by Abbott {et al.}~\cite{Abbott:2025snz}.

An important issue when applying this method to lattice QCD data is the uncertainty of that data. In order for Nevanlinna-Pick interpolation to be applied, the input data must obey an analytic constraint needed for a bounded analytic interpolating function to exist.  This constraint, known as the Pick criterion, is so restrictive that it is highly unlikely that a specific set of $N$ lattice results for $G(z)$ will even be allowed. Here we develop a method to sample, with reasonable uniformity, possible data values that obey the Pick criterion and that lie within the known statistical and systematic errors of those lattice results. Such a large sample of ``Pick-consistent'' values for $G(z)$ at each of the $N$ values of $z$ for which lattice results were obtained can then be used to determine both $G(z)$ and the allowed variation of $G(z)$ within the bounds found for each of these Pick-consistent points. We propose that these variations provide the combined lattice and Nevanlinna-Pick interpolation errors.

In order to make this discussion more concrete and establish a relation between the errors being discussed and a possible physical prediction, we focus on a particular application of this approach, the inclusive decays of heavy particles. In the inclusive decay of a $\tau$ lepton into a $\tau$ neutrino plus a sum over multi-hadron states, a calculation of the branching ratio depends on a collection of the densities of hadronic states $\rho_\alpha(E)$ as a function of the center-of-mass (CoM) energy $E$ of those hadrons. Here the label $\alpha$ distinguishes, for example, the contributions from $J=1$, $J=0$ final states as well as those created by a weak vector or axial-vector current. Likewise the branching ratio for the inclusive decay of a hadron containing a charm or bottom quark into a lepton and lepton-neutrino pair plus hadrons can be determined from a similar collection of densities of states which will depend not only on the CoM energy of the hadronic state but also the momentum of that state in the rest system of the decaying particle.  

The analytic function $G(z)$ of interest in these cases is given by
\begin{equation}
    G(z) = \int_0^\infty d\omega\, \frac{\rho(\omega)}{\omega-z}
    \label{eq:G-def}
\end{equation}

where $\rho(\omega)$ is the density function of interest and $G(z)$ can be computed from lattice QCD in the case where $z$ lies on the positive imaginary axis. (For a more complete discussion of the association between possible Green's functions and a spectral density see Refs.~\cite{Bergamaschi:2023xzx,Abbott:2025snz}.) The real, non-negative spectral density $\rho(\omega)$ is defined on the real axis and will vanish below some positive threshold energy $E_{\mathrm{min}}$. The function $\rho(E)$ can also be used to determine an inclusive decay rate $\Gamma$ by evaluating
\begin{equation}
    \Gamma = \int_0^{E_{\max}} W(\omega) \rho(\omega) d\omega,
    \label{eq:Gamma}
\end{equation}

where the weight function $W(\omega)$ is typically a positive polynomial in the variable $\omega$ that vanishes when $\omega=E_{\max}$, the largest CoM energy possible for the emitted hadrons.

\begin{figure}
    \centering
    \includegraphics[width=0.7\linewidth]{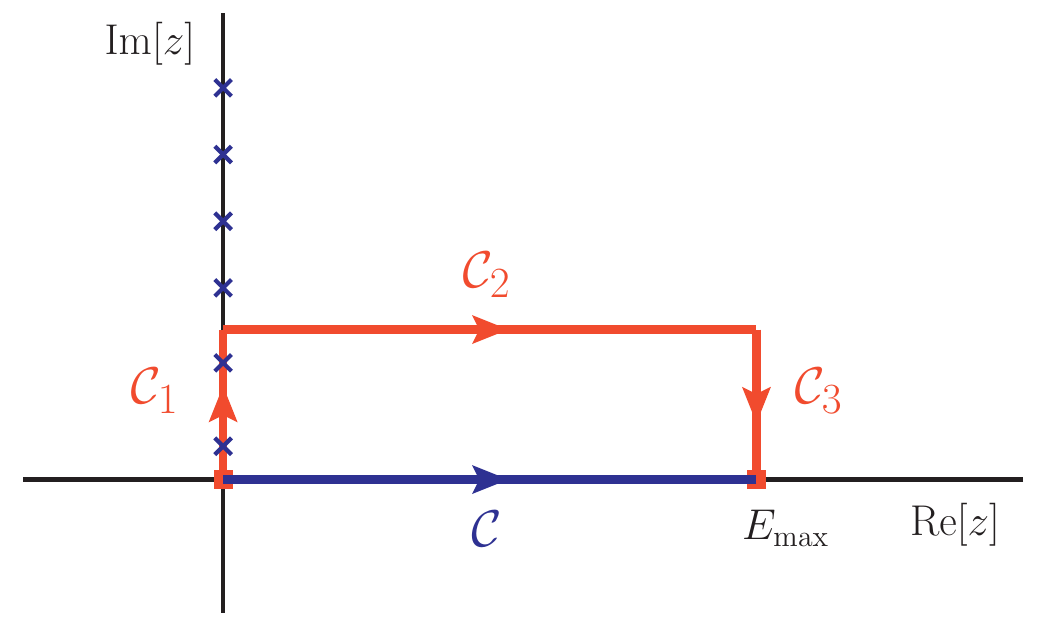}
    \caption{Sketch of the contour $\mathcal{C'} = \mathcal{C}_1\cup \mathcal{C}_2\cup\mathcal{C_3}$ in the complex plane that might be used in Eq.~\eqref{eq:contour2} to calculate inclusive heavy particle decay.  The crosses on the imaginary axis represent values of $z$ for which lattice data is available.}
    \label{fig:Contour}
\end{figure}

Here we consider determining $\Gamma$ by recognizing that $\rho(\omega) = \frac{1}{\pi}\,\operatorname{Im} G(\omega + i\epsilon)\big|_{\epsilon = 0^+}$ so that Eq.~\eqref{eq:Gamma} can be rewritten
\begin{eqnarray}
     \Gamma =\frac{1}{\pi}\int_0^{E_{\max}} W(\omega)\,\IM\left[ G(\omega+i\epsilon)\right]_{\epsilon = 0^+}
\, d\omega 
\label{eq:contour1}  \\
           &=&  \IM\left[\frac{1}{\pi}\int_{\mathcal{C}'} W(z) G(z) dz\right]  \label{eq:contour2}
\end{eqnarray}
where $\mathcal{C'} = \mathcal{C}_1\cup \mathcal{C}_2\cup\mathcal{C}_3$ is an integration contour in the complex plane constructed from the three contours, $\{\mathcal{C}_k\}_{1\le k \le 3}$ shown in Fig.~\ref{fig:Contour} while the original contour $\mathcal{C}$ is also shown in that figure. Cauchy's theorem implies that the integrals in Eqs.~\eqref{eq:contour1} and \eqref{eq:contour2} are equal.

It is the resulting errors for the integral over the contour $\mathcal{C}_2$ when evaluated using Nevanlinna-Pick interpolation that is the focus of this paper.  The integral over $\mathcal{C}_1$ is performed along the imaginary axis where the integrand might be directly computed using lattice QCD. The integral over the contour $\mathcal{C}_3$ is performed at the largest energy $E_{\max}$. Since the contour $\mathcal{C}_3$ touches the real axis, Nevanlinna-Pick interpolation will result in bounds for the integral which may diverge logarithmically, preventing its application here. However, this is the region in which QCD perturbation theory can be applied. 

This reference to heavy lepton or hadron decay is intended to focus on a quantity of physical importance and to identify a promising first target for the method being developed here.  Of course, the decay of the $\tau$ lepton, for example, has been successfully analyzed by a number of theoretical methods. The finite energy sum rule approach of Shifman, Vainshtein and Zakharov~\cite{Shifman:1978bx, Shifman:1978by, Shifman:1978bw, Braaten:1991qm, LeDiberder:1992zhd} uses QCD perturbation theory and a similar application of Cauchy's theorem to obtain results for $\tau$ decay that are accurate to a few percent and refined methods using lattice data~\cite{RBC:2018uyk} are even more precise. Additional relevant work on inclusive $\tau$ decays using the  Hansen-Lupo-Tantalo method is found in Ref.~\cite{ExtendedTwistedMass:2024myu}.

This paper is organized as follows. In Sec.~\ref{sec:NP-overview} we provide, for completeness, a brief overview of the Nevanlinna–Pick interpolation method described in Ref.~\cite{Bergamaschi:2023xzx}. Section~\ref{sec:PickCriterion} examines the challenges involved in interpolating the entire $2N$-dimensional volume determined by the lattice results and the errors on those results. We present a prescription for identifying points in this error-allowed region that can be used in the interpolation procedure and characterize the geometry of region in which these must points lie. Next, in Sect.~\ref{sec:error} we focus on determining the integral over the contour $\mathcal{C}_2$ as the result of this Nevanlinna-Pick interpolation. We present a procedure for computing the result of the interpolation and its associated error -- an error which combines the errors present in the lattice data with those that are introduced by the uncertain interpolation. Then in Sec.~\ref{sec:demo} we demonstrate this procedure using a simple example and study in some detail how the propagated errors depend on the size of the input lattice errors, the number of lattice results that enter the interpolation and the location on the imaginary axis of the lattice data that is used. Special attention is paid to the statistical properties of the errors resulting from the Nevanlinna-Pick interpolation. Finally, in Sect.~\ref{sec:conclusion} we conclude with a further discussion and summary of the results and suggestions for the next steps in applying this method.

\section{Nevanlinna-Pick interpolation}
\label{sec:NP-overview}

In this section we summarize the method we follow to perform Nevanlinna-Pick interpolation. This discussion follows closely to the approach taken in Refs.~\cite{Bergamaschi:2023xzx, 38d0ce1ad0c24c60bb91073615e8f1a5} with minor modifications in notation, and is presented here to make our discussion more self-contained. The reader is referred to Ref.~\cite{Bergamaschi:2023xzx} for more complete explanations and insights.

Nevanlinna-Pick interpolation is economically formulated for analytic functions $f(\zeta)$ which map the complex unit disk, $\mathbb{D}$, into itself. Such functions are said to belong to the Schur class $\mathcal{S}$. These methods can be applied to our problem because the Green's function $G(z)$ is analytic and Im$[G(z)] \ge 0$ in the upper half plane. This implies that $G(z)$ can be viewed as a map from the upper half of the complex plane, $\mathbb{C}^+$, into itself. We can use a Cayley transformation $C$ that relates $\mathbb{C}^+$to $\mathbb{D}$ to relate $G(z)$ to a corresponding mapping of $\mathbb{D}$ into itself. The needed Cayley transformation $C:\mathbb{C}^+ \to \mathbb{D}$ and its inverse are given by:
\begin{equation}
C(z)=\frac{z-i}{z+i} \quad \mathrm{and}\quad C^{-1}(\zeta) = -i \frac{\zeta+1}{\zeta-1}.
  \label{eq:Cayley}  
\end{equation}
Here we adopt the notation of Ref.~\cite{Bergamaschi:2023xzx} where $z \in \mathbb{C}^+$ and $\zeta\in\mathbb{D}$. Thus, to our Green's function $G(z)$ we can then associate a Schur class function $f(\zeta)$ according to
\begin{equation}
    f(\zeta) = C\left(G\left(C^{-1}(\zeta)\right)\right) \quad \mathrm{or} \quad 
    G(z) = C^{-1}\left(f\left(C(z)\right)\right).
\end{equation}
and the interpolation problem posed by the $N$ points $z_n \in \mathbb{C}^+$ for $1\le n \le N$ and the corresponding $N$ values $G_n = G(z_n)$ for $1\le n \le N$ corresponds to points $\zeta_n$ and $\Gamma_n$ in the disk:
\begin{equation}
    \zeta_n = \frac{z_n-i}{z_n+i} \quad \mathrm{and} \quad 
    \Gamma_{n} = \frac{G_n-i}{G_n+i}.
\end{equation}

Now working with Schur-class functions defined on the disk, we ask for all such functions $f(\zeta)$ which obey the $N$ conditions:
\begin{equation}
 f(\zeta_n) = \Gamma_n \quad \text{for} \quad n=1,2...N.  
\label{eq:interpolation}
\end{equation}
Such a function $f(\zeta)$ is an interpolation of the $N$ pairs $(\Gamma_n,\zeta_n)$, $1 \le n \le N$.

Perhaps the first question to answer is whether there are any Schur-class functions which obey Eq.~\eqref{eq:interpolation}?  This question is answered by the Pick criterion.  A Schur-class function $f(\zeta)$ obeying the $N$ conditions in Eq.~\eqref{eq:interpolation} exists if the Pick matrix:
\begin{equation}
    \Lambda= \begin{pmatrix}
\frac{1-{\Gamma_1}\overline{\Gamma_1}}{1-\zeta_1 \overline{\zeta_1}} & ...&  \frac{1-{\Gamma_1}\overline{\Gamma_N}}{1-\zeta_1 \overline{\zeta_N}}\\
\vdots& & \vdots\\
\frac{1-\Gamma_N\overline{\Gamma_1}}{1-\zeta_N \overline{\zeta_1}} & ... & 
\frac{1-\Gamma_N\overline{\Gamma}_N}{1-\zeta_N \overline{\zeta_N}}
\end{pmatrix}
\label{eq:pick_matrix}
\end{equation}
is positive semi-definite. Equivalently, each eigenvalue of this hermitian matrix must be non-negative. If all the eigenvalues are positive then there are an infinite number of solutions which are determined by the Nevanlinna-Pick interpolation procedure. If $K$ eigenvalues are zero then there is only a single solution which is a rational function whose degree equals the rank of the
Pick matrix and this function is determined by the Nevanlinna-Pick interpolation procedure.

The interpolation procedure relies heavily on the properties of Blaschke factors, where a Blaschke factor is a M\"obius transformation of the form:
\begin{equation}
b_a(\zeta)=\frac{|a|}{a}\frac{a-\zeta}{1-\overline{a} \zeta}
    \label{blaschke}.
\end{equation}
The Blaschke factor $b_a(\zeta)$ is an analytic function of $\zeta$ which maps $\mathbb{D}$ into itself and which vanishes at $\zeta = a$. Exploiting the property $b_a(a) = 0$, one can construct a collection of Schur-class functions $f(\zeta)$ that obey the condition $f(\zeta_1) = \Gamma_1$ by writing:
\begin{equation}
\label{eq:mobius_transform}
  f(\zeta)=  \frac{b_{\zeta_1}(\zeta)f_1(\zeta)+\Gamma_1}{1+\overline{\Gamma_1}b_{\zeta_1}(\zeta)f_1(\zeta)}.
\end{equation}where $f_1(\zeta)$ is an arbitrary Schur-class function. It is clear that $f(\zeta_1)=\Gamma_1$ and possible to show that all Schur-class functions obeying $f(\zeta_1)=\Gamma_1$ can be written in this form.

Since Eq.~\eqref{eq:mobius_transform} expresses $f(\zeta)$ as a M\"obius transformation of $f_1(\zeta)$ and M\"obius transformations map circles into circles, the largest possible values of $f_1(\zeta)$, the circle $|f_1(\zeta)|=1$, map into a circle inside the disk implying that the range of $f(\zeta)$ is explicitly reduced by Eq.~\eqref{eq:mobius_transform}. This restriction might be viewed as a first example of Nevanlinna-Pick interpolation.  

Since the transformation in Eq.~\eqref{eq:mobius_transform} can be inverted, a second condition that $f(\zeta_2) = \Gamma_2$ can be written as a condition that the new function $f_1(\zeta)$ must obey at $\zeta=\zeta_2$. We can then repeat the previous procedure and write $f_1(\zeta)$ as an analogous function of $f_2(\zeta)$. This will restrict the range of $f_1(\zeta)$ which will further restrict the range of $f(\zeta)$. Our $N$ conditions can then be applied sequentially and an explicit formula written for $f(\zeta)$ which incorporates all $N$ constraints.

This nesting of products of M\"obius transformations can be reduced to $2\times2$ matrix products by exploiting the relation between M\"obius transformations and regular complex $2\times2$ matrices. We adopt the notation that a M\"obius transformation, defined by the 4 complex variables $a$, $b$, $c$ and $d$, transforming a variable $\zeta$ can be written as:
\begin{equation}
    \left(\begin{array}{cc} a & b \\ c & d \end{array}\right)\, \circledstar\,\zeta = \frac{a\zeta +b}{c + d\zeta}
    \label{eq:Mob-equiv-mat}
\end{equation}
where we introduce the symbol $\circledstar$ to indicate the nonlinear M\"obius transformation determined by our four parameters and written out on the right-hand side of Eq.~\eqref{eq:Mob-equiv-mat}.  It is easy to show that:
\begin{equation}
    \left(\begin{array}{cc} a' & b' \\ c' & d' \end{array}\right)\,\circledstar\,\left\{ \left(\begin{array}{cc} a & b \\ c & d \end{array}\right)\, \circledstar\,\zeta\right\}
    =
    \left\{\left(\begin{array}{cc} a' & b' \\ c' & d' \end{array}\right)\,\times\, \left(\begin{array}{cc} a & b \\ c & d \end{array}\right)\right\}\, \circledstar\,\zeta
\end{equation}
where the $\times$ symbol on the right-hand side indicates ordinary matrix multiplication, showing the equivalence of composing M\"obius transformations and multiplying the corresponding complex matrices.

Following the conventions in Ref.~\cite{Bergamaschi:2023xzx} we can rewrite Eq.~\eqref{eq:mobius_transform} as:
\begin{equation}
    f(\zeta) =  U_1(\zeta)\, \circledstar\, f_1(\zeta)
\end{equation}
where the matrix $U_1$ is given by
\begin{equation}
    U_1(\zeta) =  \frac{1}{\sqrt{1-\lvert \Gamma_1 \lvert ^2}} \begin{pmatrix} 
b_{\zeta_1}(\zeta) & \Gamma_1 \\
\overline{\Gamma_1}b_{\zeta_1}(\zeta) & 1 
\end{pmatrix}.
\end{equation}
Note the convenient normalization introduced in this definition of $U_1$ implies that $\mathrm{det} \left[U_1(\zeta)\right] = b_{\zeta_1}(\zeta)$ but has no effect on the M\"obius transformation specified by $U_1$.

Using this notation we can then inductively determine the sequence of matrices $\{U_n\}_{1\le n \le N}$ needed to explicitly impose the $N$ normalization conditions obeyed by $f(\zeta)$ by writing:
\begin{equation}
    \label{eq:f_U}
f(\zeta)=\left\{U_1(\zeta)U_2(\zeta)...U_n(\zeta)\right\}\,\circledstar\,f_n(\zeta),
\end{equation}
for $1\le n \le N$.
Here the matrix $U_n(\zeta)$ is given by:
\begin{equation} \label{eq:u_n}
   U_n(\zeta)= \frac{1}{\sqrt{1-\lvert \Gamma_n^\prime \rvert ^2}} \begin{pmatrix} 
b_{\zeta_n} (\zeta) & \Gamma_n^\prime \\
\overline{\Gamma_n^\prime}b_{\zeta_n} (\zeta)& 1 
\end{pmatrix}.
\end{equation}
where
\begin{equation}
    \Gamma_n^\prime = f_{n-1}(\zeta_n) =  \left\{U_{n-1}^{-1}(\zeta_n)U_{n-2}^{-1}(\zeta_n)\ldots U_{1}^{-1} U_0^{-1} (\zeta_n)\right\}\, \circledstar\,\Gamma_n
\end{equation}
with $U_0 = I$.

It is useful and customary to re-write Eq.~\eqref{eq:f_U} for the case $n=N$ by parameterizing the product of the $N$, $U_n$ matrices appearing in that equation as follows:
\begin{equation}
   f(\zeta)= \begin{pmatrix} 
P_N(\zeta) & Q_N(\zeta)  \\
R_N(\zeta)& S_N(\zeta) 
\end{pmatrix}\, \circledstar\,
f_N(\zeta)\label{eq:Wert-1}
\end{equation}
where the polynomials $P_N(\zeta), Q_N(\zeta), R_N(\zeta)$, and $S_N(\zeta)$ are the Nevanlinna coefficients. The values taken by the remaining unknown Schur-class function $f_N(\zeta)$ at a fixed value of $\zeta$ must cover the unit disk while the M\"obius transformation in Eq.~\eqref{eq:Wert-1} maps this disk onto a second disk contained in the unit disk for each value of $\zeta$. This second disk is labeled $\Delta_N(\zeta)$. These values, consistent the $N$ interpolation conditions, are explicitly parameterized by the complex function $T_{N,\zeta}(\eta)$ where $\eta \in \mathbb{D}$, a possible value for $f_N(\zeta)$ and 
\begin{equation}
    T_{N,\zeta}(\eta) = \frac{P_N(\zeta)\eta +Q_N(\zeta)}{R_N(\zeta)\eta+S_N(\zeta)}.
    \label{eq:Wert2}
\end{equation}

One can show that the center of $\Delta_N(\zeta)$ is given by
\begin{equation}
    c_N(\zeta)=\frac{Q_N(\zeta) \overline{S_N(\zeta)} -P_N(\zeta)\overline{R_N(\zeta)}}{\lvert S_N(\zeta) \rvert ^2 -\lvert R_N(\zeta) \rvert ^2}
\end{equation}
and the radius, $r_N(\zeta)$ is
\begin{equation}
    r_N(\zeta)=\frac{\lvert P_N(\zeta)S_n(\zeta)-Q_N(\zeta)R_n(\zeta) \rvert}{\lvert S_N(\zeta) \lvert ^2-\lvert R_N(\zeta) \rvert ^2}.
\end{equation}
This disk at the point $\zeta$ describes the range of allowed values for all functions $f(\zeta)$ which obey the interpolation condition and is called the Wertevorrat (pl. Wertevorr\"ate) corresponding to the point $\zeta$. The Wertevorrat provides a bound on all possible analytic continuations of the given finite data set, without needing to know anything about the function $f_N(\zeta)$ beyond its lying in the unit disk. Thus, we can solve for $c_n(\zeta)$ and $r_N(\zeta)$ using the above equations to explicitly determine a set of points that lie on the boundary of the Wertevorrat for each desired choice of $\zeta$. 

The final step is to determine the corresponding bounds for our desired function in the complex plane. We use the inverse Cayley transformation of Eq.~\eqref{eq:Cayley},
\begin{equation}
    \zeta \mapsto  z
\end{equation}
    
\begin{equation}
    \Delta_N^\mathbb{D} (\zeta) \mapsto\Delta_N^\mathbb{C} (z) .
\end{equation}
Where $\Delta_N^\mathbb{D}$ and $\Delta_N^\mathbb{C}$ denote the Wertevorrat defined in the complex disk and plane respectively. In practice, we apply the transformation to a finite set of points sampled along the boundary of each Wertevorrat.

The final topic discussed in this section is the character of numerical work which exploits the Nevanlinna-Pick interpolation. If $N$ data values from a lattice QCD calculation are to be interpolated using these methods, we must first establish that this data obeys the Pick criterion, that is that the $N\times N$ matrix $\Lambda$ given in Eq.~\eqref{eq:pick_matrix} is positive semidefinite. This is a remarkable condition. When the number of points $N$ is more than a few, the Pick-allowed region can easily be 30 orders of magnitude smaller in some directions than in others as is shown in Sec.~\ref{sec:PickCriterion}. Correct evaluation of this condition then requires extraordinarily precise arithmetic. Higher precision requirements were also found in Ref.~\cite{PhysRevLett.126.056402}. 

Consequently all calculations performed when working with complex variables defined in the unit disk, including the Cayley transformations between the complex unit disk and the complex plane were conducted using extended precision arithmetic to an accuracy of 150 decimal digits. This extreme numerical precision was used to ensure that precision was not a factor in our findings. High numerical precision was achieved through the use of the mpmath \cite{mpmath} library in Python as it supports arbitrary-precision complex floating-point arithmetic.  We verified that this level of precision was sufficient by increasing the precision to 160 decimal digits and finding consistent results.

\section{Satisfying the Pick criterion within computational errors}
\label{sec:PickCriterion}

In this section we first describe the simplified example spectral density which we study. After choosing the location of $N$ mock lattice data points $\{z_n\}_{1 \le n \le N}$ that lie on the imaginary axis we assume that our lattice data at these $N$ points lie within an error volume centered at each of the exact lattice results $G(z_n)$. Second, we describe the challenge of finding a representative sample of possible lattice results that lie within this error volume and also obey the Pick criterion so that they are consistent with the assumed analytic properties of $G(z)$. Third, we propose a method to find such a set of Pick-consistent results whose distribution reasonably samples the assumed error volume. Finally, we synthesize the results obtained in this section to describe the geometry of the Pick-consistent region which we are trying to sample.

\subsection{Example spectral density}

We analyze a single example with a known continuous spectral density function $\rho(\omega)$ given by
\begin{equation}
\label{eq:example-rho}
    \rho(\omega)= \left\{\begin{array}{cl} \sum_{i=1}^2 \frac{1}{\sqrt{2\pi}\sigma} e^{\frac{-(\omega -\mu_i)^2}{2\sigma^2}} & \omega \ge \omega_\mathrm{min} \\
    0 & \omega < \omega_\mathrm{min} \end{array} \right.
\end{equation}
with $\sigma$ = 0.1 and $\mu_i=\{ 0.25, 0.75 \}$ as in Ref.~\cite{Bergamaschi:2023xzx}.  In contrast to the paper of Bergamaschi, {\it et al.} we introduce a gap $\omega_\mathrm{min}=0.1$, separating the energy spectrum from zero as would appear in a physical case.

This spectral density $\rho(\omega)$ appears in a Euclidean-time correlation function $\mathcal{C}(\tau)$:
\begin{equation}
   \mathcal{C}(\tau) = \int_0^\infty d\omega \rho(\omega) e^{-\tau\omega}
   \label{eq:EuclidCorr}
\end{equation}
that can be approximately calculated using lattice QCD. 
The Green's function $G(z)$ introduced in Eq.~\eqref{eq:G-def} evaluated on the positive imaginary axis can then be obtained by performing a Laplace transform on the function $\mathcal{C}(\tau)$:
\begin{eqnarray}
\int_0^\infty \mathcal{C}(\tau) e^{i\nu\tau} d\tau
    &=& \int_0^\infty d\omega \frac{\rho(\omega)}{ \omega - i\nu} \label{eq:Laplace1}\\
    &=& G(i\nu).  \label{eq:Laplace2}
\end{eqnarray}
We will assume that the Laplace transform on the left-hand side of Eq.~\eqref{eq:Laplace1} has been evaluated at the locations $z_n = i \nu_n$ of our $N$ mock lattice data points, for the $N$ positive energies $\nu_n$. We assume that the actual $n^{th}$ data value corresponding to the energy $\nu_n$ will lie within an error volume centered at the exact result $G_n = G(i\nu_n)$. (In the following we will refer to the results $G_n$ as the exact values.) We then want to interpolate the function $G(z)$ from the $N$ conditions $G_n = G(z_n)$ to provide results for $G(z)$, also with known uncertainties, for values of $z$ away from the imaginary axis in the upper half-plane $\mathbb{C}^+$.  

We assign an error to each of the $N$ mock lattice data points by assuming that each has the same absolute error.  This common error, $\sigma_{\mathrm{Lat}}$, is assigned to both the real and imaginary parts of each component of the $N$ complex Green's function values is obtained as follows.  We first find the average magnitude of the $N$ exact values $G_n$.  The common error $\sigma_{\mathrm{Lat}}$ is obtained by multiplying that average magnitude by a fixed scale factor $\xi$, which we will refer to as the error scale.  Below we use the error scale $\xi$ to characterize the assigned error.  We view $\xi = 0.01$ as giving a $1\%$ error.  Thus, for the purpose of these studies, we model the error volume for our lattice data as a simple hypercube with sides of length $2\sigma_{\mathrm{Lat}}$. While we choose this description for this initial study, in reality the correlations between the errors on actual lattice results will make a more complicated 2N-dimensional solid which may have interesting effects on the results presented here.

In an actual calculation, the exact values $G_n$ will not be known and the results of the calculation should be uniformly distributed within our assumed error volume. Thus, we generate samples by choosing $M$ random, $N$-dimensional complex vectors $\widetilde{G}^m$ uniformly distributed within the specified error volume. We apply the Cayley transform of Eq.~\eqref{eq:Cayley} to each component $\widetilde{G}^m_n$ of this sample lattice result and the eigenvalues of the Pick matrix are calculated to determine if that sample satisfies the Pick criterion. However, we will usually work with the sample data transformed back to the complex plane to analyze the results in a more familiar, physical environment.

\subsection{Existence of solutions}

It should be expected that as the number of lattice data points $N$ grows, the volume of the Pick consistent region within our error volume will decrease.  However, this decrease is precipitous, and even for $N=10$ very few of our random samples of possible lattice data will obey the Pick criterion. In order to understand this phenomena better, we begin by examining a less extreme case of interpolating from results obtained at four lattice data points and show graphically how the error volume is divided into a region which is Pick-consistent and one which is not. For the purpose of this study, we stochastically vary only two of the $2N$ real parameters within the error volume while the remaining $2N-2$ components are kept fixed at their exact values. 

\begin{figure}[hbt!]
\centering
\includegraphics[width=0.32\linewidth]{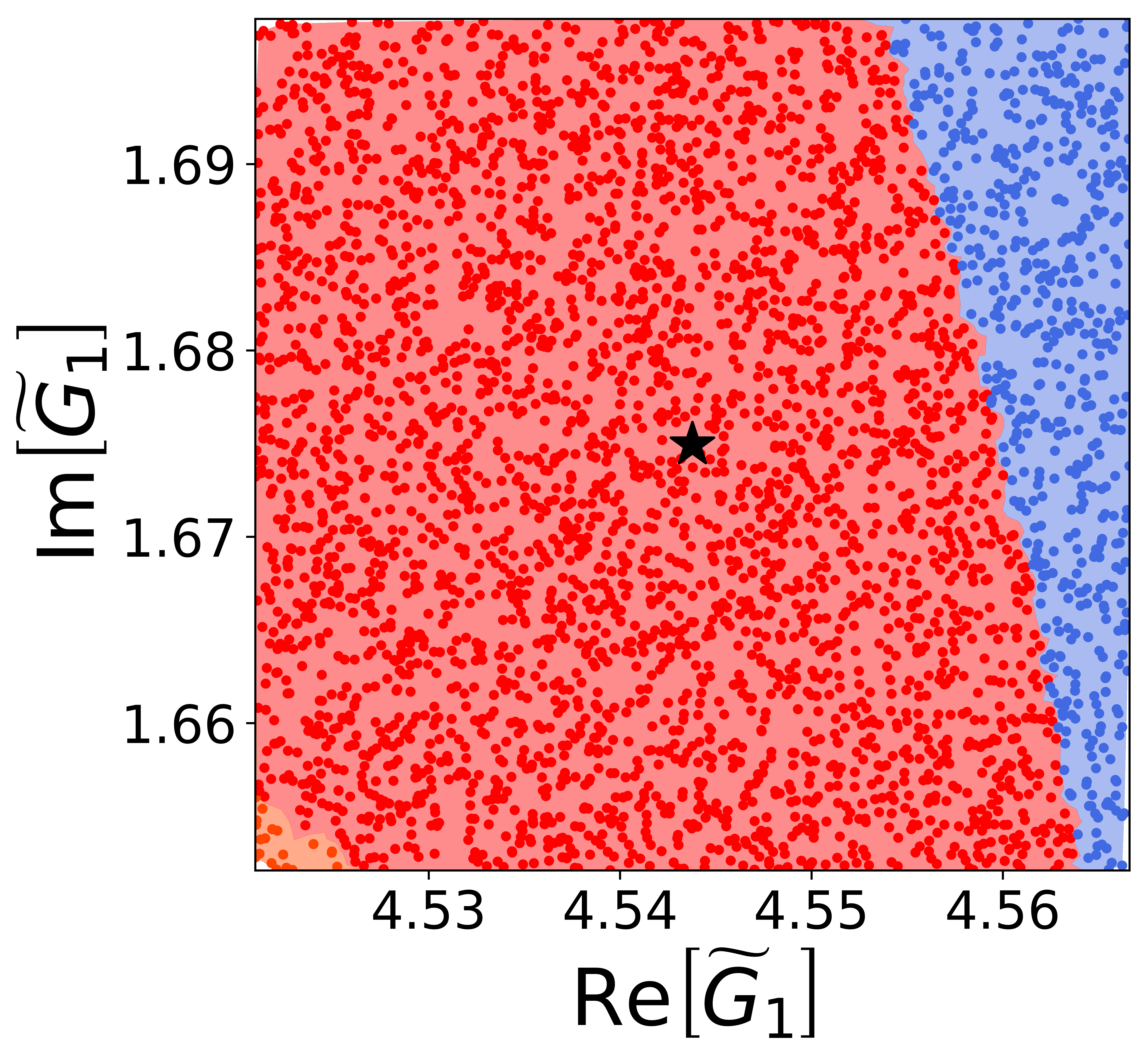}
\includegraphics[width=0.32\linewidth]{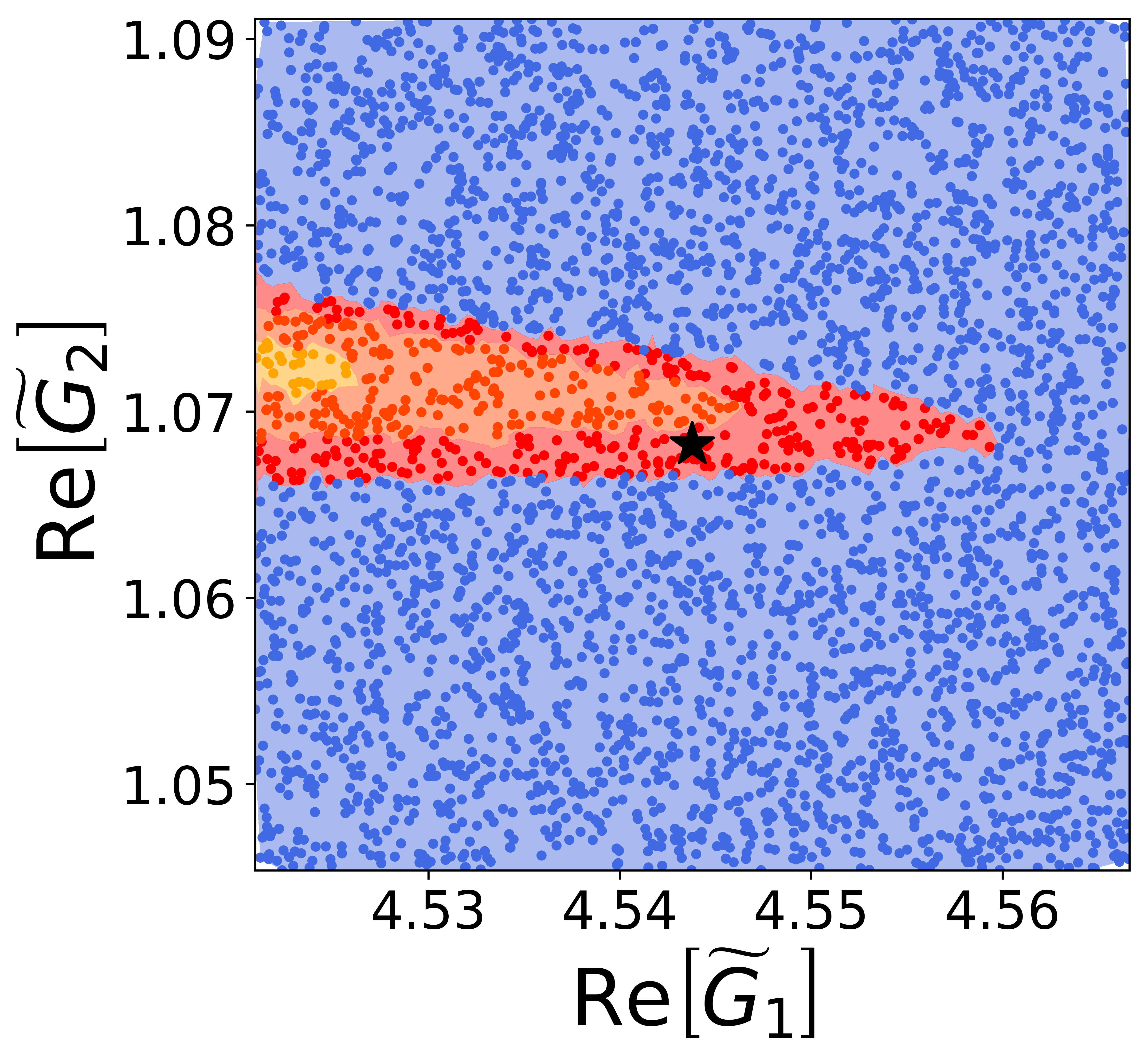}
\includegraphics[width=0.32\linewidth]{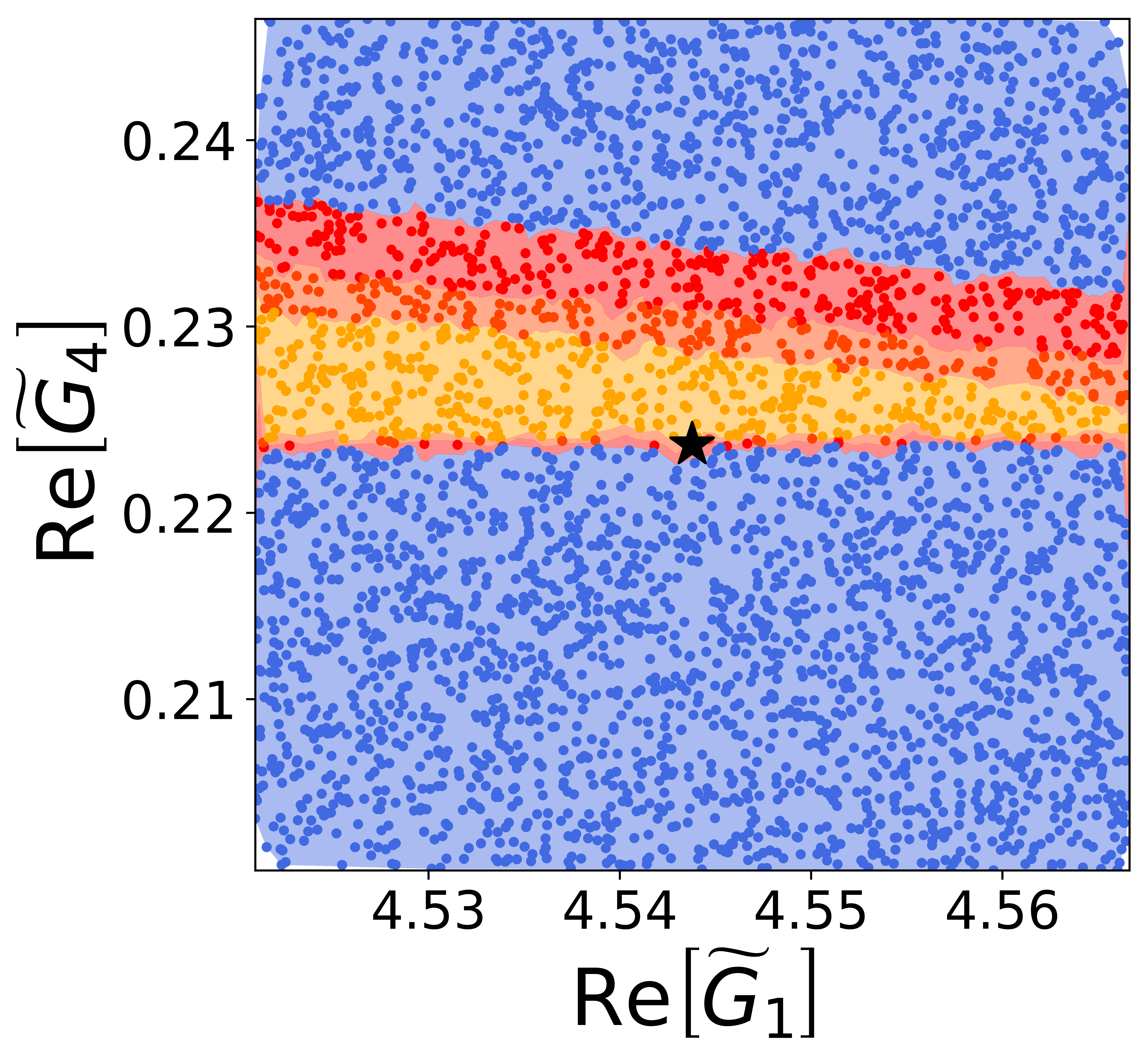}
\includegraphics[width=0.32\linewidth]{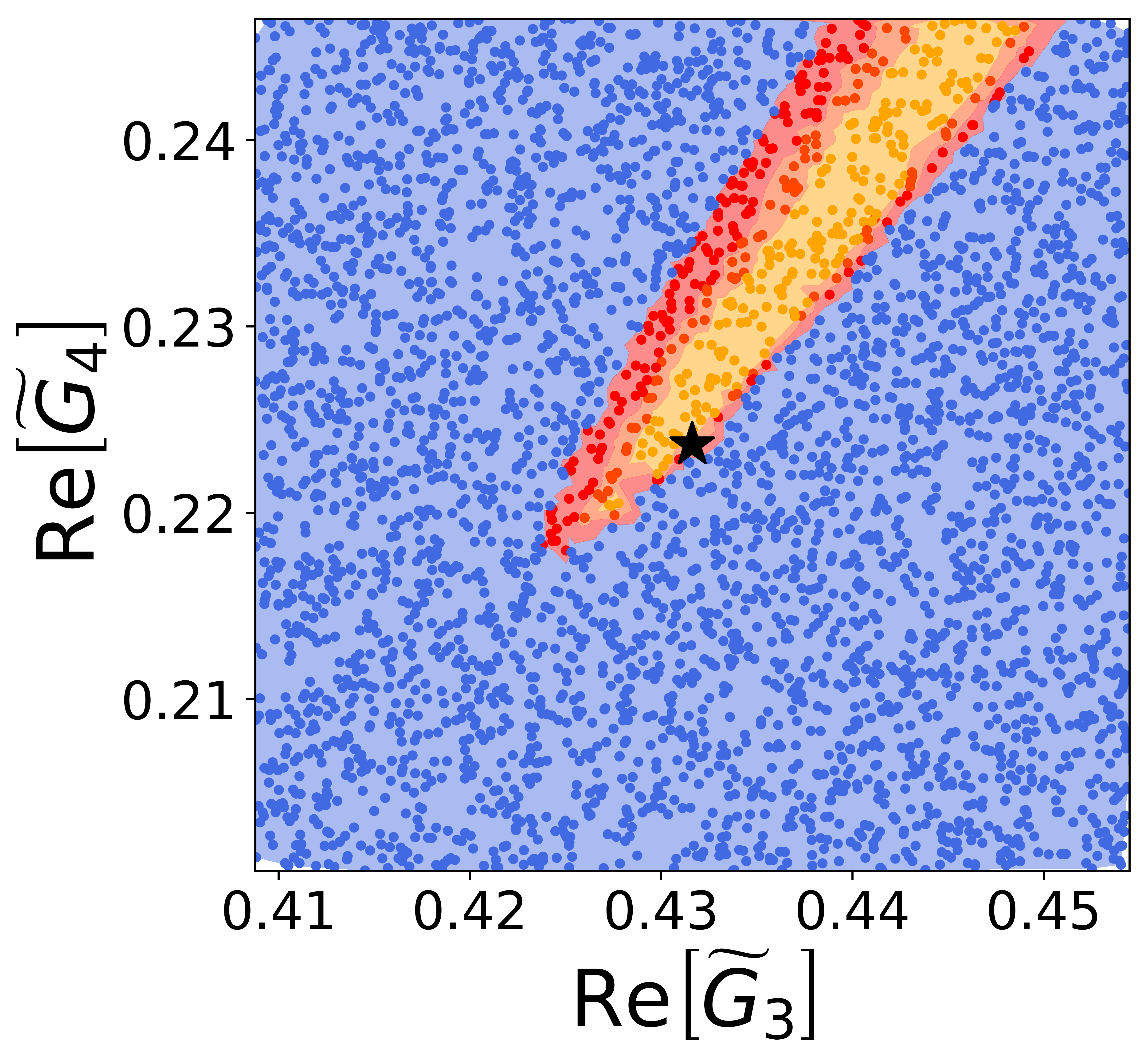}
\includegraphics[width=0.32\linewidth]{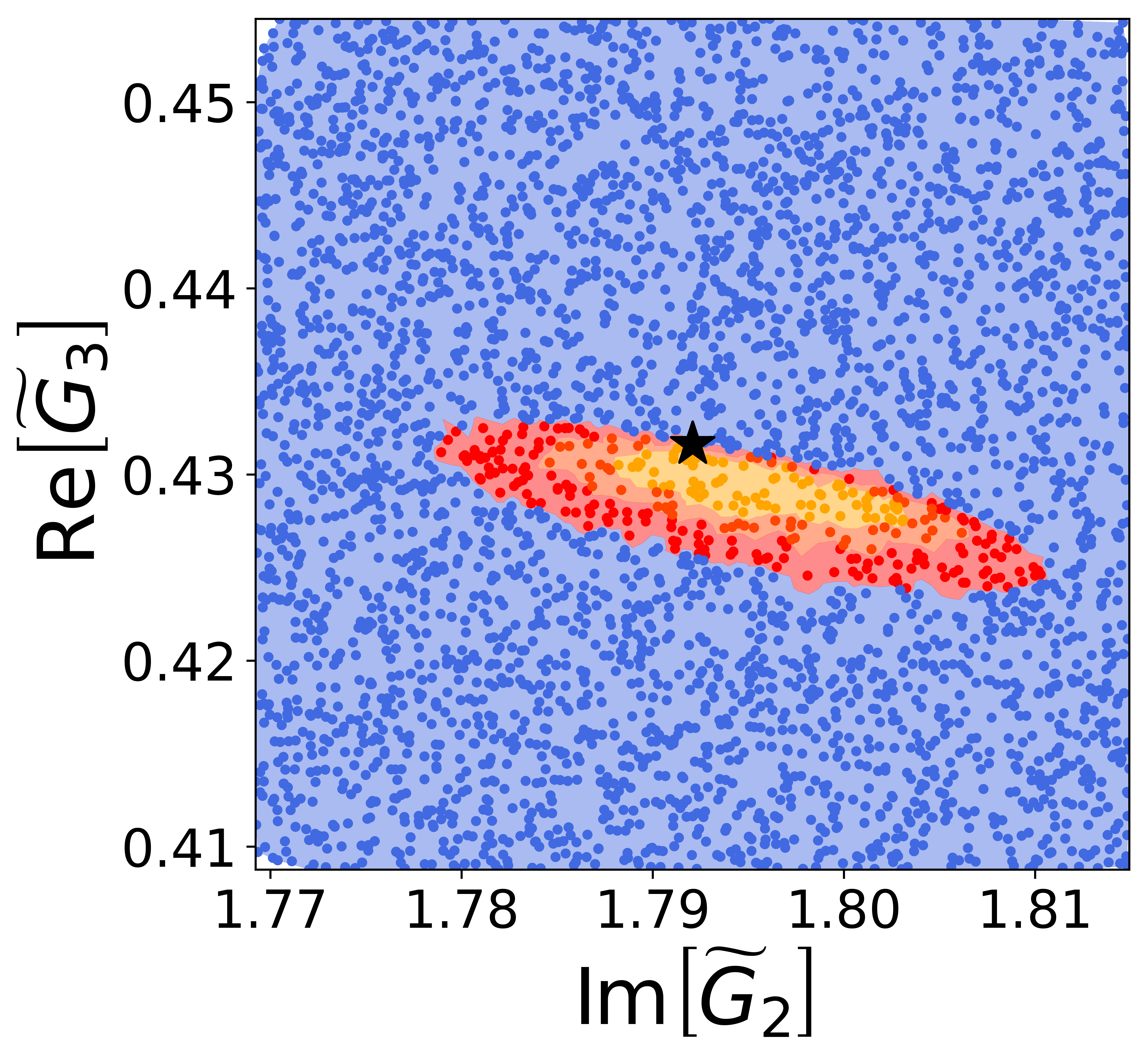}
\includegraphics[width=0.32\linewidth]{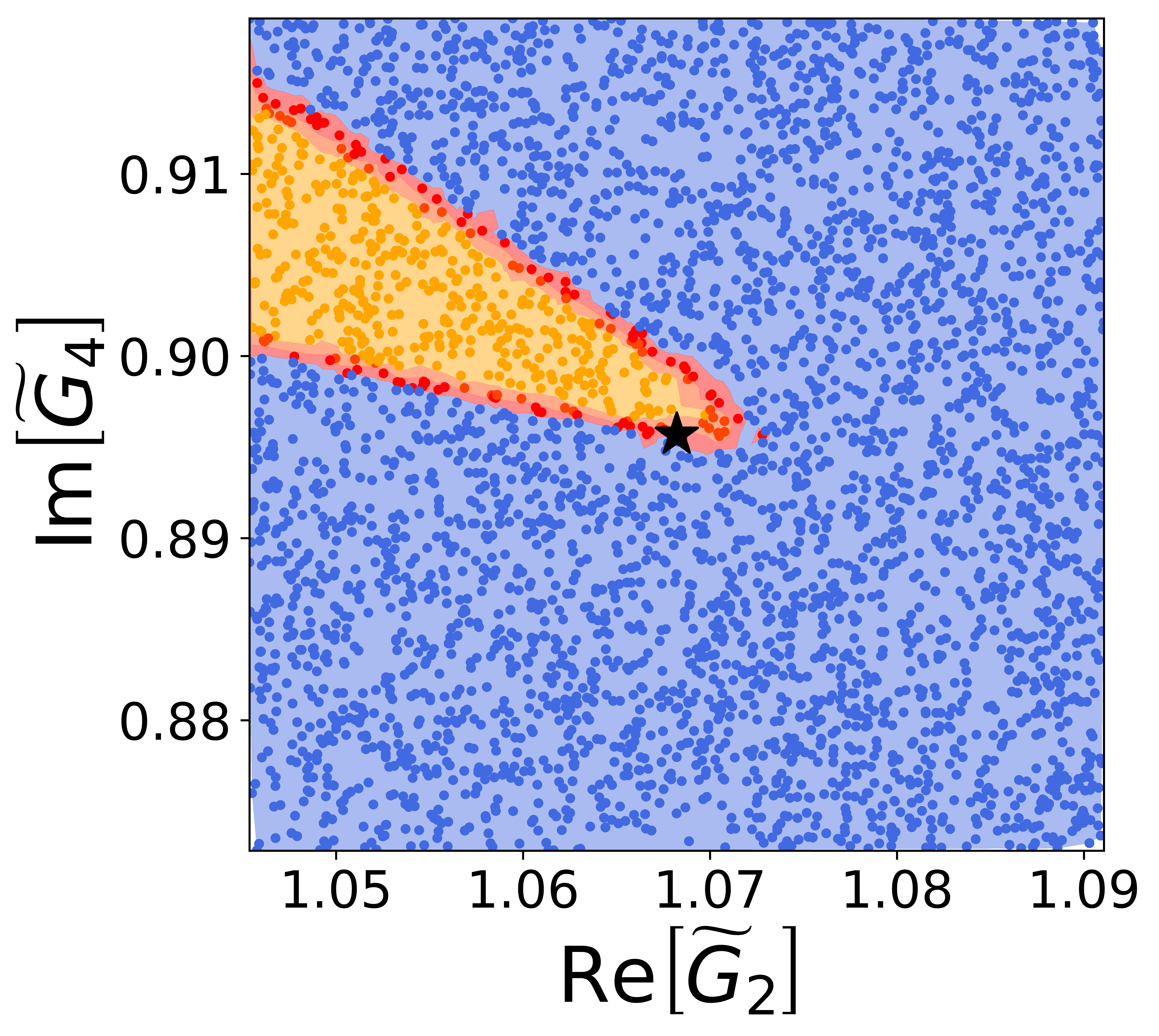}
\includegraphics[width=0.32\linewidth]{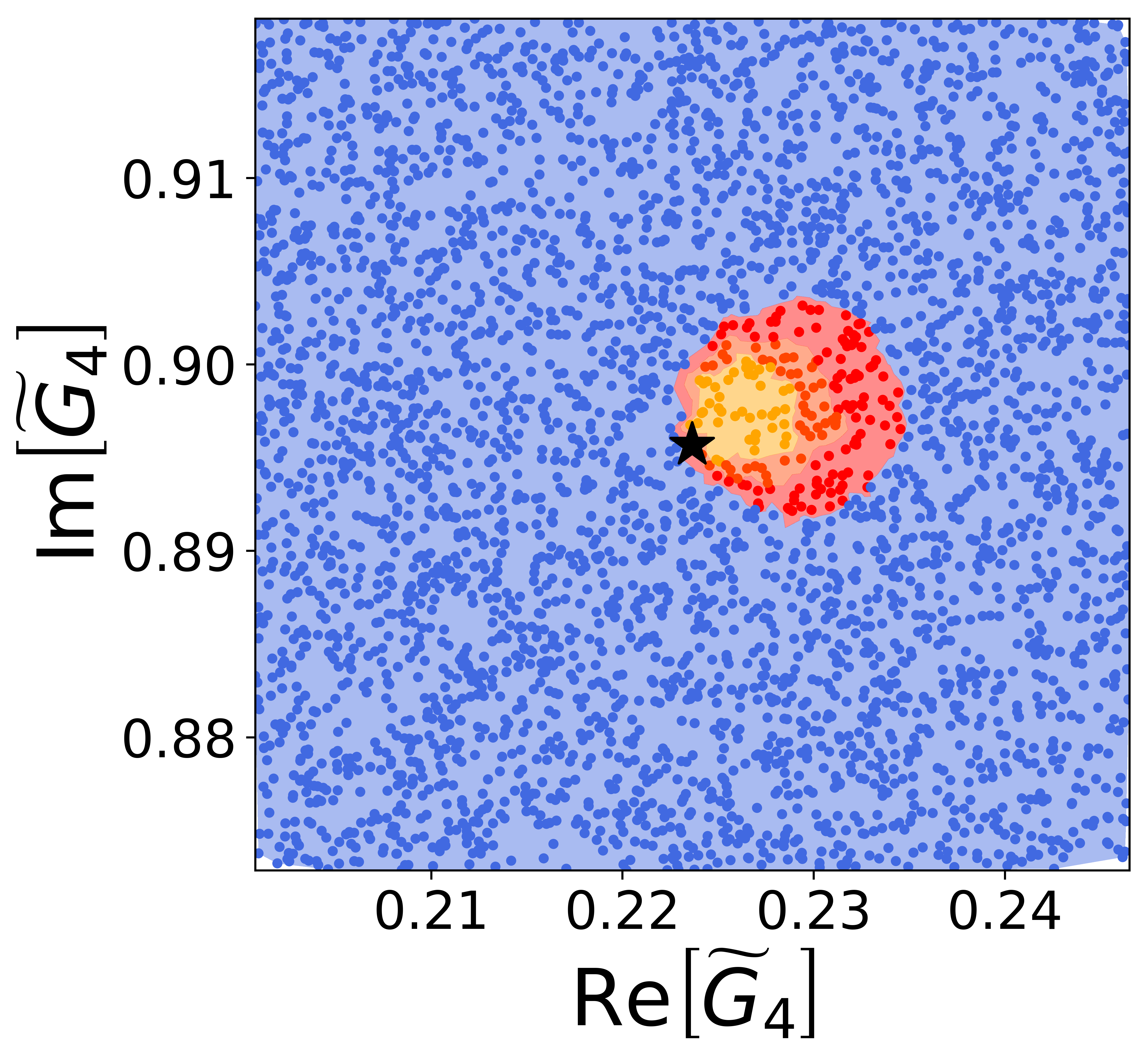}
\raisebox{.6\height}{\includegraphics[width=0.5\linewidth]{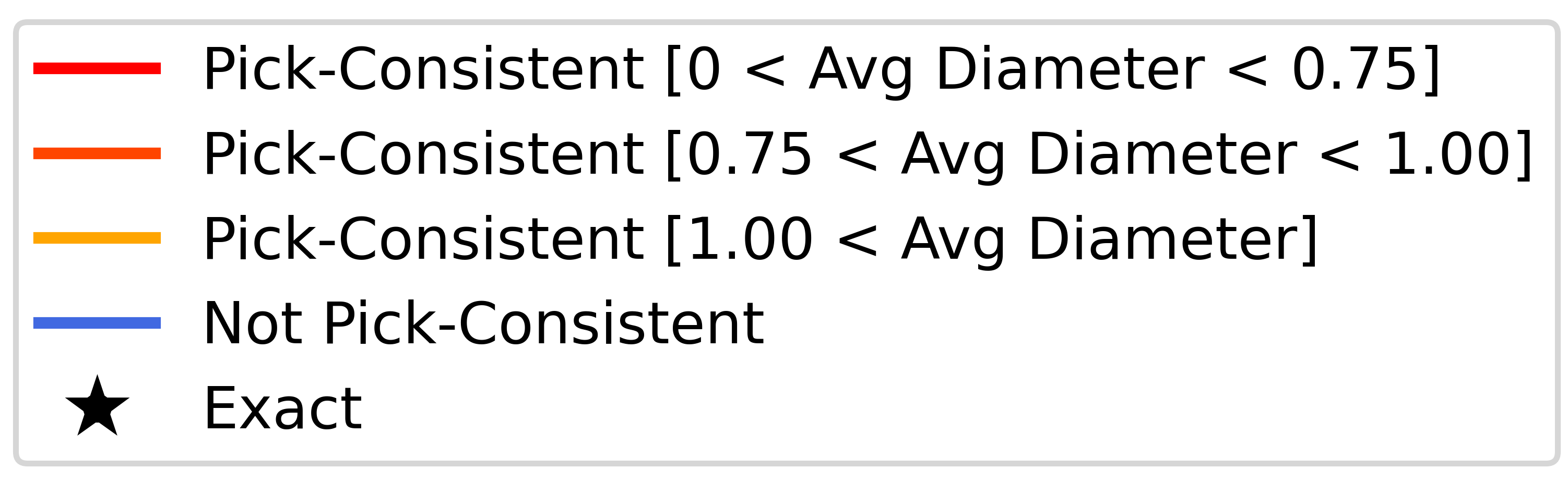}}
    
    \caption{Cross-section plots from tests of Pick consistency for the case of interpolating from four lattice data points. The plots show seven independent tests in which pairs of trial Green's functions values were varied randomly within an error volume corresponding to an error scale $\xi=0.01$ while the other six results were held fixed at their exact values. Trial data which did not obey the Pick criterion are blue. Pick-consistent points were given three different colors depending on the average diameter of the Wertevorr\"ate for the vector of complex Green's function value $\widetilde{G}^m$ corresponding to that point. Specifically, for each $\widetilde{G}^m$, we evaluated the Wertevorr\"ate on $z \in [0 + 0.1i,\, 1.5 + 0.1i]$ and computed the average diameter over the segment. Each plot shows 4000 random samples.}
    \label{fig:cross_section_plots}
\end{figure}

Figure~\ref{fig:cross_section_plots} presents examples of cross-section plots for this $N=4$ case in which two of the 8 real parameters within the error volume were stochastically sampled. For each of the seven plots shown, the two components of the complex vector $\widetilde{G}^m$ that were varied are identified as a point in the graph. Thus, the coordinate axes for the graph show which pair of indices were randomly varied in each of the seven independent studies. The color of the plotted point indicates whether or not that sample was Pick-consistent. 

As can be seen, even for this case of a small number of energies on the imaginary axis at which lattice results were obtained, there is a substantial variation in the width of the Pick-consistent region among the $2N$ dimensions. For the case of the real and imaginary parts of $\widetilde{G}_1$ corresponding to the location $z_1=0.1i$ the Pick-consistent region fills almost the entire error volume.  However, for the case of coordinate 4 and $\widetilde{G}(z_4)$ ($z_4=2.0i$) only $\approx5\%$ of the error volume is Pick-consistent.

For the Pick-consistent points shown in Fig.~\ref{fig:cross_section_plots} we use color to provide additional information about the average diameter of the Wertevorr\"ate that would correspond to the values of that $m^{th}$ sample value $\widetilde{G}(z_n)^m$. One can see that near the boundary of the Pick-consistent region the Wertevorr\"ate are the smallest as should be expected since when the determinant of the Pick matrix vanishes (which by definition will be at the boundary) the interpolated function is unique.

\begin{figure}[hbt!]
    \centering
\includegraphics[width=0.32\linewidth]{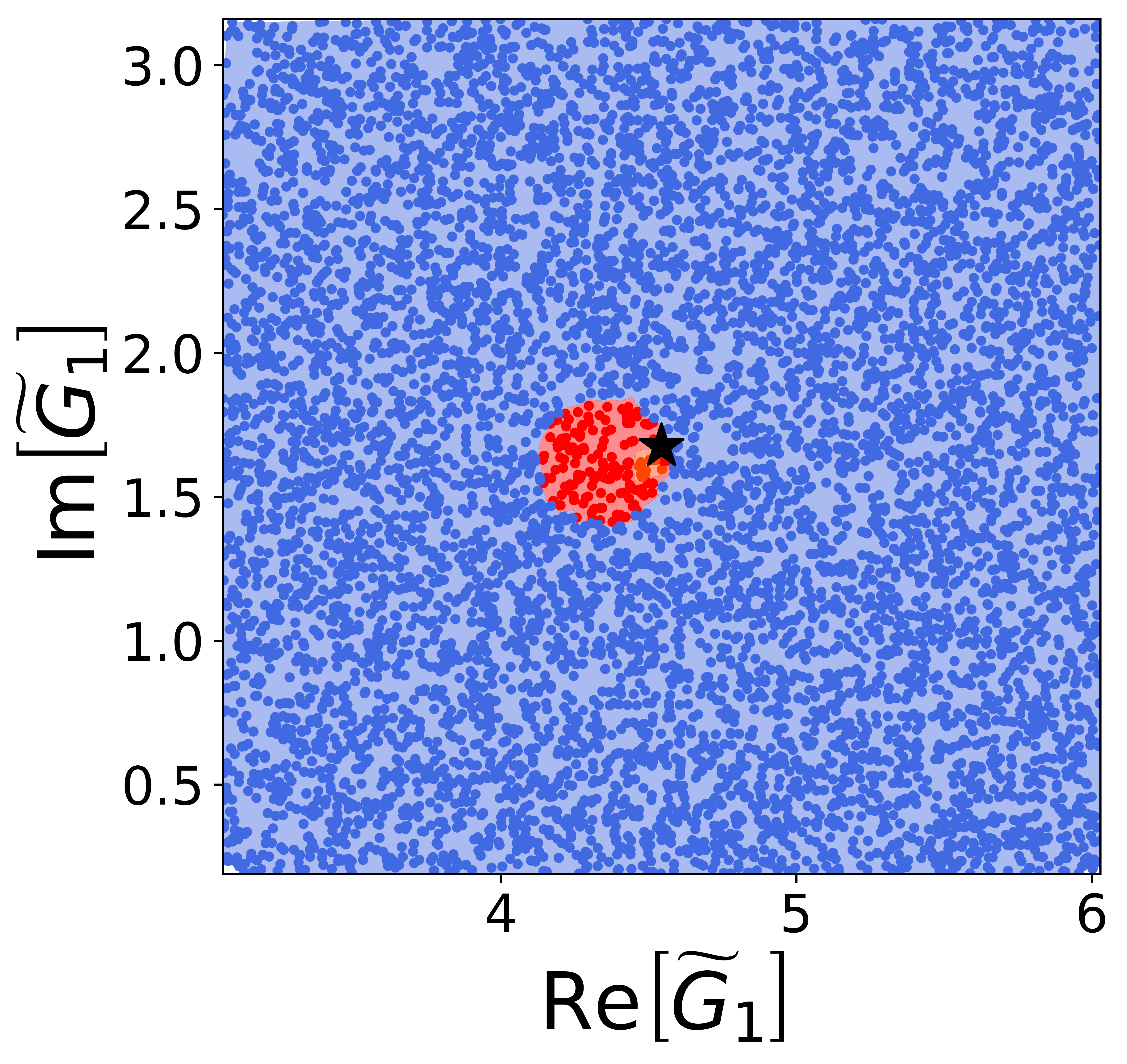}
\includegraphics[width=0.32\linewidth]{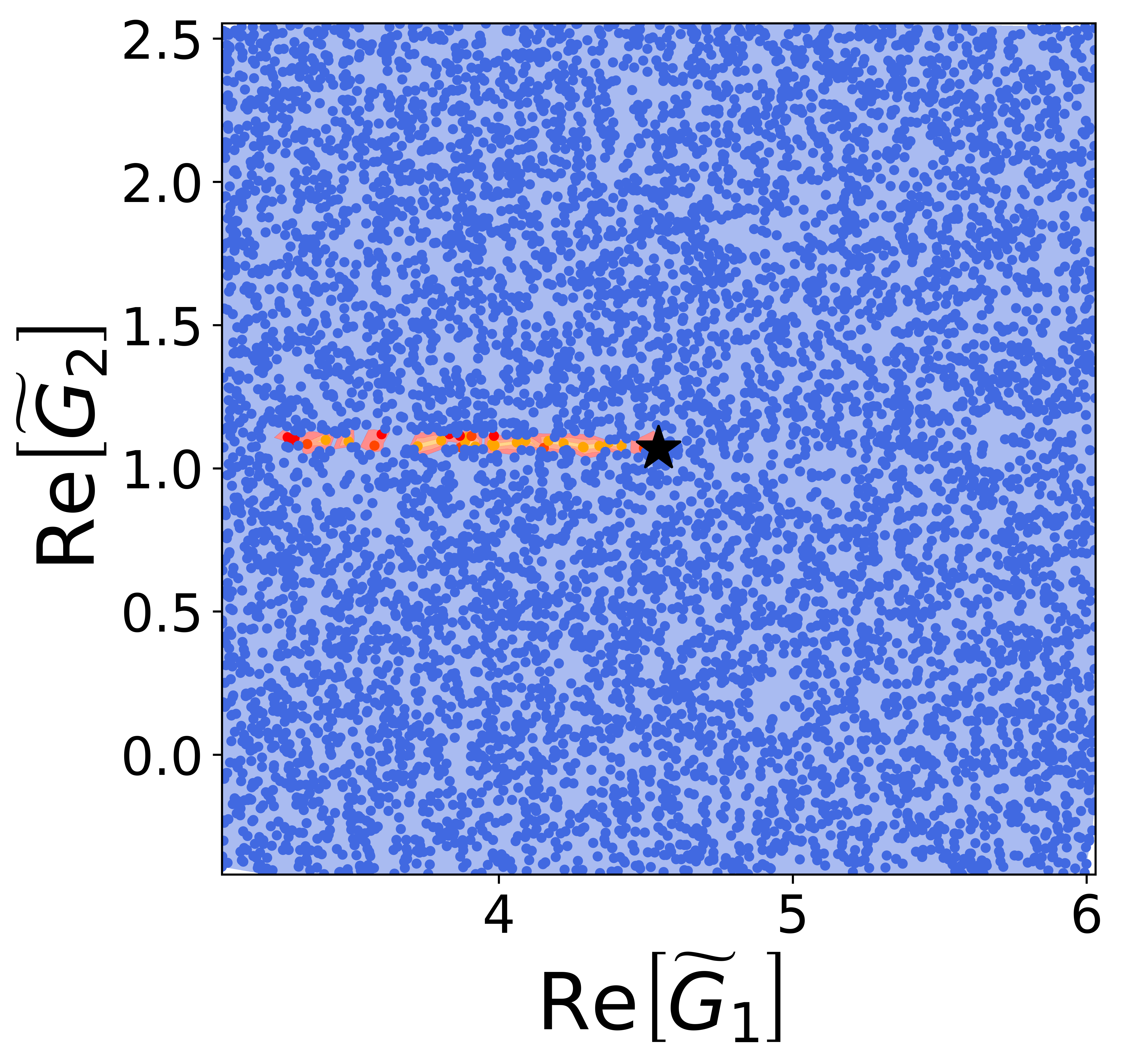}  \includegraphics[width=0.32\linewidth]{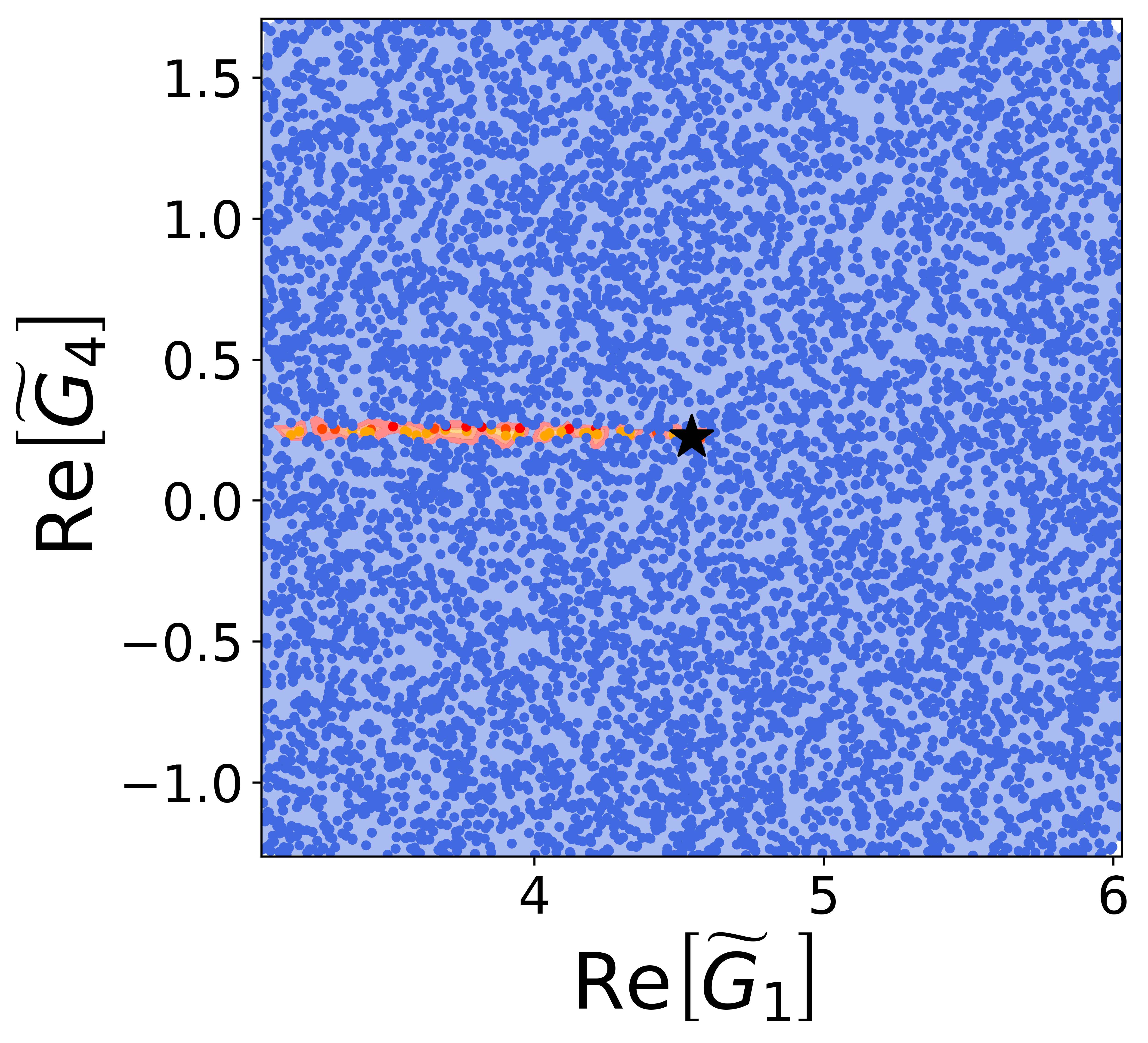}
\includegraphics[width=0.32\linewidth]{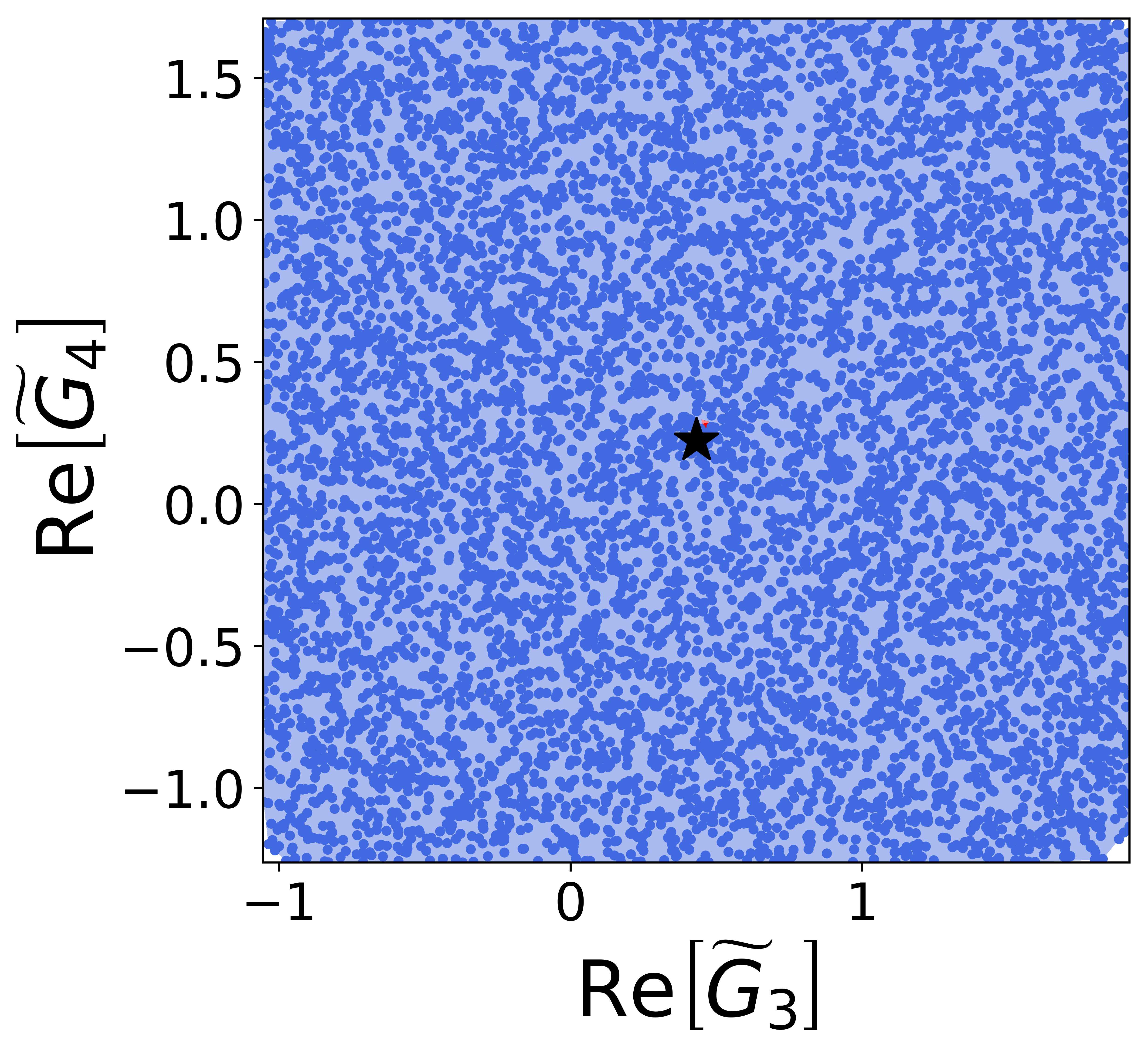}
\raisebox{.97\height}{\includegraphics[width=0.5\linewidth]{plots/cross_section_plots/4_points_0.01/legend.png}}

    \caption{Cross-section plots corresponding to the top row of plots shown in Fig.~\ref{fig:cross_section_plots}. Here we have expanded the $x$ and $y$ scales by a factor of 65 to capture the complete Pick-consistent region explored in this study where only two of the eight possible lattice data values are randomly sampled in the error volume. This was accomplished by increasing the error scale from $\xi=0.01$ to $\xi = 0.65$ region. Each plot shows 8000 random samples.}
    \label{fig:zoomed_out_cross_section_plots}
\end{figure}

Figure~\ref{fig:zoomed_out_cross_section_plots} offers a wider view of these cross-section plots for those cases where the Pick-consistent region was not contained within plots shown in Fig.~\ref{fig:cross_section_plots}. Now the scale of the $x$ and $y$ axes has been increased by a factor of 65 to allow the full Pick-consistent region to be seen. 

We would like to apply this methodology to interpolate from lattice results for a larger number of energies on the imaginary axis since this will more tightly constrain the set of interpolated functions, reducing the size of the Wertevorr\"ate. We therefore performed a similar test to the one described above, but increased the number of measured lattice energies from 4 to 10 while keeping the range of those input energies at $\{0.1i-2.0i\}$ and the error scale at $\xi = 0.01$. For each of the 190 pairs of distinct real and imaginary parts of the 10 complex amplitudes $\{G_n\}_{1\le n\le10}$ we choose 1000 random samples within the error volume while keeping the remaining 18 components fixed at their exact values. Of the resulting 190,000 samples, we found no sample that satisfied the Pick criteria. This was repeated with an error scale of 0.001 so that all 190,000 samples were $10\times$ closer to the exact, Pick-consistent values and again found no Pick-consistent results. 

This was done with only two of the twenty dimensions being varied from the exact result. In the case of an actual lattice calculation, we expect the resulting lattice data to be uniformly distributed in the full $2N$-dimension error volume. This suggests the need for a better method than random sampling to find Pick-consistent points which could be used to represent possible Pick-consistent results lying within the error volume surrounding those lattice data. Recall that our strategy is to use the variation of the Nevanlinna-Pick extrapolations from such a sample of mock lattice data to estimate the error on our interpolated result.

\subsection{Method of finding Pick-consistent sampled lattice data}

In order to reliably estimate the errors introduced by the fluctuations resulting when Nevanlinna-Pick interpolation is performed on mock lattice data that lies within our error volume, we need to find a reasonable sample of Pick-consistent lattice data in this volume. As demonstrated above, we cannot simply assign these points in the error volume at random because only an extreme fraction of these values will be Pick-consistent.

Instead, we start with $M$ data samples uniformly distributed within the $2N$ real-dimensional error volume. Recall that these $M$ samples are $N$-dimensional complex vectors $\{\widetilde{G}^m\}_{1\le m \le M}$ that lie with the error volume centered at the actual lattice result. We then adjust each of these values making them Pick-consistent while also requiring that they remain within the error volume. This is done as follows. For the $m^{th}$ sample we determine the least positive eigenvalue, $\lambda_\mathrm{min}(\widetilde{G}^m)$ of the Pick matrix and use the $2N$-dimensional gradient of this function to determine the path along which $\lambda_\mathrm{min}$ increases most rapidly. We follow this path to locate values for the $2N$ components of $\widetilde{G}^m$ for which $\lambda_\mathrm{min}$ has become positive so that the Pick criterion is obeyed. This gradient ascent procedure is implemented numerically using an adaptive step size and can be used effectively on a standard workstation.

It was discovered that during the gradient ascent procedure the components $\widetilde{\Gamma}_n$ in directions for which the Pick-consistent region was the widest (the smallest values of $n$) moved the farthest and were far more likely to leave the error volume. Thus, following the full gradient would often lead to a Pick-consistent point that was outside the original error volume for those dimensions, most frequently $\widetilde{\Gamma}_1$ and less often $\widetilde{\Gamma}_2$. Following the full gradient would also result in Pick-consistent regions that were localized clusters in the remaining dimensions, resulting in small clusters of points instead of a uniform sample filling the error volume. (Note, this gradient ascent procedure was applied in the unit disk and we have therefore replaced $\widetilde{G}$ with $\widetilde{\Gamma}$ in this discussion.)

In order to avoid losing points that move out of the error volume and to more uniformly sample the values of $\widetilde{\Gamma}_n$ for large $n$, we construct a hand-tuned ``constrained'' gradient accent in which we hold certain dimensions fixed. Specifically, we do not allow the first two complex components $\widetilde{\Gamma}_1$ and $\widetilde{\Gamma}_2$ to vary, fixing them at their random sampled values. In addition, as we move from evaluating the gradient ascent from the $m^{th}$ $\widetilde{\Gamma}^m$ to the next, $\widetilde{\Gamma}^{m+1}$ we sequentially loop over a third complex coordinate $3 \le n \le N$ of $\widetilde{\Gamma}_n^m$ which is also held fixed at its initial random value for the entire gradient ascent. Thus, we now follow a (2N-6) dimensional gradient ascent to find a Pick-consistent point closer to our random starting point than would result from following the complete gradient. Looping through the additional constraining of the coordinates $n \ge 3$ further ensures that the initial uniformity in our random sampling is not lost. We then verify that the point at the end of our trajectory in the 2N-dimensional space satisfies the Pick criterion, and check that the point is within the required error volume.   

Following this gradient ascent procedure allows for points that were originally not Pick-consistent to be adjusted until they reach a critical point at which the consistency criteria is now satisfied. This naturally results in samples that lie on the boundary of the Pick-consistent region. As shown in Fig~\ref{fig:cross_section_plots}, points that lie on the boundary result in Wertevorr\"ate with the smallest widths.  Thus, properly sampling the full Pick-consistent region requires points that lie between these boundaries. In practice we found that continuing to follow the gradient ascent procedure in order to find a maximum of $\lambda_\mathrm{min}(\widetilde{G}^m)$ became computationally costly since the norm of the gradient is already extremely small, for example on the order of $10^{-6}$ for the case of 10 lattice points by the time the smallest eigenvalue has become non-negative. 

Instead, we exploit the convexity of Pick-consistent region, which is discussed more thoroughly in \ref{subsec:Geometry of Pick-consistent region}, and draw lines between pairs of boundary points. To achieve uniform sampling, we select points at one-quarter, one-half, and three-quarters of the distance along each such line. 

Our complete procedure aims to robustly sample the entire Pick-consistent region by approaching its boundary from diverse directions using gradient ascent and then distributing points through the interior. Specifically, we randomly select 50 samples from our set of boundary samples and connect them pairwise with lines. Along each line, we take intermediate points at $25\%, 50\%$, and $75\%$ of the length of the line, resulting in a total of 3725 Pick-consistent samples. 

The gradient ascent portion of the calculation is computationally expensive because of the small fraction of the gradient ascent trajectories which ultimately become Pick-consistent while still inside the error volume and the amount of extended-precision arithmetic that must be performed. For the case of 10 initial lattice points and 600 data samples uniformly distributed within the $2N$-dimensional error volume, following the constrained gradient accent resulted in 254 Pick-consistent points that lie within the bounds of the original error volume. 187 of the 600 trajectories landed in the Pick-consistent region, but outside the original error volume while the remaining 159 samples reached the maximum iteration count (introduced to limit computational time) but never reached the Pick-consistent region.

\begin{figure} [hbt!]
    \centering
    \includegraphics[width=0.30\linewidth]{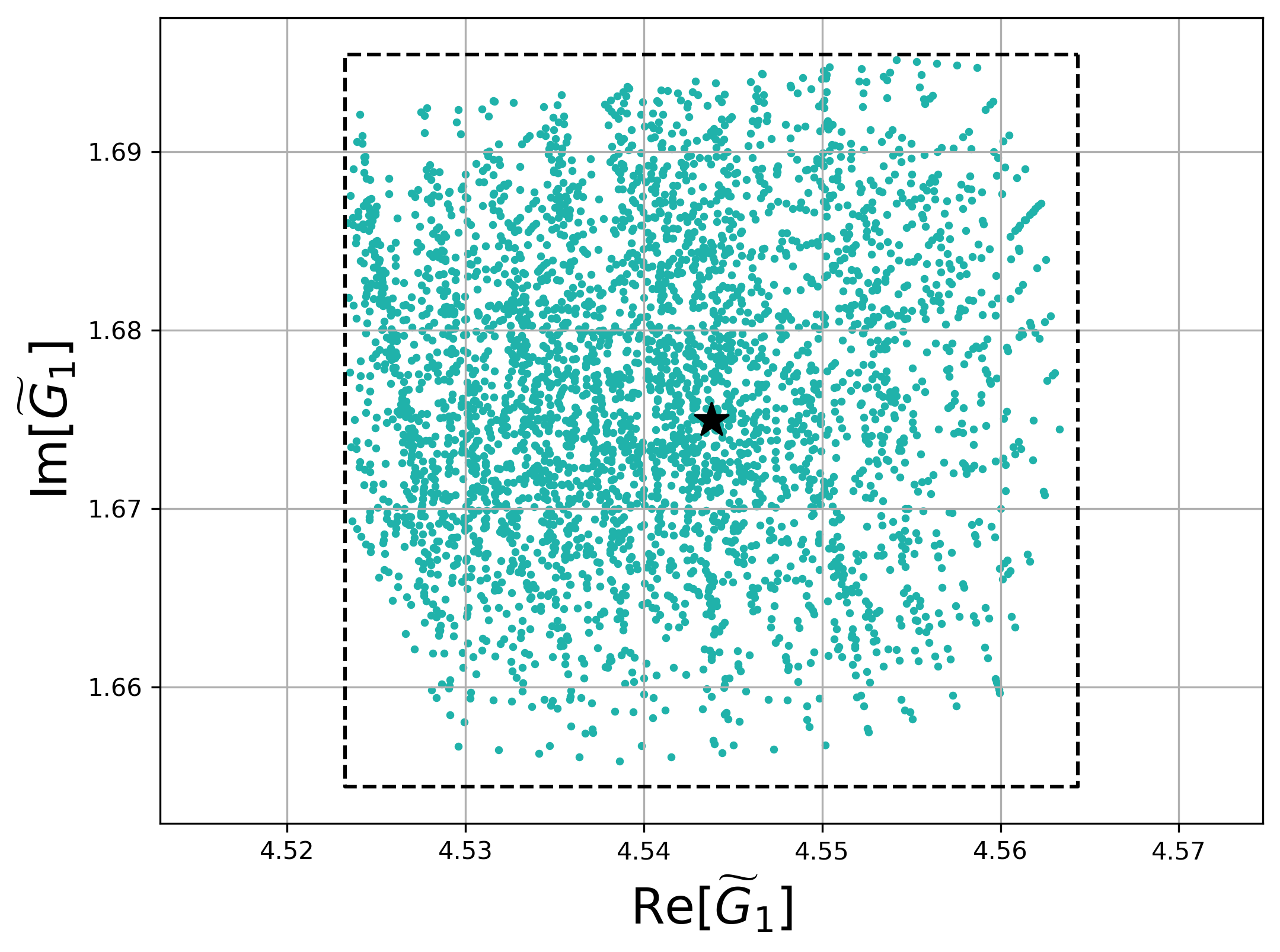}
    \includegraphics[width=0.30\linewidth]{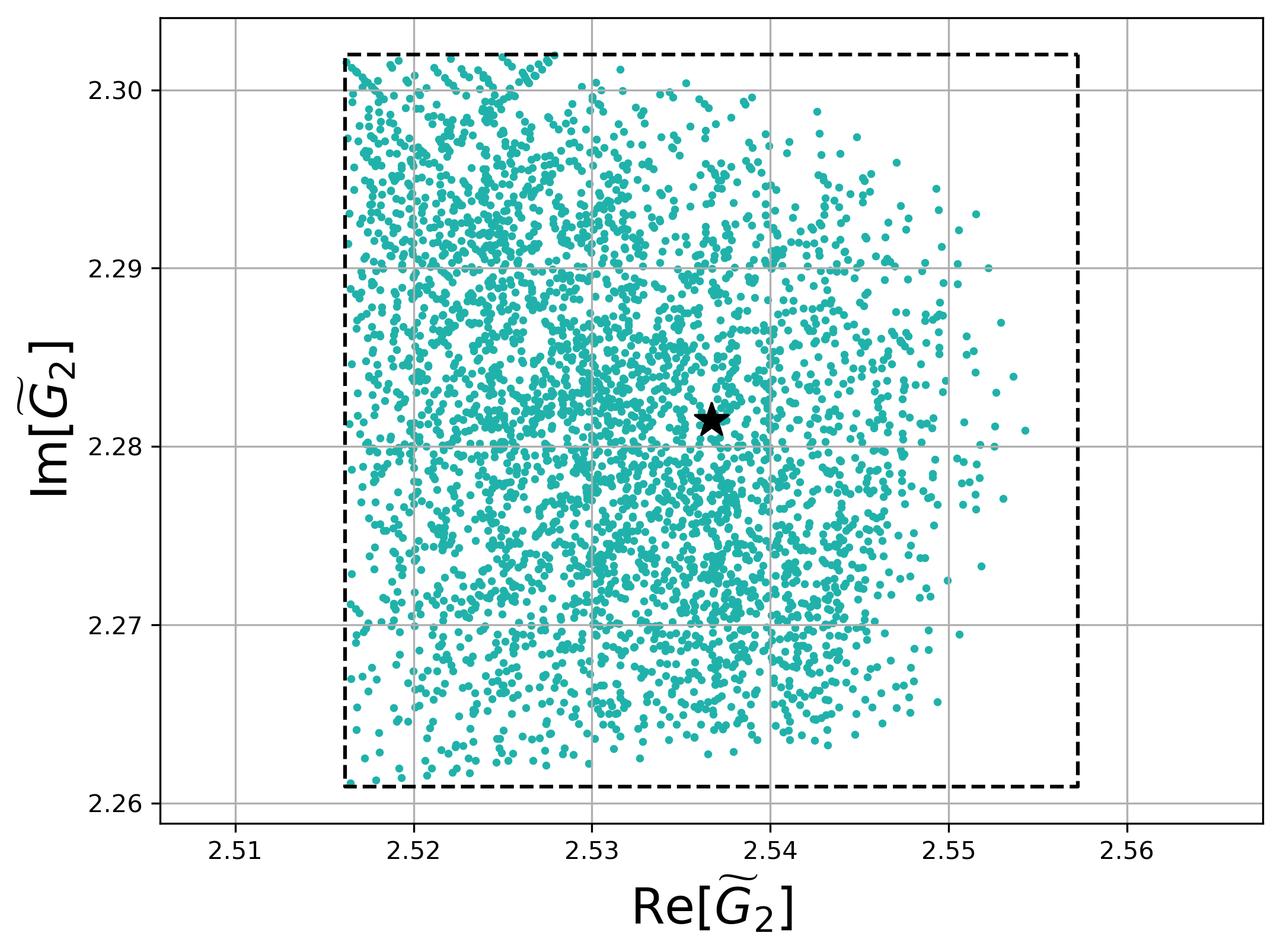}
    \includegraphics[width=0.30\linewidth]{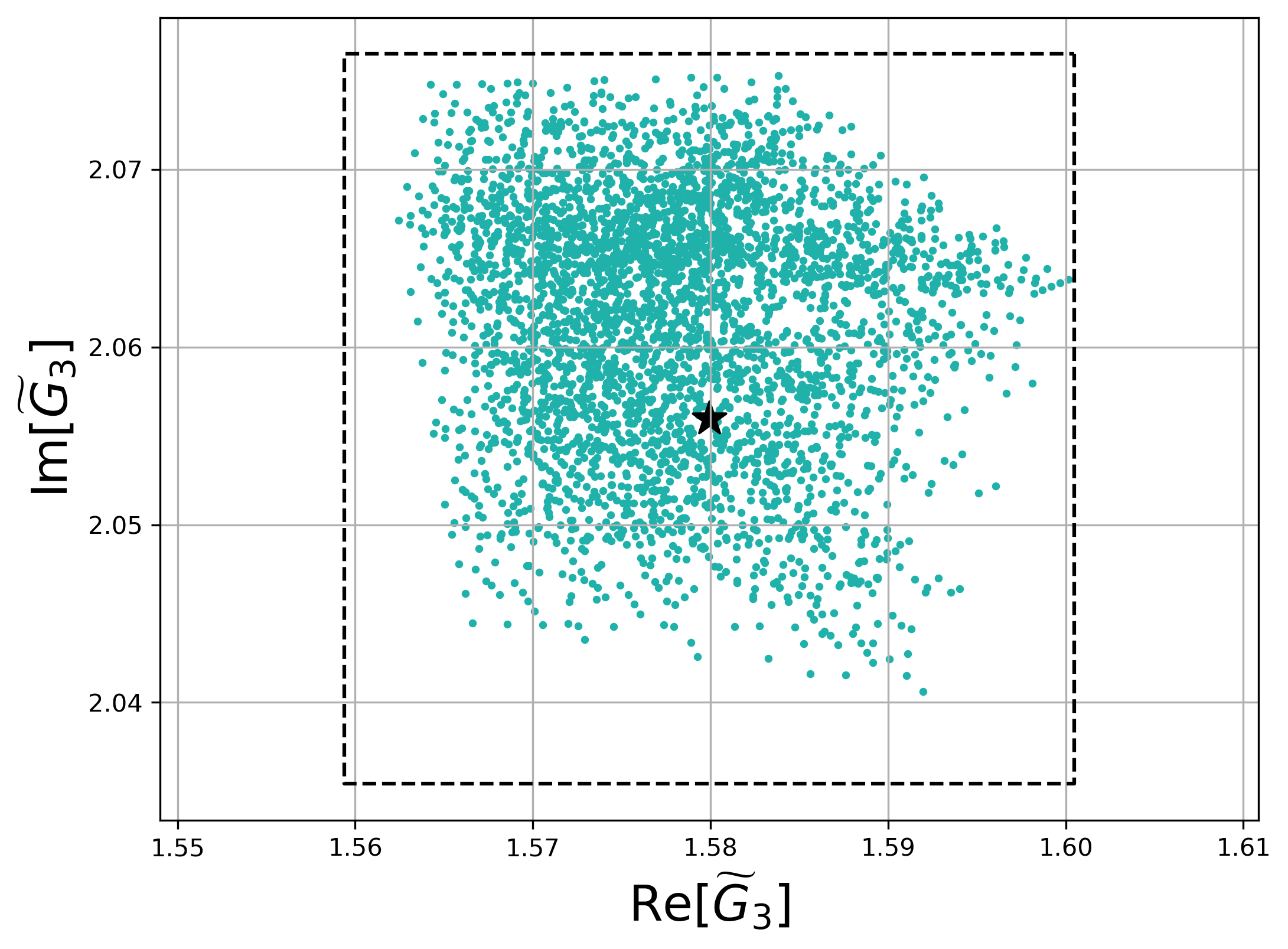}
    \\
    \includegraphics[width=0.30\linewidth]{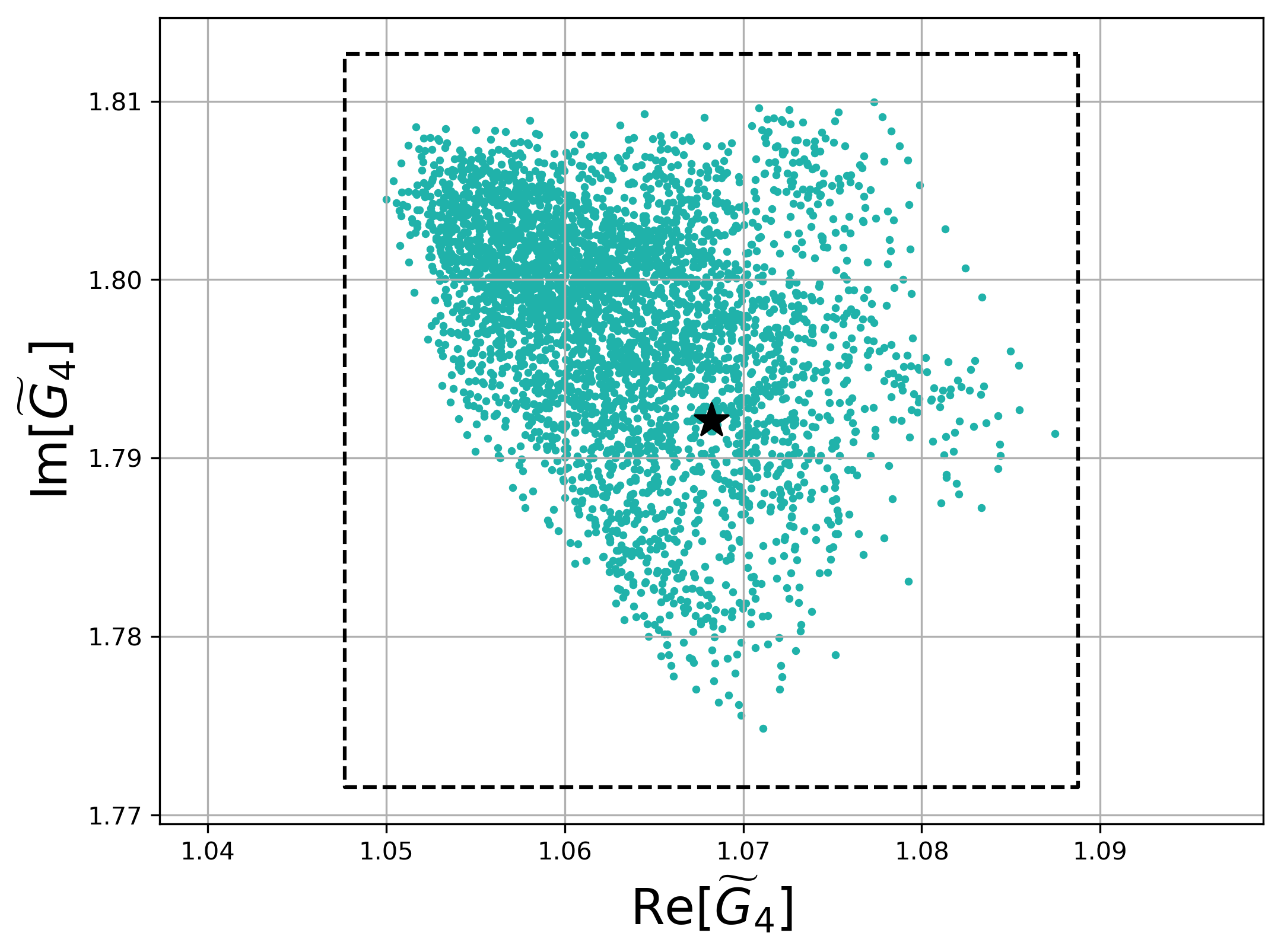}
    \includegraphics[width=0.30\linewidth]{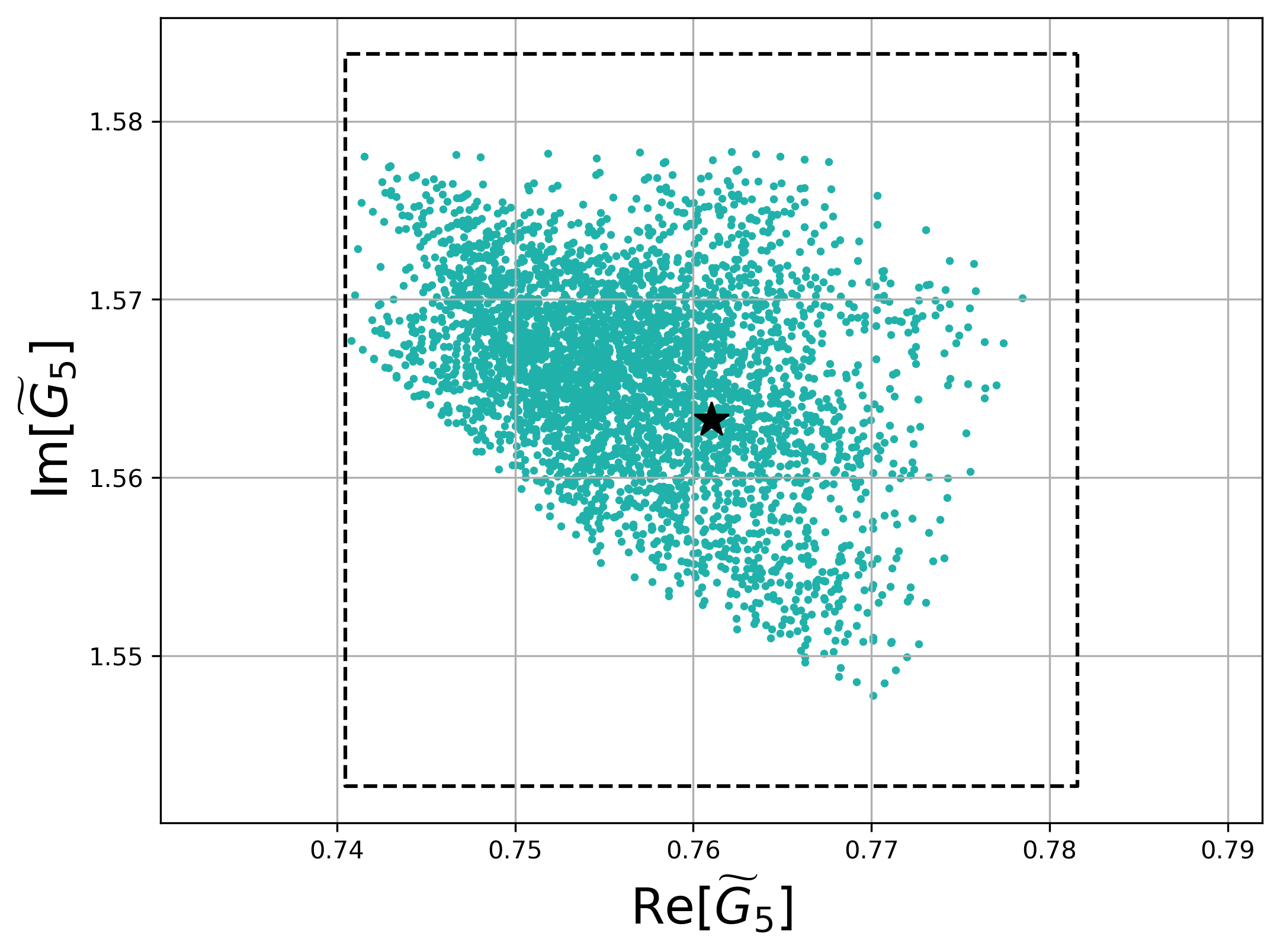}
    \includegraphics[width=0.30\linewidth]{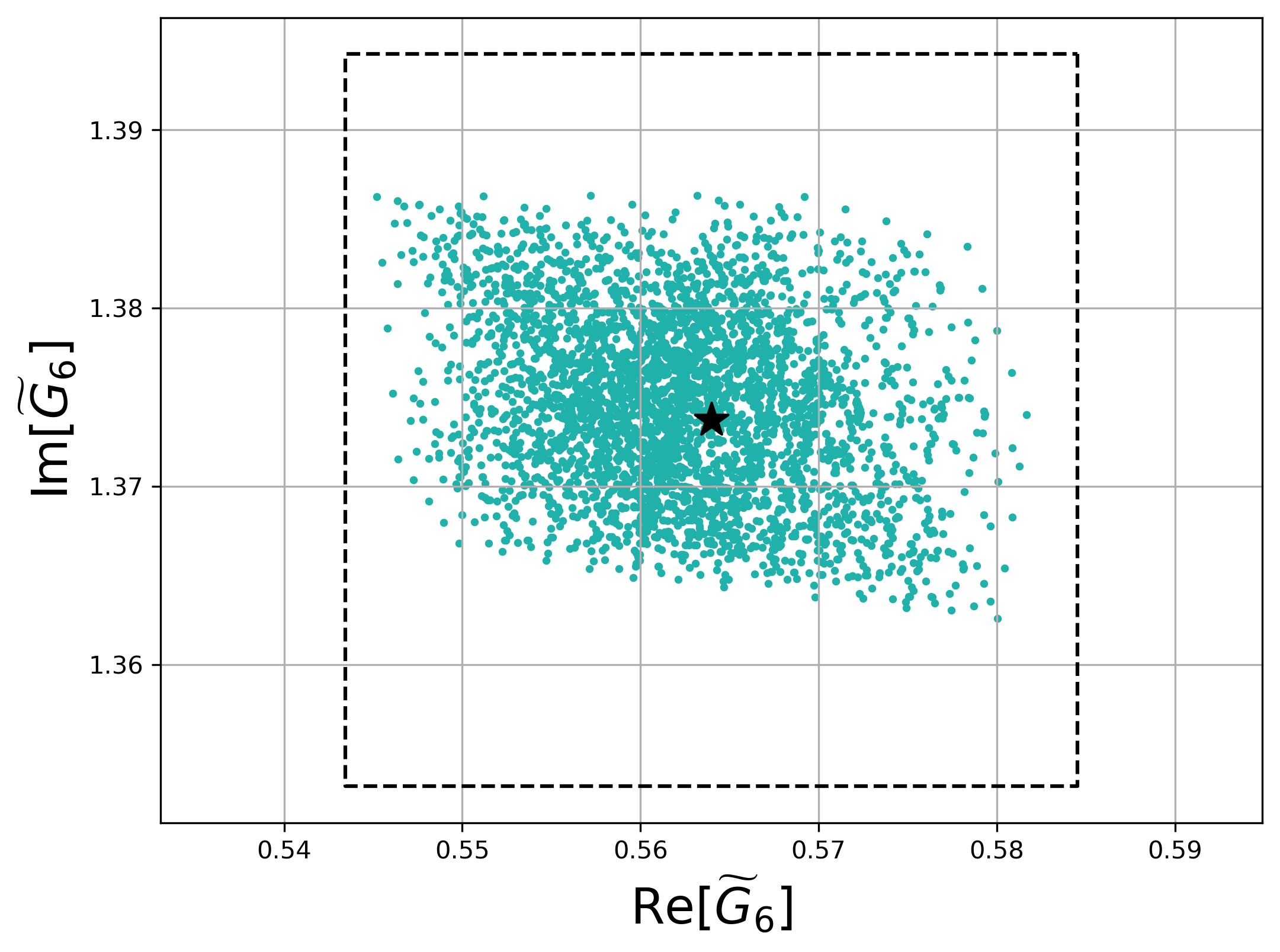}
    \\
    \includegraphics[width=0.30\linewidth]{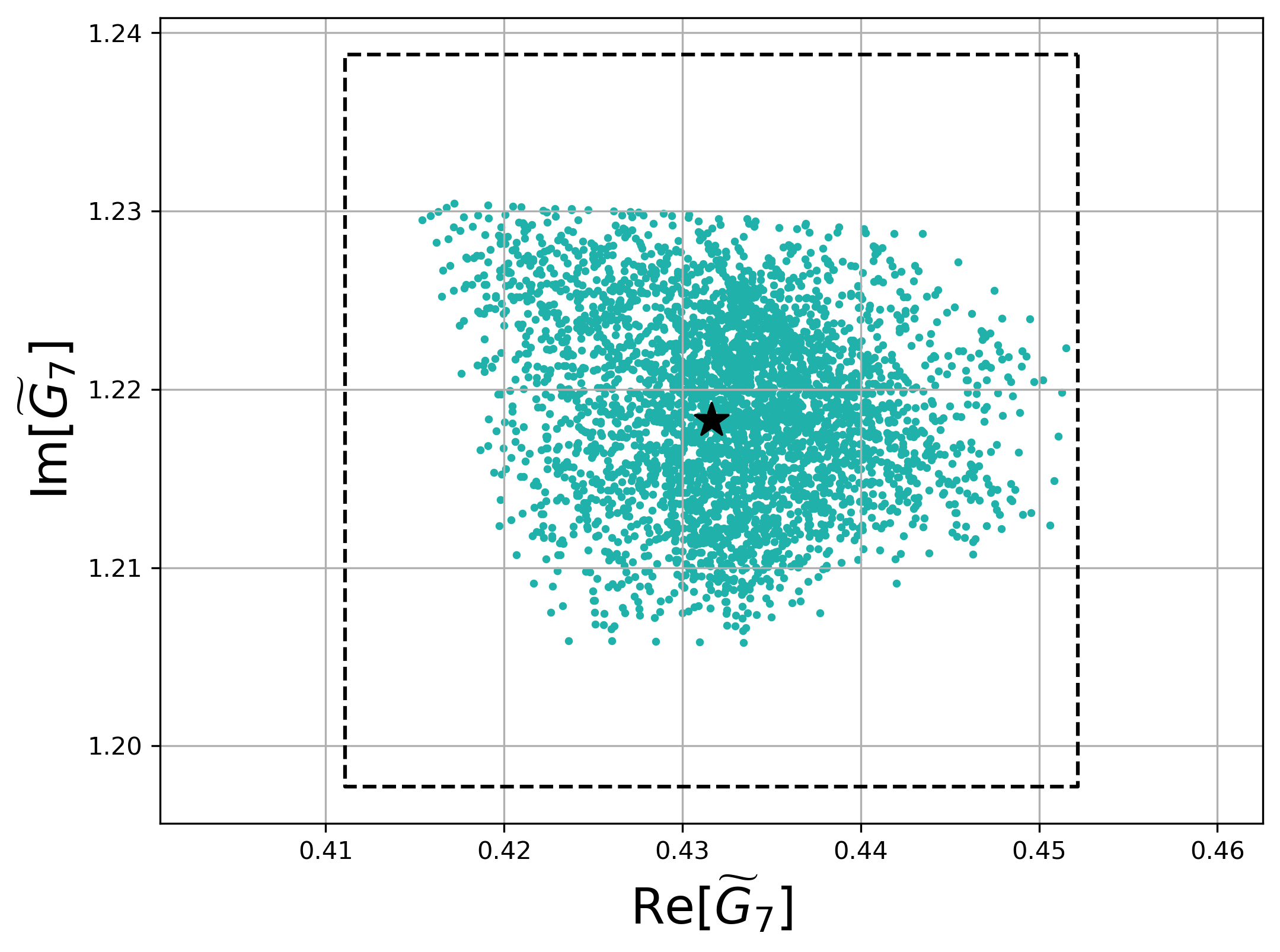}
    \includegraphics[width=0.30\linewidth]{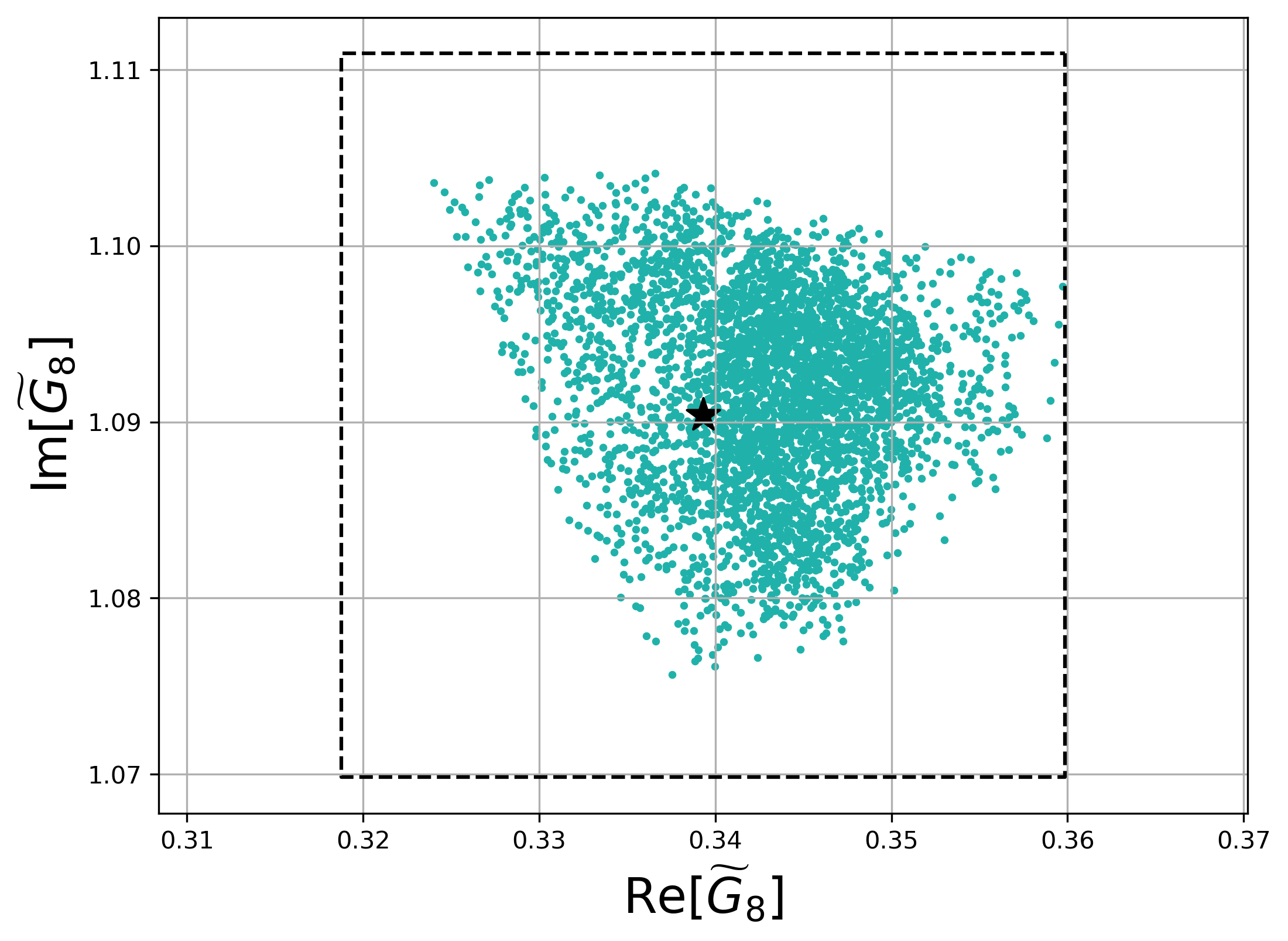}
    \includegraphics[width=0.30\linewidth]{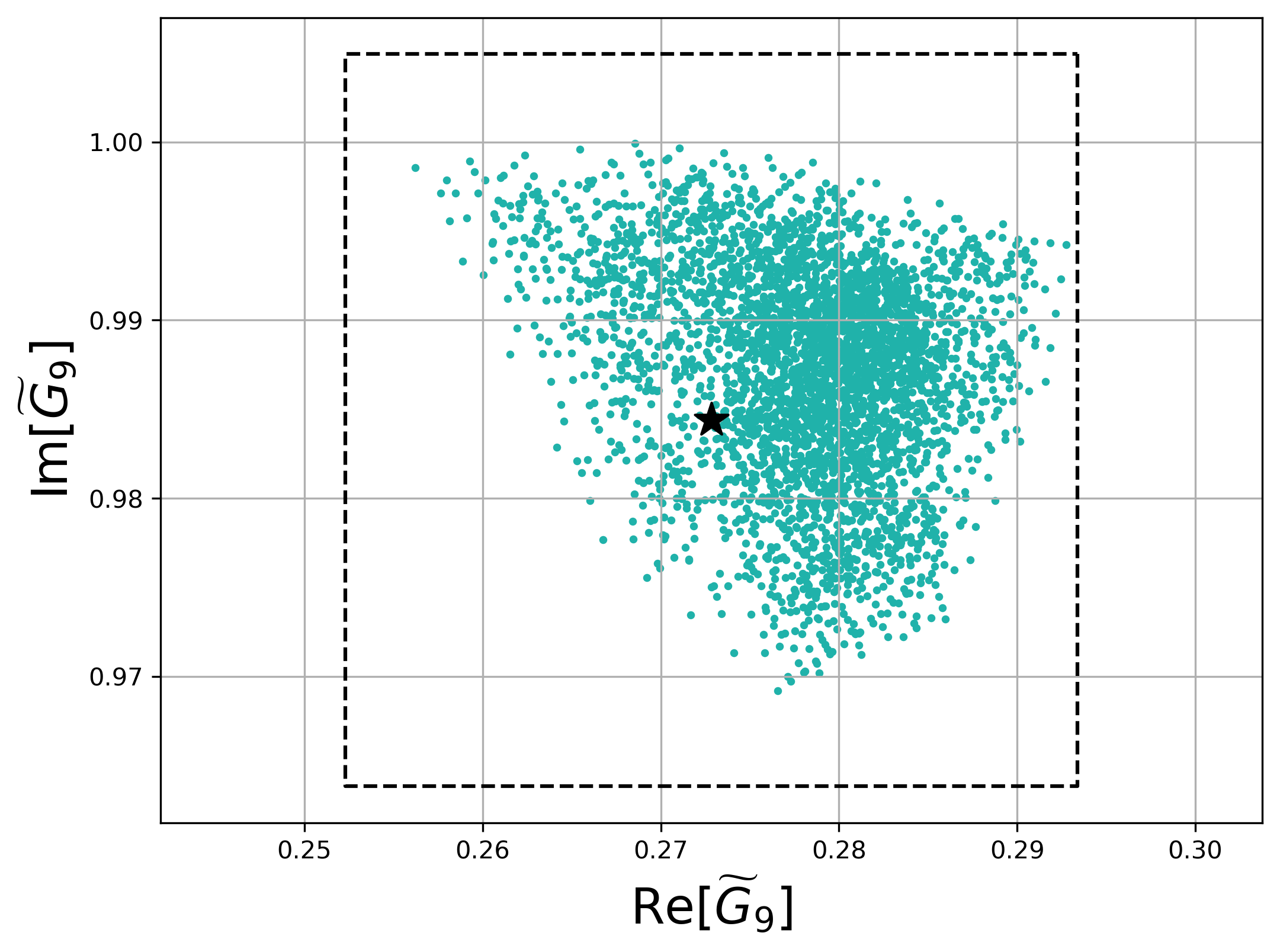}
    \\
    \hspace{0.2\linewidth}
    \includegraphics[width=0.30\linewidth]{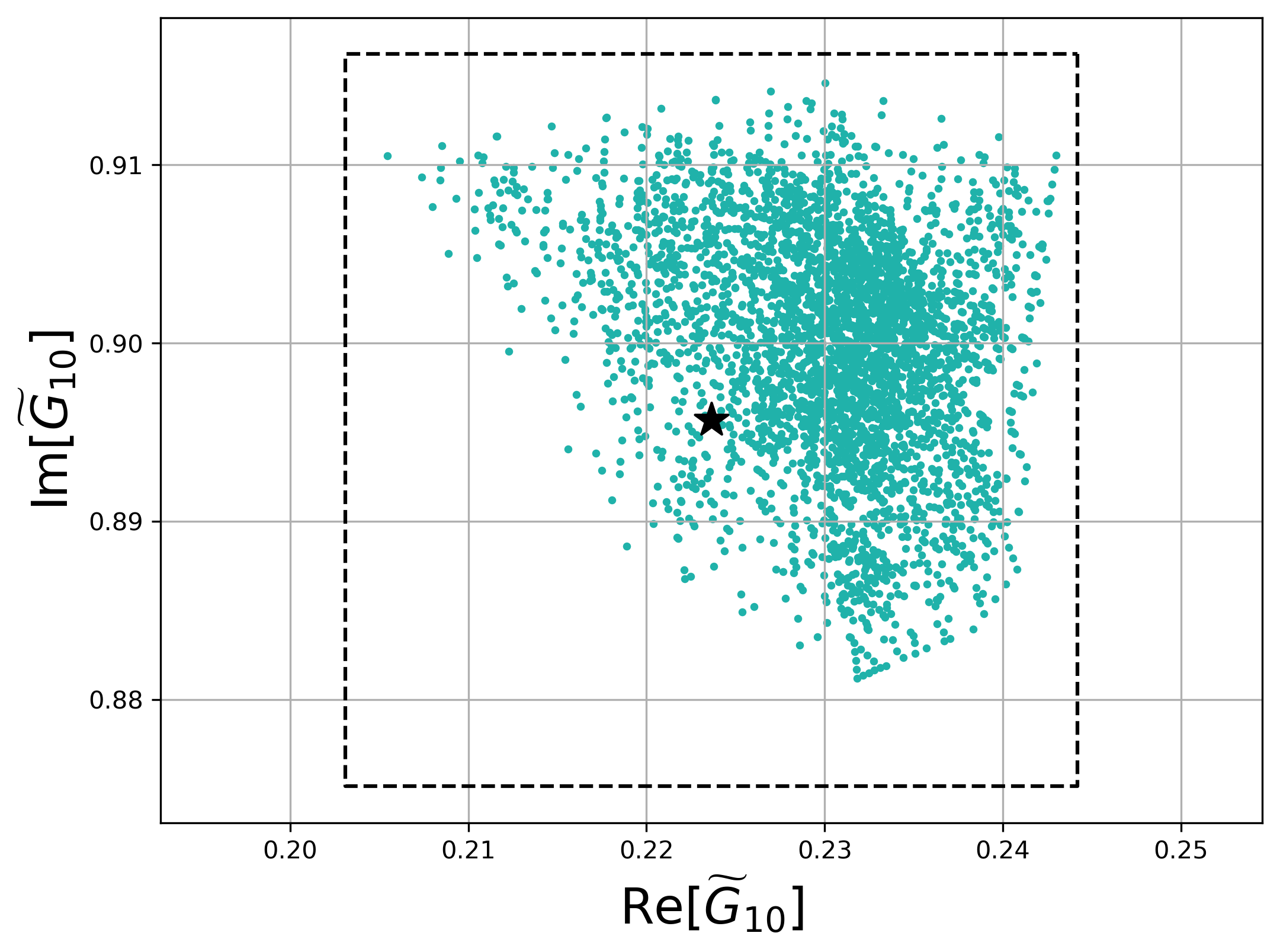}
    \raisebox{.9\height}{\includegraphics[width=0.2\linewidth]{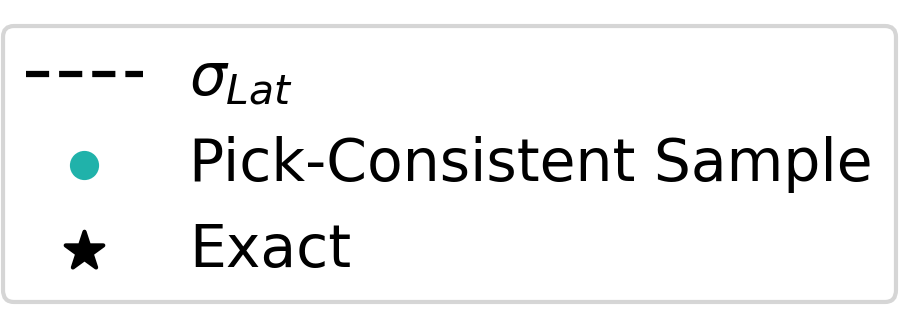}}

    \caption{Cross-section plots indicating the location of 3725 Pick-consistent Green's functions values where the sample contains 50 random boundary points found through the constrained gradient ascent procedure, and the remaining points were gathered by following the lines connecting all pairs of the boundary points. The imaginary energies $z_n$ to which the lattice correspond were chosen as 10 equally spaced values along the interval $\{0.1i,2.0i\}$ and the error scale was $\xi=0.01$. The gradient ascent procedure was performed in the unit disk, and then all values were transformed back onto the complex plane to obtain the ten components $(1\le n \le 10)$  $\widetilde{G}_n^m$ for each 3725 samples $1\le m \le 3725$. }
    \label{fig:location_of_pick_consitant_points_N=10}
\end{figure}

Figure~\ref{fig:location_of_pick_consitant_points_N=10} plots the distribution of Pick-consistent points for the N=10 case that were found using the method described above. Here it appears that our efforts to achieve uniformity were not entirely successful. The coordinates with larger values of $n$ appear to favor the central region of the plot with more than random white space near the boundaries. However, this may reflect the actual distribution of Pick consistent points which need not be uniform. Under these circumstances, we believe it is reasonable to assume this sampling is adequately uniform and that the distribution of these 3725 samples will provide a reasonable estimate of the final error introduced by this interpolation method.

\subsection{Geometry of Pick-consistent region}
\label{subsec:Geometry of Pick-consistent region}

The singular geometry of the Pick-consistent region plays a central role in our strategy 
to choose representative samples that can be used to estimate the errors introduced by the Nevanlinna-Pick interpolation and the central value of our result. In this Section we examine that geometry as the number $N$ of measured lattice energies increases. We use $S_\mathrm{NP}(z_1, z_2,\ldots z_N)$ to represent the set of Pick-consistent Green's functions values $\widetilde{G}$ given the $N$ locations of the lattice data being used in the interpolation.

We demonstrate the singular structure of $S_\mathrm{NP}$ by studying the width of this set along the $2N$ distinct real directions at a known Pick-consistent point. We vary each component of the $N$-dimensional complex vector $\widetilde{\Gamma}$ along the real or imaginary axis while holding the other $2N-1$ values fixed. Here we extend the notation introduced above, distinguishing the exact Green's function's values $\Gamma \in \mathbb{D}$ from the $N$-dimensional complex vector $\widetilde{\Gamma} \in\mathbb{D}$ which we allow to vary away from $\Gamma$ as we study the behavior of the Pick matrix. Thus:
\begin{equation}
    \Gamma_n = C\left(G_n\right)\quad \mathrm{and}\quad
    \widetilde{\Gamma}_n = C\left(\widetilde{G}_n\right), \quad
    1 \le n \le N.
\end{equation}

Each of the $2N$ real variables Re$\left[\widetilde{\Gamma_n}\right]$ and Im$\left[\widetilde{\Gamma_n}\right]$ are independently varied to identify how far we can move in each of these $2N$ dimensions while still maintaining the validity of the Pick criterion. Once the boundary points of Pick consistency in each direction have been identified, they are mapped back to the complex plane. The distance between them determines the length of the Pick-consistent region in each dimension, now expressed in the more physical complex plane. We adopt $\delta\mathrm{Re}[\widetilde{\Gamma_n}]$ and 
$\delta\mathrm{Im}[\widetilde{\Gamma_n}]$ to specify these $2N$ widths.

Table \ref{table:0_1_2_10_points_pick_distances} lists these widths for the $N=10$ case, illustrating the eight-order-of-magnitude variation in these widths between the mock data point obtained closest to the real axis $(n=1)$ and that farthest away $(n=10)$ as well as how small the allowed region is for those data points located far from the real axis. This observation is consistent with the results shown in the previous cross-section plots in Fig.\ref{fig:cross_section_plots} but has become much more singular for this $N=10$ case. These larger widths for mock data points closer to the real axis are also consistent with the increased movement found during gradient ascent for these points, as discussed in the previous section. 

Next we repeat the study shown in Table~\ref{table:0_1_2_10_points_pick_distances} but increase the number of points being used for the interpolation from 10 to 20, while keeping the range of the 20 imaginary locations $\{z_n\}_{1 \le n \le 20}$ of the interpolated points the same: between $0.10i$ and $2.00i$. The $N=20$ results are shown in Table~\ref{table:0_1_2_20_points_pick_distances}. As should be expected, the width of the Pick-consistent region around the exact solution has decreased. However, the decrease is by a dramatic factor of $\approx 10^{-7}$ for the case of $z_1 = 0.10i$, closest to the real axis. This reduction in width between the $N=10$ and 20 cases falls by a factor of $\approx10^{-15}$ for $z = 2.00i$, the point farthest from the real axis.

\begin{table}[hbt!]
    \centering 
    {\footnotesize
    \begin{tabular}{|c|c|c|c|}
    \hline
    $z_n$ & $G_n$ & $\delta\mathrm{Re}[\widetilde{G}_n]$  & $\delta\mathrm{Im}[\widetilde{G}_n]$  \\
    \hline
    0.100i & 4.544 + 1.675i & 9.272e-6  & 2.128e-5 \\
    0.311i & 2.537 + 2.281i & 4.318e-8  & 6.956e-8   \\
    0.522i & 1.580 + 2.056i & 5.568e-10 & 1.142e-9   \\
    0.733i & 1.068 + 1.792i & 4.816e-11 & 2.873e-11  \\
    0.944i & 0.761 + 1.563i & 2.559e-12 & 4.997e-12  \\
    1.156i & 0.564 + 1.374i & 1.068e-12 & 3.979e-13   \\
    1.367i & 0.432 + 1.218i & 3.346e-13 & 2.787e-13  \\
    1.578i & 0.339 + 1.090i & 1.099e-13 & 2.887e-13  \\
    1.789i & 0.273 + 0.984i & 1.519e-13 & 4.085e-13   \\
    2.000i & 0.224 + 0.896i & 9.967e-13 & 1.199e-12   \\
    \hline
    
\end{tabular}
}\caption{The width of the Pick-consistent region in the real and imaginary directions for each $z_n$ for $N=10$. Shown are the choices of $z_n$, the corresponding exact Green's function values, and the real and imaginary widths for all 10 values of $z_n$. The widths are reported as distances in the complex plane. The real and imaginary components of each Green's function value were varied independently. This table illustrates a very large variation in these widths among the various components of $\widetilde{G}_n$. The table shows that Green's function values corresponding to points located nearer to the real axis exhibit a larger Pick-consistent region.}
\label{table:0_1_2_10_points_pick_distances}
\end{table}

\begin{table}[hbt!]
    \centering
    {\footnotesize
 \begin{tabular}{|c|c|c|c|}
    \hline
    ${z_n}$ & $G_n$  & $\delta\mathrm{Re}[\widetilde{G}_n]$ & $\delta\mathrm{Im}[\widetilde{G}_n]$   \\
    \hline
    0.100i & 4.544 + 1.675i & 1.138e-12 & 1.346e-12\\
    0.200i & 3.419 + 2.220i & 1.968e-15 & 2.951e-15 \\
    0.300i & 2.609 + 2.286i & 9.922e-18 & 1.283e-17 \\
    0.400i & 2.051 + 2.206i & 7.357e-20 & 1.310e-19 \\
    0.500i & 1.653 + 2.084i & 1.244e-21 & 2.236e-21\\
    0.600i & 1.358 + 1.956i & 3.624e-23 & 6.243e-23\\
    0.700i & 1.132 + 1.832i & 2.852e-24 & 1.493e-24 \\
    0.800i & 0.955 + 1.715i & 1.195e-25 & 1.777e-25\\
    0.900i & 0.814 + 1.608i & 1.726e-26 & 1.110e-26\\
    1.000i & 0.701 + 1.510i & 1.185e-27 & 2.494e-27\\
    1.100i & 0.608 + 1.420i & 4.713e-28 & 2.036e-28\\
    1.200i & 0.532 + 1.338i & 8.669e-29 & 9.736e-29\\
    1.300i & 0.468 + 1.264i & 1.732e-29 & 4.131e-29\\
    1.400i & 0.415 + 1.196i & 1.696e-29 & 1.187e-29\\
    1.500i & 0.370 + 1.135i & 1.229e-29 & 4.654e-30\\
    1.600i & 0.331 + 1.078i & 7.819e-30 &  8.408e-30\\
    1.700i & 0.298 + 1.027i & 4.658e-30 &  1.335e-29\\
    1.800i & 0.270 + 0.979i & 1.207e-29 &  2.378e-29\\
    1.900i & 0.245 + 0.936i & 6.427e-29 & 5.460e-29 \\
    2.000i & 0.224 + 0.896i & 5.526e-28 & 2.048e-28 \\
    \hline
    
\end{tabular}
}\caption{The width of the Pick-consistent region in the real and imaginary directions for each $z_n$ for the case $N=20$. Except for the larger number of interpolation points, this is the same as Table~\ref{table:0_1_2_10_points_pick_distances}. A comparison with Table~\ref{table:0_1_2_10_points_pick_distances} shows an eight order-of-magnitude decrease in the width of the largest Pick-consistent regions for smallest value of $z_1 = 0.01i$ and an even more dramatic increase in the range between the largest and smallest width from eight to fifteen orders-of-magnitude for the largest $z_{10/20}=2.00i$. For both tables, $z$ varies between these two fixed limits.}
\label{table:0_1_2_20_points_pick_distances}
\end{table}

Both of the studies just described examined the widths of the Pick-consistent region measured along $2N$ lines, parallel to our $2N$ real or imaginary axes, that passed through the known, exact values $G_n$ of the example Green's function. In this third comparison, we examine the $N=10$ case and the width of the Pick-consistent region as measured by the length of 20 lines also drawn within the Pick-consistent region but now passing through a $2N$-dimensional point that was found by taking the midpoint of two boundary points obtained through our gradient ascent procedure. The shifted values of these new Pick-consistent data, $\widetilde{G}_n$ found by our procedure and the corresponding 20 widths of the Pick-consistent region at this point are given in Table~\ref{table:0_1_2_10_points_pick_distances_non_exact}.

As can be seen in Table~\ref{table:0_1_2_10_points_pick_distances_non_exact}, for this point that should be in a more central portion of the Pick-consistent region, the widths are substantially larger than for the point that corresponds to the exact Green's function whose values we are trying to interpolate. These larger widths likely reflect that fact that the exact Green's function values $\{G_n\}_{1 \le n \ln N}$ are found much closer to the boundaries of the Pick-consistent region than a point chosen to be close to the center of that region, as can be seen in Fig.~\ref{fig:cross_section_plots}.

\begin{table}[hbt!]
    \centering
    {\footnotesize
    \begin{tabular}{|c|c|c|c|}
    \hline
    ${z_n}$ & $\widetilde{G}_n$  & $\delta\mathrm{Re}[\widetilde{G}_n]$ & $\delta\mathrm{Im}[\widetilde{G}_n]$   \\
    \hline
    0.100i & 4.542 + 1.675i & 9.318e-3  & 8.055e-3 \\
    0.311i & 2.531 + 2.276i & 2.930e-4  & 2.896e-4   \\
    0.522i & 1.580 + 2.067i & 1.773e-5 & 1.803e-5   \\
    0.733i & 1.059 + 1.807i & 1.962e-6 & 2.124e-6  \\
    0.944i & 0.746 + 1.575i & 3.929e-7 & 4.580e-7  \\
    1.156i & 0.549 + 1.382i & 1.372e-7 & 1.727e-7   \\
    1.367i & 0.418 + 1.225i & 8.146e-8 & 1.106e-7  \\
    1.578i & 0.327 + 1.096i & 8.436e-8 & 1.212e-7  \\
    1.789i & 0.263 + 0.989i & 1.619e-7 & 2.438e-7   \\
    2.000i & 0.215 + 0.901i & 7.485e-7 & 1.182e-6   \\
    \hline
    
\end{tabular}

}
\caption{The width of the Pick-consistent region in the real and imaginary directions for each $z_n$ and for $N=10$. However, in contrast to Table~\ref{table:0_1_2_10_points_pick_distances}, these widths are determined by lines passing though the Pick-consistent points $\widetilde{G}$, determined by taking the midpoint between two boundary points found by our gradient ascent procedure, not the $2N$ values of the example Green's function at the locations $z_n$. Shown are the choices of $z_n$, the corresponding central Green's function values $\widetilde{G}$ about which the widths are being computed, and the real and imaginary widths for all 10 values of $z_n$ determined as in Table~\ref{table:0_1_2_10_points_pick_distances}. The widths are reported as distances in the complex plane. Compared to the results in that table, the largest widths have increased by a factor of $10^3$ and the range of widths has decreased by nearly a factor of $10^4$.}
\label{table:0_1_2_10_points_pick_distances_non_exact}
\end{table}

Additional significant information about the geometry of the Pick-consistent region was obtained by drawing lines between all pairs from a set of 50 points drawn at random from all Pick-consistent points obtained by our gradient ascent procedure. We evaluated the eigenvalues of the Pick matrix at points along these 1225 lines and found them to all be positive and therefore Pick-consistent. This suggests that the Pick-consistent region should be convex which in fact is a known result.~\footnote{The authors thank Ryan Abbott for pointing this out.}

The convexity of the Pick consistent region combined with its dramatically varying widths would appear to give a clear picture of the geometry of the Pick-consistent region: an ellipse-like $2N$-dimensional volume of dramatically varying widths in the dimensions corresponding to the allowed values at each of the imaginary locations $z_n$ of the $N$ lattice data. The smallest widths corresponding to those locations $z_n$ farthest from the real axis.

\section{Propagating computational errors through Nevanlinna-Pick interpolation}
\label{sec:error}

In the previous Section we proposed a method to represent the statistical and systematic errors on the data obtained from a lattice QCD calculation by a collection of $M$ samples of possible results that lie within what we defined as the error volume centered on the actual lattice result. These $M$ samples of possible results obey the Pick criterion and will be used in this section to: (i) estimate the final result of the Nevanlinna-Pick extrapolation of that lattice data and (ii) determine the error on that result.

As explained in Sec.~\ref{sec:intro} we make the discussion concrete by assuming that the quantity which we wish to compute is the integral of the interpolated Green's function along the contour $C_2$ shown in Fig.~\ref{fig:Contour}. Thus, for a Green's function $G(z)$ we will study the integral $\mathcal{I}$:
\begin{eqnarray}
    \mathcal{I} = \frac{1}{\pi}\int_{C_2} dz G(z) 
                 = \frac{1}{\pi}\int_0^{E_\mathrm{max}} d\omega G(\omega + i\epsilon)
                 \label{eq:target-integral}
\end{eqnarray}

Where all work in this paper was done with $E_\mathrm{max}=1.5$.  We can interpolate each sample to determine the Wertevorrat for each value of $z$ on that contour. The real and imaginary parts of the Wertevorrat correspond to the allowed range for the real and imaginary values of all possible Green's functions $\widetilde{G}(z)$ evaluated at $z$ which take the values of that sample when evaluated at the $N$ locations $z_n$ on the imaginary axis and which have the required analyticity properties.  

We use the bounds of the Wertevorrat for each $z$ and each of the $M$ samples to define $6M$ functions: $\bigl(\Delta^{\mathbb{C},X}_\mathrm{max}(z)\bigr)^m$, $\bigl(\Delta^{\mathbb{C},X}_\mathrm{min}(z)\bigr)^m$ and $\bigl(\Delta^{\mathbb{C},X}_\mathrm{avg}(z)\bigr)^m$. Here $X$ indicates the real or imaginary part of $\widetilde{G}_m$. The index $m$, $1 \le m \le M$, identifies the sample and `max' or `min' indicates the upper or lower edge of the Wertevorrat. Finally, `avg' indicates the average of these upper (`max') and lower (`min') limits.

For the sample $m$ we will evaluate six integrals over the contour $C_2$:
\begin{eqnarray}
    \mathcal{I}^{X,m}_Y = \frac{1}{\pi}\int_{C_2} \left(\Delta^{\mathbb{C},X}_Y(z)\right)^m dz
    \label{eq:SampleIntegral}
\end{eqnarray}
where $Y=$ `max', `min', or `avg' while $X$ is either `Re' or `Im' as described above. We interpret $\mathcal{I}^{X,m}_\mathrm{avg}$ as the result for the integral of interest coming from the $m^{th}$ sample.  We define the average over these $M$ individual results:
\begin{equation}
    \left\langle\mathcal{I}^X_\mathrm{avg}\right\rangle = \frac{1}{M}\sum_{m=1}^M \mathcal{I}^{X,m}_\mathrm{avg},
    \label{eq:NP-average}
\end{equation}
as the final result of our calculation of the integral $\mathcal{I}$ defined in Eq.~\eqref{eq:target-integral}.  

If this were a traditional calculation, we would then assign the root-mean-square of the fluctuations in the values $\mathcal{I}^{X,m}_\mathrm{avg}$ for the $M$ samples as the error. In fact, since the Nevanlinna-Pick interpolated result $\mathcal{I}^{X,m}_\mathrm{avg}$ is a differentiable function of the lattice data, collecting these $M$ samples could be avoided and we could directly evaluate $\left\langle\mathcal{I}^X_\mathrm{avg}\right\rangle$ at the central lattice result and compute its derivatives with respect to the $2N$ lattice values to propagate the lattice errors. However, because of the imprecise interpolation and the resulting Wertevorr\"ate, there are additional uncertainties that this traditional approach would omit. (In addition, a standard Taylor expansion may be challenging because of the near-singular character of the Nevanlinna-Pick interpolation.)

We choose to include the implications of the Wertevorr\"ate for the errors on the average interpolated quantity, by defining the error on the final result as:
\begin{equation}
    \mathcal{E}^X = \left[
    \frac{1}{2M}\sum_m\left(\left(\mathcal{I}^X_\mathrm{max}\right)^m-\langle \mathcal{I}_\mathrm{avg}\rangle\right)^2 
    + \frac{1}{2M}\sum_m\left(\left(\mathcal{I}^X_\mathrm{min}\right)^m-\langle \mathcal{I}_\mathrm{avg}\rangle\right)^2 \right]^\frac{1}{2}
    \label{eq:error-def}
\end{equation}
where, as above, $X$ is either `Re' or `Im'. This proposal is not as conservative as taking the error from the largest deviation between the boundaries of the $M$ Wertevorr\"ate and the average. Instead, this definition includes a weighting by the frequency of occurrence for such outliers among our $M$ samples and takes the root-mean-square average of the difference of the result, $\left\langle\mathcal{I}^X_\mathrm{avg}\right\rangle$, and the result when the integral is evaluated using the largest and smallest of the Wertevorrat limits, which we propose as more reasonable. 

This assignment of errors is similar in many ways to the usual assignment of statistical errors to the results of a lattice calculation.  In the typical jackknife determination of lattice errors, it is the width of distribution of jackknife samples that determines the statistical error not the deviation of the sample that is most discrepant from the average.  However, in the case of the statistical evaluation of the Feynman path integral, one relies on the law of large numbers to guarantee that with sufficiently many Monte Carlo samples, the distribution of the resulting average will be Gaussian so that the assigned error has a precise frequentist interpretation.  

For the errors defined in Eq.~\eqref{eq:error-def}, the resulting distribution is not known but may be a valuable goal of mathematical research.  While not (yet) known analytically, we can study this distribution empirically as is done below.  Certainly the statistical distribution of lattice results over the Pick-consistent region will be Gaussian and hence close to uniform within our $\pm 1\, \sigma$ limits. While our procedure of finding points on the boundary of the Pick-consistent region by constrained gradient ascent and then using chords drawn between pairs of these points will not be precisely uniform, it will be sufficiently close to uniform to provide a reasonable sample.  It is the distribution of errors implied by these samples, obtained as described above, that we can determine empirically.  That distribution then gives a meaning to our quoted errors equivalent to the implications of the Gaussian distribution for conventional lattice statistical errors. (A similar discussion of systematic errors is hindered by a lack of knowledge of the distribution of those errors.)

With the definition proposed in Eq.~\eqref{eq:error-def} the error arises from two sources. The first is the fluctuations coming from the varying average interpolation as the uncertain values from the lattice calculation fluctuate. We define this traditional error, coming from the fluctuations of the average or mean of the Wertevorr\"ate bounds about our final result as $\mathcal{E}_\mathrm{mean}^X$:
\begin{equation}
    \mathcal{E}_\mathrm{mean}^X = \left[  \frac{1}{M}\sum_m\left(\frac{\left(\mathcal{I}^X_\mathrm{max}\right)^m+\left(\mathcal{I}^X_\mathrm{min}\right)^m}{2}-
    \langle\mathcal{I}_\mathrm{avg}\rangle\right)^2 
    \right]^\frac{1}{2}.
\end{equation}
  
 In addition to this traditional error, Eq.~\eqref{eq:error-def} includes a second error associated with the average width of the Wertevorr\"ate, the bound on the uncertainty inherent in this procedure when interpolating from a single sample. This second source of error, resulting from the existence of the Wertevorr\"ate, defined as $\mathcal{E}_\mathrm{W}^X$, is given by 
\begin{equation}
    \mathcal{E}_\mathrm{W}^X = \left[
    \frac{1}{M}\sum_m\left(\frac{\left(\mathcal{I}^X_\mathrm{max}\right)^m-\left(\mathcal{I}^X_\mathrm{min}\right)^m}{2}\right)^2 
    \right]^\frac{1}{2}.
\end{equation}
Here the additional standard factor of $1/2$ is introduced because if a quantity lies between two bounds, it might be presumed to lie at the midpoint between those bounds with an error given by one-half their difference.

With the definition of the total error given in Eq.~\eqref{eq:error-def}, a little algebra shows that this error is actually the square root of the sum of the squares of the Wertevorrat and mean errors identified above, as the two components of that total error:
\begin{equation}
    \mathcal{E}^X = \left[\left(\mathcal{E}_\mathrm{W}^X\right)^2 + 
    \left(\mathcal{E}_\mathrm{mean}^X\right)^2\right]^\frac{1}{2}.
\end{equation}

\begin{figure} [t!]
    \centering
    \includegraphics[width=0.45\linewidth]{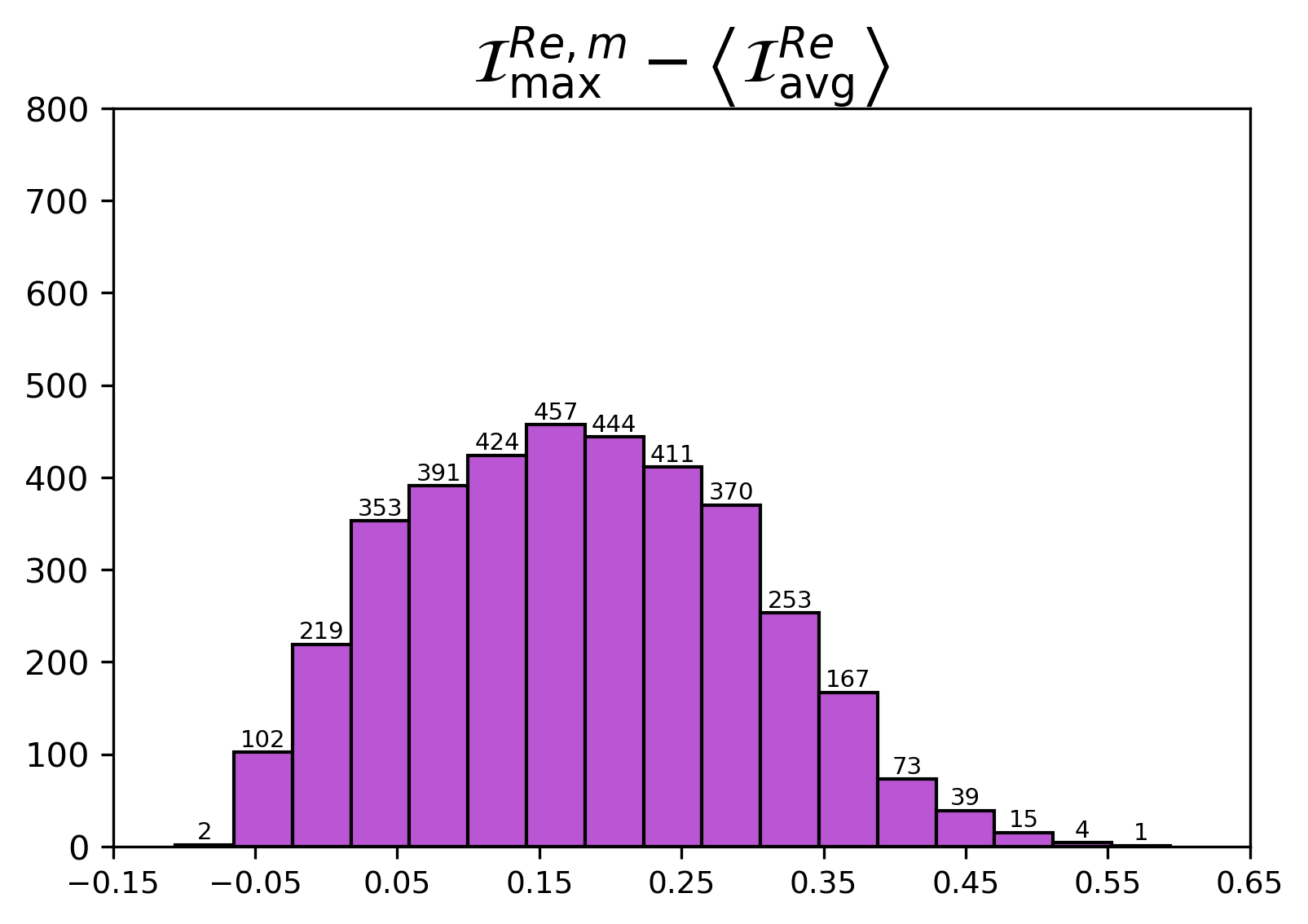}
    \includegraphics[width=0.45\linewidth]{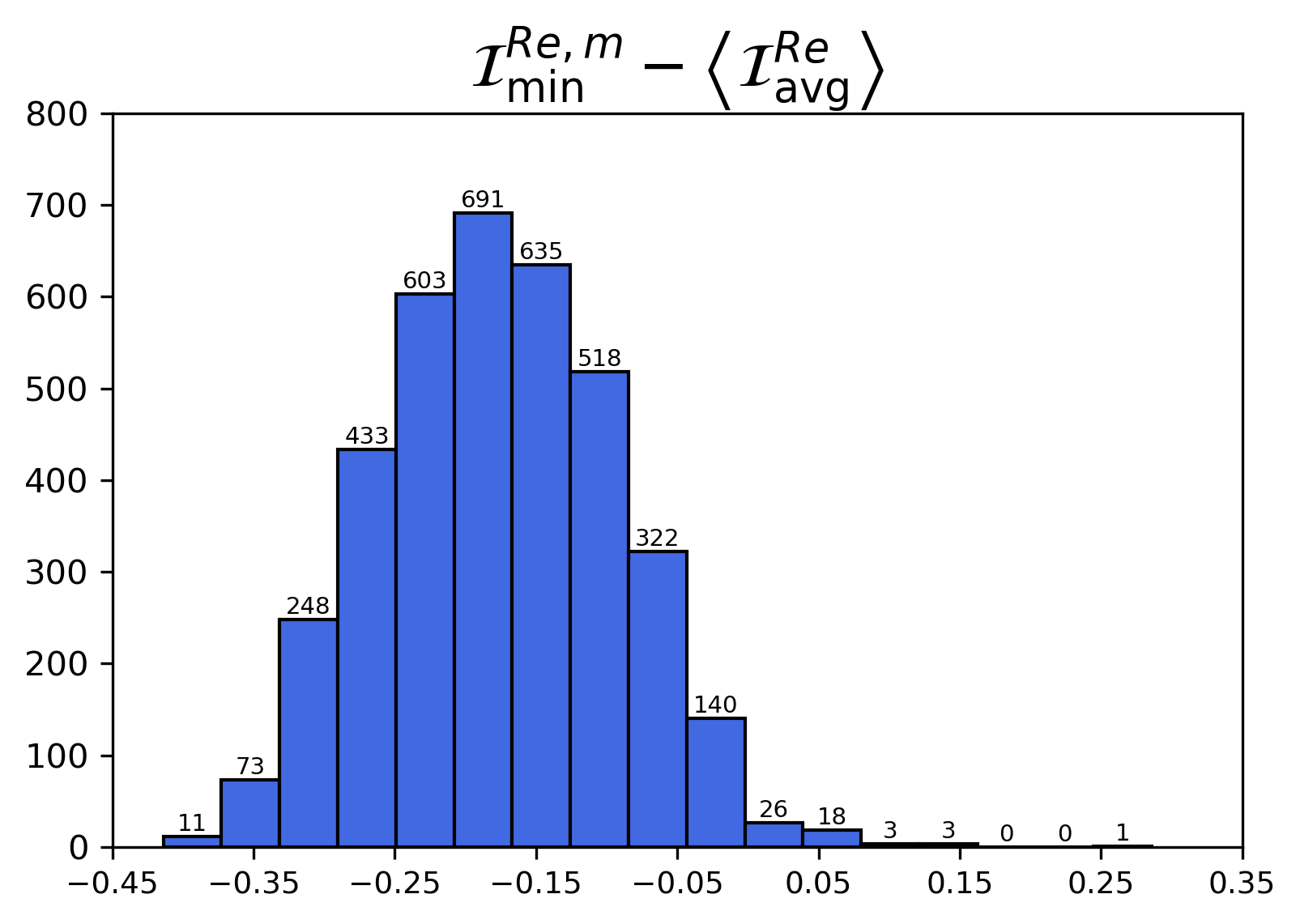}
    \includegraphics[width=0.45\linewidth]{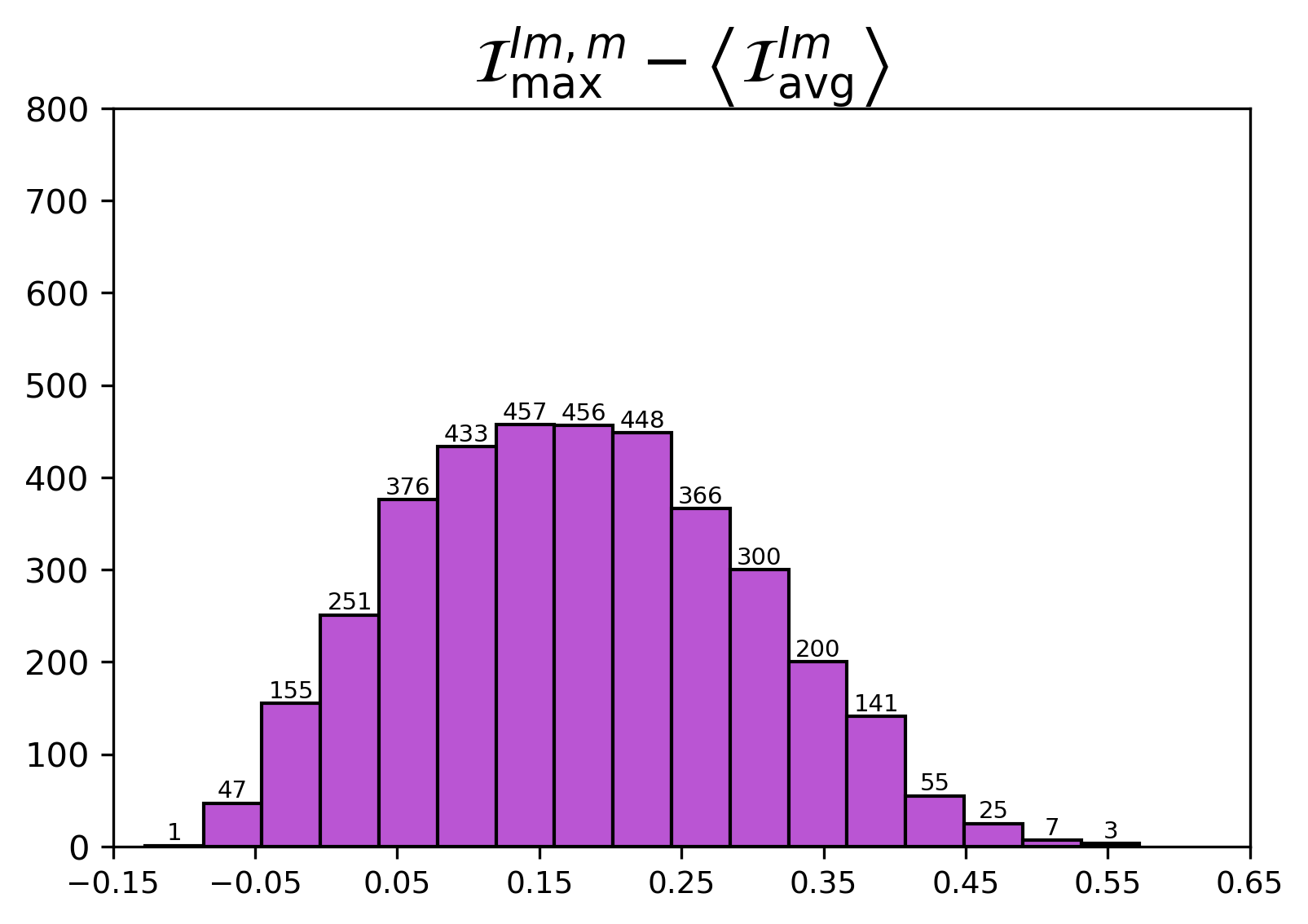}
    \includegraphics[width=0.45\linewidth]{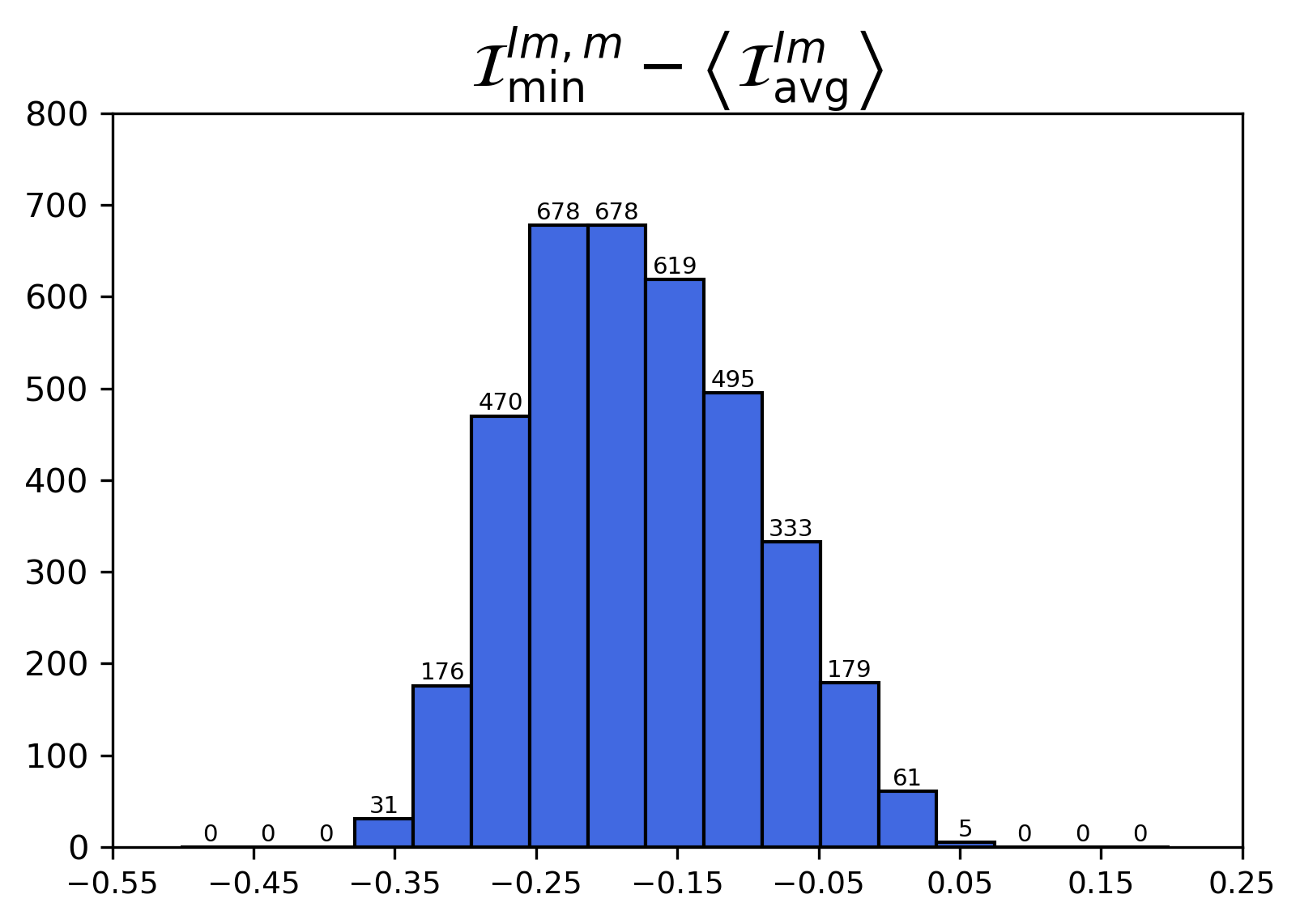}
    \caption{Histograms of the real (top) and imaginary (bottom) distributions of the $M$ differences $\mathcal{I}^{X,m}_\mathrm{max} - \left\langle\mathcal{I}^X_\mathrm{avg}\right\rangle$ (left) and similarly for the $M$ differences $\mathcal{I}^{X,m}_\mathrm{min} - \left\langle\mathcal{I}^X_\mathrm{avg}\right\rangle$ (right). The initial points $z_n$ were chosen as 10 equally spaced values along the interval $\{0.1i,2.0i\}$ and the error scale was $\xi = 0.01$.}
    \label{fig:max_and_min_varation_from_ave_histagrams}
\end{figure}

\begin{figure} [t!]
    \centering
    \includegraphics[width=0.45\linewidth]{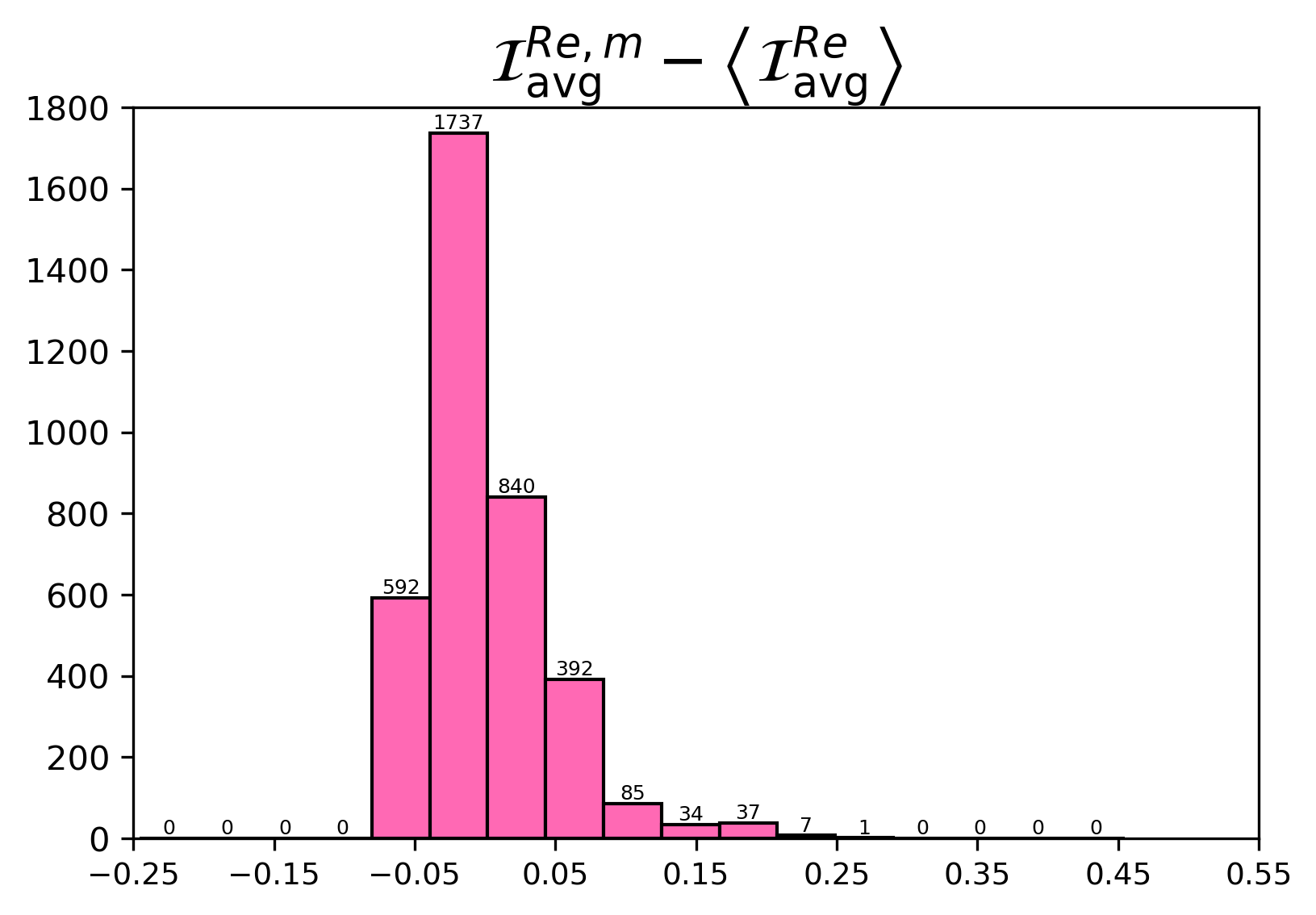}
    \includegraphics[width=0.45\linewidth]{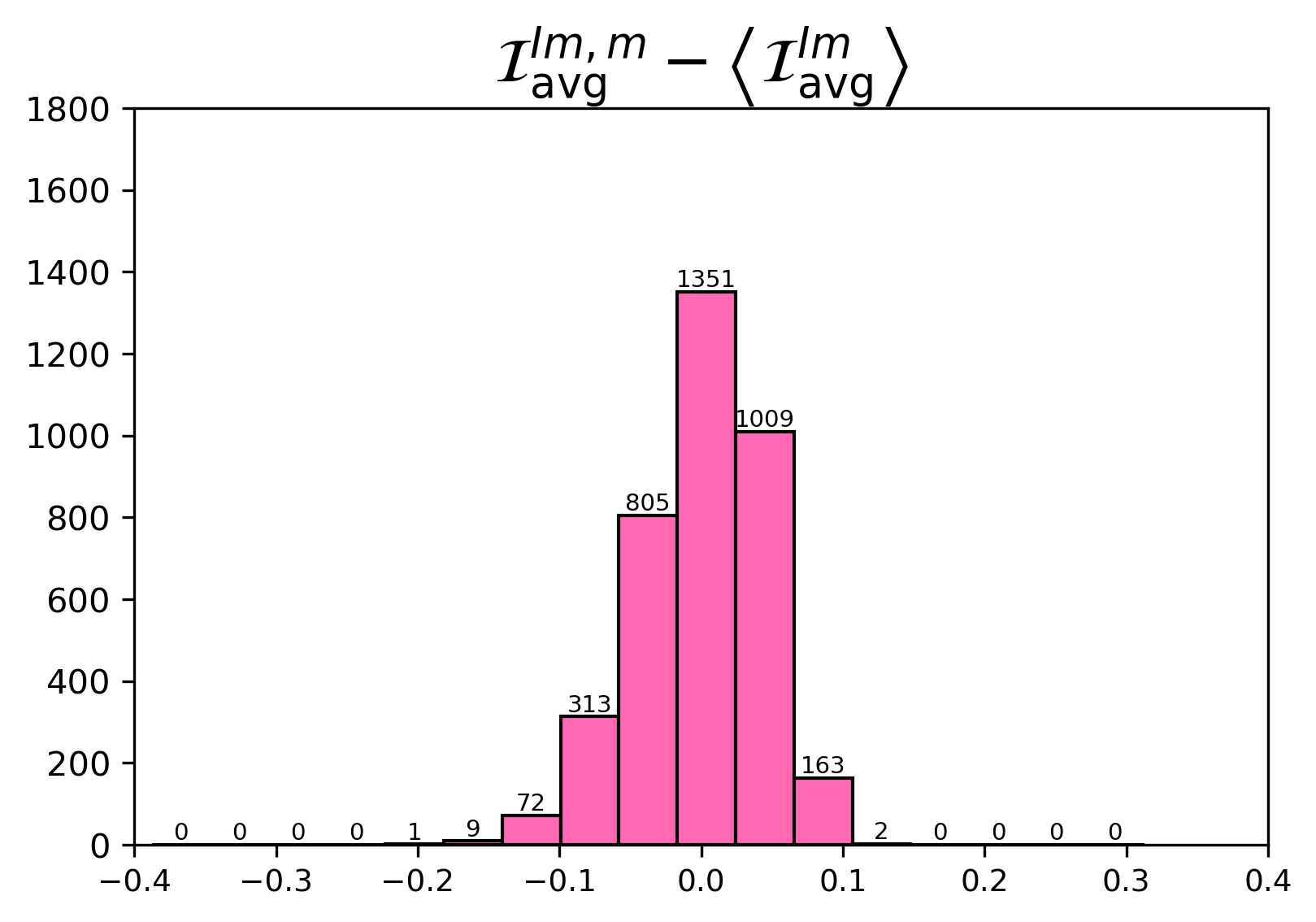}
   
    \caption{Histograms of the real (left) and imaginary (right) distributions of the integral of the $M$ differences $\mathcal{I}^{X,m}_\mathrm{avg} - \left\langle\mathcal{I}^X_\mathrm{avg}\right\rangle$. The initial points $z_n$ were chosen as 10 equally spaced values along the interval $\{0.1i,2.0i\}$ and the error scale was $\xi = 0.01$.}
    \label{fig:varation_from_ave_histagrams}
\end{figure}

We can better understand our estimate for the error by studying the statistical properties of the differences 
\begin{equation}
    \mathcal{I}^{X,m}_\mathrm{max} - \left\langle\mathcal{I}^X_\mathrm{avg}\right\rangle
    \quad \mathrm{and} \quad
    \mathcal{I}^{X,m}_\mathrm{min} - \left\langle\mathcal{I}^X_\mathrm{avg}\right\rangle
    \label{eq:max-avg_min-avg}
\end{equation}
for $X =$ `Re' or `Im' and our $M$ values of $m$.
Histograms of these differences obtained from our $M=3725$ samples for $N=10$ are shown in Fig.~\ref{fig:max_and_min_varation_from_ave_histagrams}. These distributions show a single peak and, while asymmetric, show only a small fraction of the samples lying in the tails. Similar behavior is seen in the histograms in Fig.~\ref{fig:varation_from_ave_histagrams} which show the distribution in the average of the upper and lower bounds of the Wertevorrat for each sample minus the average of those quantities over the entire sample:
\begin{equation}
    \mathcal{I}^{X,m}_\mathrm{avg} - \left\langle\mathcal{I}^X_\mathrm{avg}\right\rangle
    \label{eq:sample-of-averages}
\end{equation}
for $X =$ `Re' or `Im'. Again, these distributions appear statistically manageable with relatively small tails. Note that the width of the distributions in Fig.~\ref{fig:varation_from_ave_histagrams} are approximately two times smaller than those shown in Fig.~\ref{fig:max_and_min_varation_from_ave_histagrams} suggesting the presence of fluctuations in the upper and lower limits that cancel in their average for the case of $N=10$.

\begin{figure} [h!]
    \centering
    \includegraphics[width=0.45\linewidth]{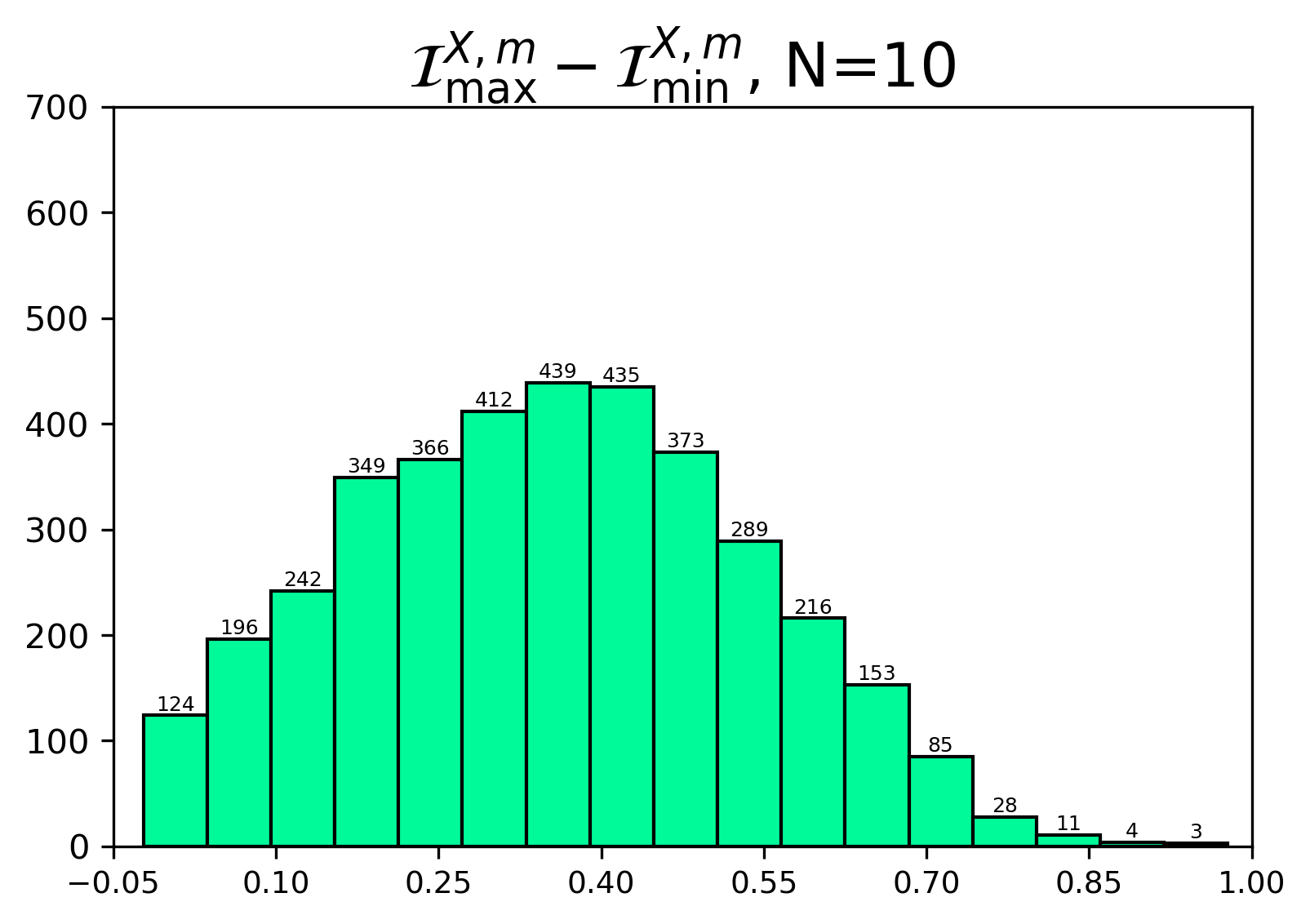}
    \includegraphics[width=0.45\linewidth]{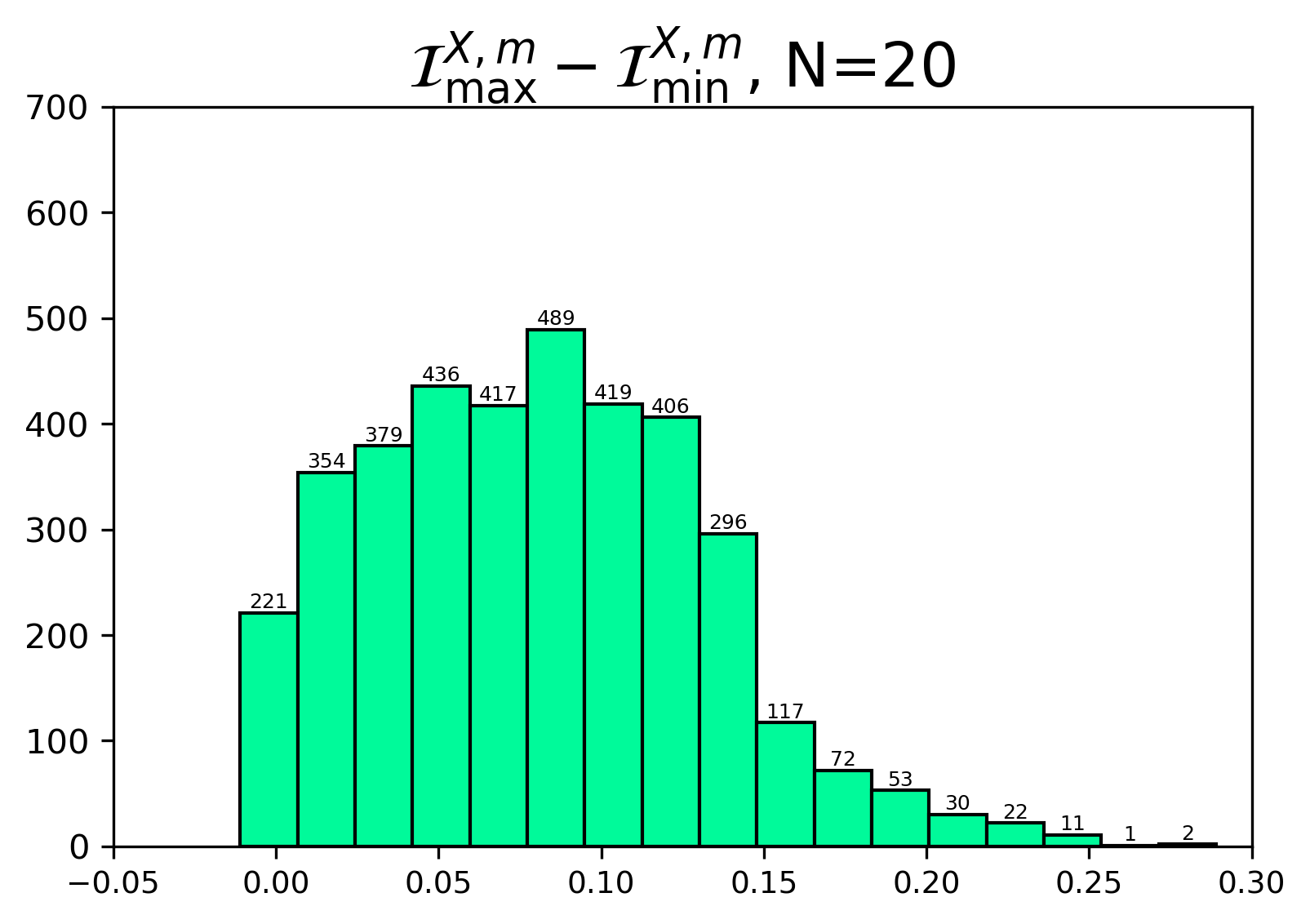}
    \includegraphics[width=0.45\linewidth]{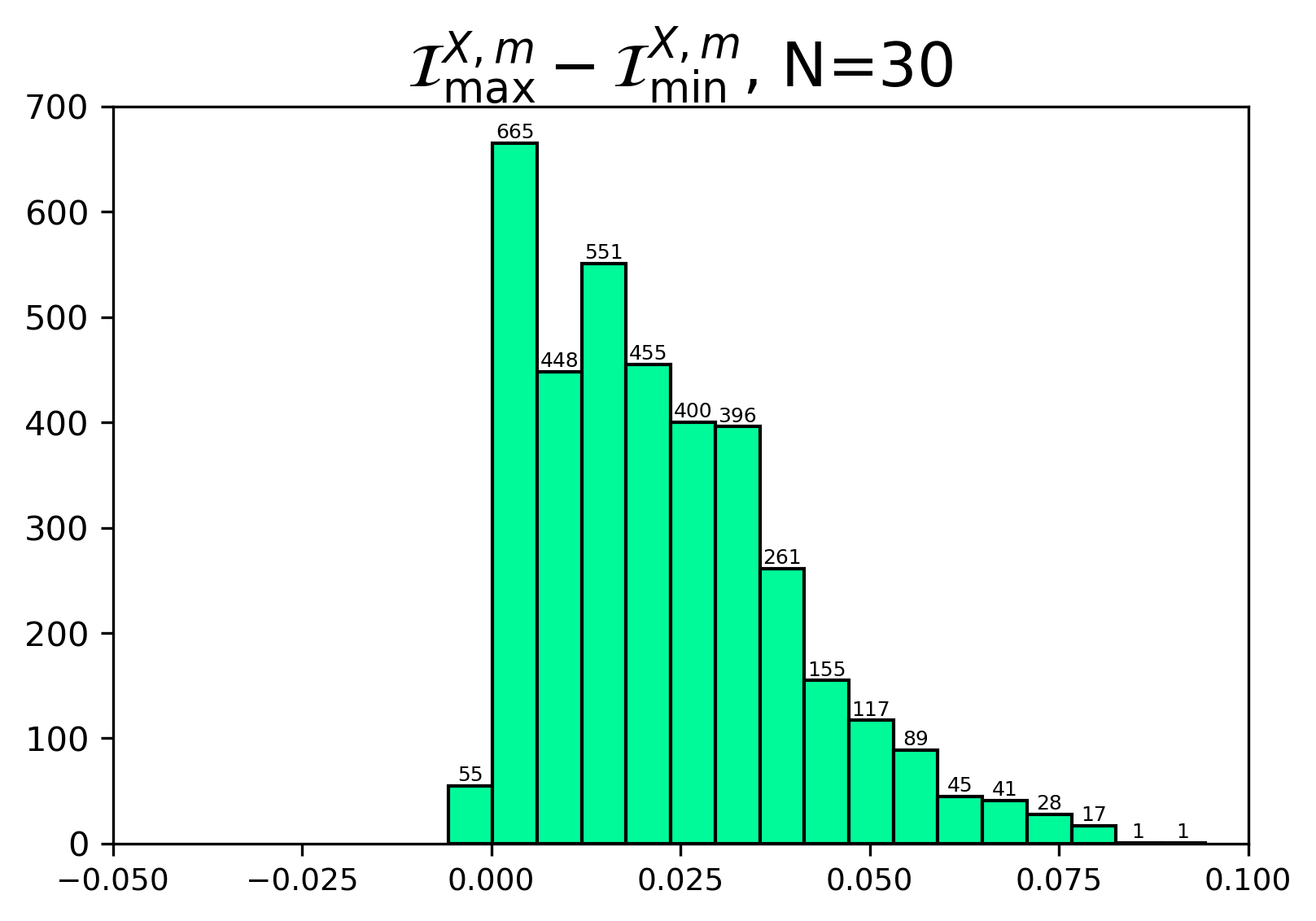}
    \caption{Histograms of the $M$ differences 
     $\mathcal{I}^{X,m}_\mathrm{max} - \mathcal{I}^{X,m}_\mathrm{min}$ showing the distribution of the effects of the Wertevorrat bounds on the integral of interest. The $N$ initial points $z_n$ were chosen to be equally spaced in the interval $\{0.1i,2.0i\}$ on the imaginary axis and the error scale was $\xi=0.01$. The case of $N=10$ is shown on the top left, $N=20$ on the top right, and $N=30$ on the bottom. It is important to note the significantly smaller range of the $x$ axis in $N=20$ and $N=30$ plots.}
    \label{fig:width_hist}
\end{figure}

The distribution of the $M$, (`max' + `min')/2 averages appearing in Eq.~\eqref{eq:sample-of-averages} represents the fluctuation among the results from which a traditional error would be determined. In our case, as discussed above, the error defined in Eq.~\eqref{eq:error-def} also receives contributions from the widths of Wertevorr\"ate. Thus, it is also of interest to examine the distribution of the quantity
\begin{equation}
    \mathcal{I}^{X,m}_\mathrm{max} - \mathcal{I}^{X,m}_\mathrm{min}
    \label{eq:sample-of-diameters}
\end{equation}
which is shown in Fig.\ref{fig:width_hist} for the interpolation from $N=10$, 20 and 30 lattice data points. 

Although we discuss the dependence of the interpolation results on the structure of the input lattice data in the next section, we make one such comparison here. We examine how the difference between the Wertevorrat-determined upper and lower bounds, $\mathcal{I}^{\mathrm{Im},m}_\mathrm{max}-\mathcal{I}^{\mathrm{Im},m}_\mathrm{min}$ for our 3725 samples varies as the number of lattice data points increases from $N=10$ to 30. The distribution of these differences is shown in Fig.~\ref{fig:width_hist} for $N=10$, 20 and 30. As should be expected, the average and width of this distribution decreases as $N$ is increased. This decrease is by a factor of 15 suggesting a large reduction in error from increasing the number of lattice results entering the interpolation. 

Given our statistical treatment of the Nevanlinna-Pick interpolation errors, it is natural to explore if there may be an effective way to combine these error with the usual jackknife or bootstrap resampling of the lattice data.  In fact this is not possible since our interpolation procedure cannot be applied to results from a single gauge configuration, preventing our having individual samples at the configuration-interpolation level.  These two statistical methods must be applied sequentially as we propose.

\section{Dependence of the interpolation results on the properties of the input Euclidean data}
\label{sec:demo}

In this section we examine how the interpolated result $\left\langle\mathcal{I}^X_\mathrm{avg}\right\rangle$ and the assigned error $\mathcal{E}^X$ as well as its components $\mathcal{E}^X_\mathrm{mean}$ and $\mathcal{E}^X_\mathrm{W}$ for our simple example are affected by the properties of the input data from which they are determined. To make the presentation more accessible, all results discussed are for the case $X = $`Im' since the behavior of results for the real part is similar.

We will discuss in turn the dependence on: (A) The number of input lattice data values. (B) The range on the imaginary axis of the locations of those input values. (C) The size of the errors assigned to the input lattice data and (D) The distance $\epsilon$ between the integration contour $C_2$ and the real axis.

\subsection{Dependence on $N$}
\label{sec:Depend-N}

\begin{table}[hbt!]
    \centering
    \footnotesize
    \begin{tabular}{|c|c|c|c|c|}
    \hline
    N & $\sigma_{\mathrm{Lat}}$ & $\left\langle\mathcal{I}^\mathrm{Im}_\mathrm{avg}\right\rangle$ & $\mathcal{E}_\mathrm{mean}^\mathrm{Im}$ & $\mathcal{E}_\mathrm{W}^\mathrm{Im}$\\
    \hline
    
    10 & 0.021 & 1.682  $\pm$ 0.204 & 0.044 & 0.199 \\
    20 & 0.020 & 1.669  $\pm$ 0.059 & 0.034 & 0.048 \\
    30 & 0.020 & 1.671  $\pm$ 0.035 & 0.032 & 0.014 \\
    \hline
    $\mathcal{I}^\mathrm{Im}_\mathrm{Exact}$ & - & 1.706 & - & - \\
    \hline
\end{tabular}
\caption{The result for the quantity $\left\langle\mathcal{I}^\mathrm{Im}_\mathrm{avg}\right\rangle$ and their errors obtained following the method described in the previous section for a varying number $N$ of input lattice data values. The right-most two columns show the two components that make up the error shown in the central column. The error scale was 0.01 and $\epsilon$, the displacement of the integration contour from the real axis, was 0.1. The sample size was 3725 for all cases. The final line shows the result obtained by directly integrating the exact Green's function ($\mathcal{I}^\mathrm{Im}_\mathrm{Exact}$) for our simple example.}
\label{table:increased_N}
\end{table}

Table \ref{table:increased_N} shows the results of the calculation of $\left\langle\mathcal{I}^\mathrm{Im}_\mathrm{avg}\right\rangle$ and its corresponding error computed following the interpolation method proposed in the previous section for initial lattice data sets containing a varying number $N$ of lattice results. The input data locations $z_n$ are equally spaced in the interval $[0.1i,2.0i]$ for the given number of initial data points. The assigned error $\sigma_{\mathrm{Lat}}$, determining the error volume, changes slightly because of variations in the data (whose average determines $\sigma_{\mathrm{Lat}}$) as $N$ varies). In all cases the interpolated integrated result lies within the calculated error of the known exact value. It can be seen that the error decreases significantly as the number of initial lattice points increases. As can be seen from the two right-most columns, this decrease results from a large change in the interpolation error $\mathcal{E}_\mathrm{W}^\mathrm{Im}$. The fluctuation in the mean (`max'+`min')/2 between the $N$ samples, $\mathcal{E}_\mathrm{mean}^\mathrm{Im}$, decreases much more slowly and for $N=30$ dominates the error.

\begin{figure}[t!]
    \centering
    \includegraphics[width=0.45\linewidth]{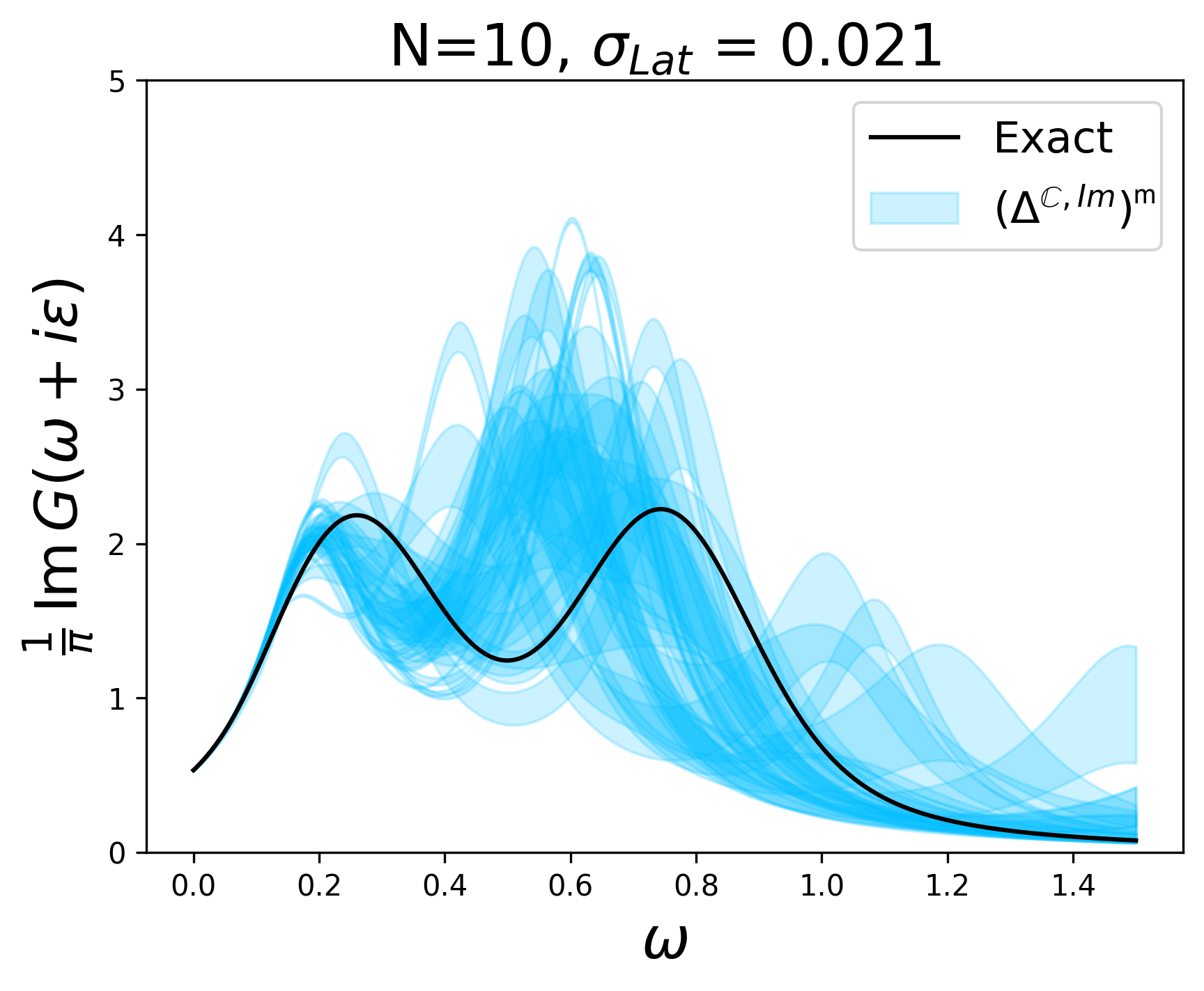}
    \includegraphics[width=0.45\linewidth]{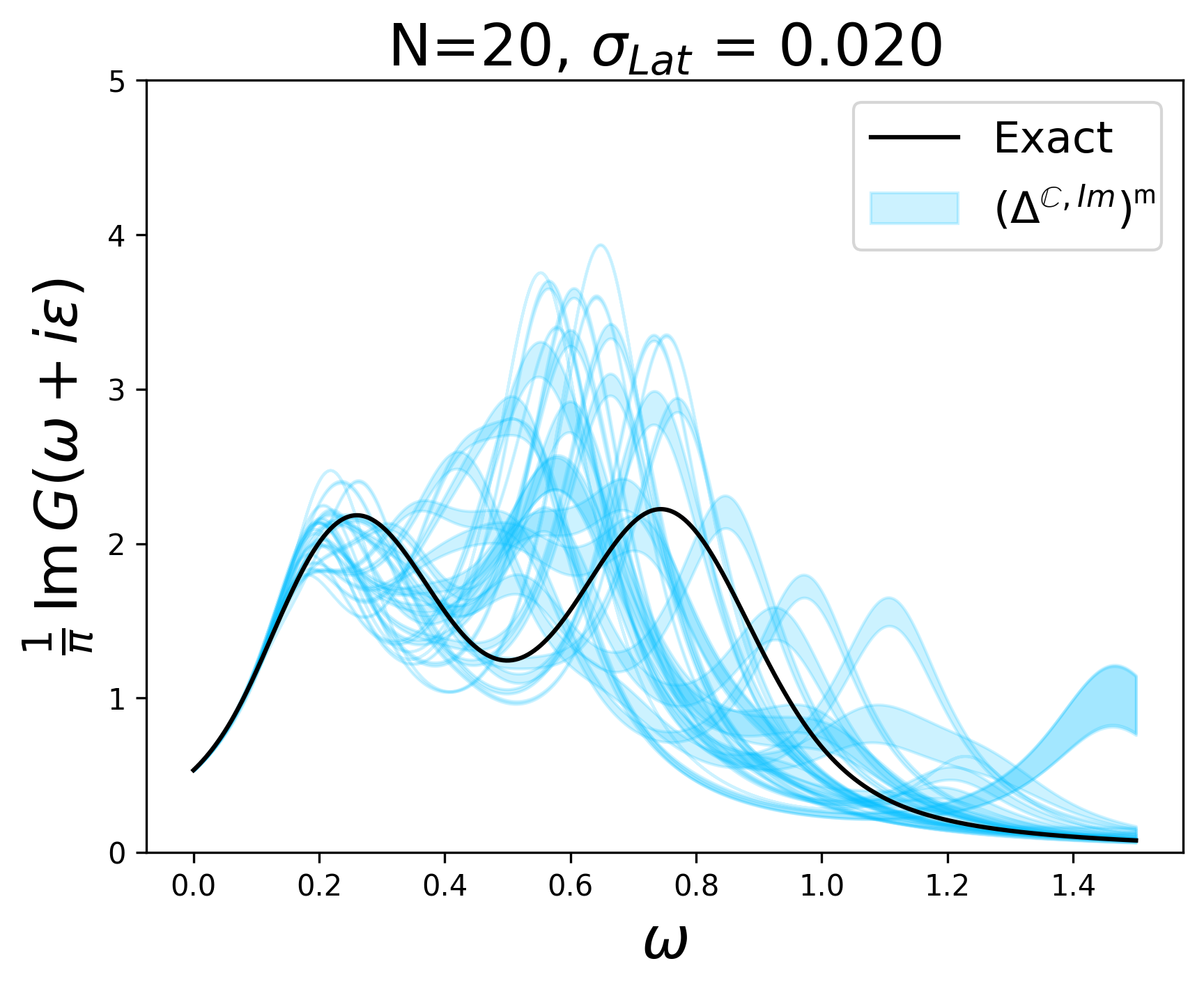}
    \includegraphics[width=0.45\linewidth]{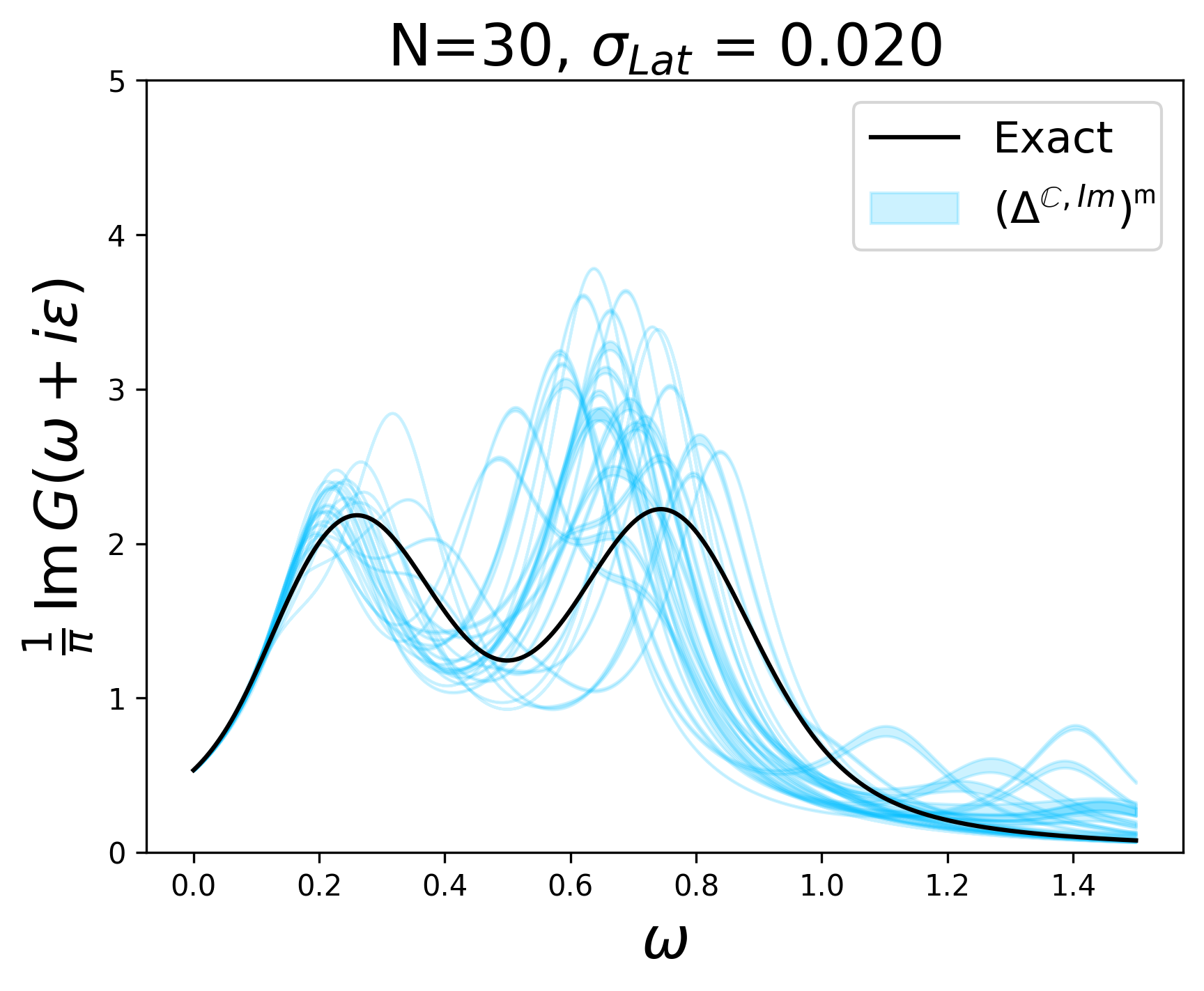}
    \caption{Plots showing the individual Wertevorrat regions for the integrand $\widetilde{G}^{\mathrm{Im},m}(z)/\pi$ as a function of $z$ for 25 random choices of samples $m$ from the total set of 3725 samples for the cases of N=10, 20 and 30. The solid black line shows the result from the exact value of the Green's function. The initial points $z_n$ were chosen as equally spaced values along the interval $\{0.1i,2.0i\}$ for the given number of points. The error scale  was 0.01 and the $\epsilon$ value was 0.1. A smaller subset sample size of 25 was chosen for consistency and for visualization purposes to maintain clarity.}
    \label{fig:Number_of_N}
\end{figure}

In Fig.~\ref{fig:Number_of_N} we show the $z$ dependence of the interpolation data that determined the integrand used in Eq.~\eqref{eq:SampleIntegral} to obtain the results shown in Table~\ref{table:increased_N} for 25 of the 3725 samples. Each of the 25 samples is represented by a shaded region corresponding to the Wertevorrat for the Green's function $\widetilde{G}^m(z)/\pi$ interpolated from the $m^{th}$ data sample, $\widetilde{G}^m$ to the point $z$ on the integration contour $C_2$.\footnote{Here we extend our notation, identifying the entire Green's function interpolated from the $N$ components $\{G^m_n\}_{1\le n \le N}$ as $\widetilde{G}^m(z)$.} The widths of these shaded regions decrease dramatically as $N$ increases so for the case of $N=30$ many of these regions appear to be rather narrow lines. This observation combined with the data in Table ~\ref{table:increased_N} demonstrates that for large N the widths of the Wertevorr\"ate become less important and the error becomes primarily driven by the variations in the averages (not the separations) of the Wertevorrat bounds. These fluctuations among the samples result from the errors in the lattice data and not from the uncertainties in the interpolation of a single sample.

\subsection{Dependence on energy range of the data to be interpolated}
\label{sec:Depend-range}

\begin{table}[t!]
    \centering
    \footnotesize
    \begin{tabular}{|c|c|c|c|c|c|}
    \hline
    N  & $\sigma_{\mathrm{Lat}}$ &  range of $z_n$ &$\left\langle\mathcal{I}^\mathrm{Im}_\mathrm{avg}\right\rangle$ & $\mathcal{E}_\mathrm{mean}^\mathrm{Im}$ & $\mathcal{E}_\mathrm{W}^\mathrm{Im}$\\
    \hline
    
    10  & 0.021 & $\{0.1i,2.0i\}$   & 1.682$\pm$ 0.204 & 0.044 & 0.199\\
    10  & 0.021 & $\{0.005i,2.0i\}$ &  1.691 $\pm$ 0.224 & 0.039 & 0.221  \\
    10  & 0.021 & $\{0.1i,4.0i\}$ & 1.690 $\pm$ 0.217 & 0.038 & 0.214\\
    20  & 0.020 & $\{0.1i,2.0i\}$   & 1.669  $\pm$ 0.059 & 0.034 & 0.048\\
    20 & 0.020 & $\{0.1i,4.0i\}$   & 1.680 $\pm$ 0.093 & 0.037 & 0.085 \\
    \hline
    $\mathcal{I}^\mathrm{Im}_\mathrm{Exact}$& - & - & 1.705 & - & - \\ 
    \hline

    \hline
\end{tabular}
\caption{Table of the interpolated results for $\left\langle\mathcal{I}^\mathrm{Im}_\mathrm{avg}\right\rangle$ and their assigned errors for sets of initial lattice data with 10 or 20 samples distributed according to three different sets of locations on the imaginary axis. The error scale was $\xi =0.01$ while $\epsilon = 0.1$ was used. In all cases the number of samples was M=3725. }
\label{table:changing_range}
\end{table}

We next study how the range of locations of the initial lattice data points from which we interpolate the Green's function affects the result of the integral and its error. Table \ref{table:changing_range} shows that there is a minimal reduction in the error as the initial data used in the interpolation covers a larger range on the imaginary axis for the case of 10 initial lattice data points. However, in a similar study for the case of 20 points, the error increased when the range of initial points was increased. The observed variations in error are small and show conflicting trends, making it difficult to draw reliable conclusions. Of course, the range of energies over which the lattice data should be provided must reflect the energy scales of the quantity being calculated. Thus, a more thorough effort to optimize the range of energies sampled in the underlying lattice calculation should be made in the future when attempting a specific physical calculation. 

\begin{figure}[ht!]
    \centering
    \includegraphics[width=0.35\linewidth]{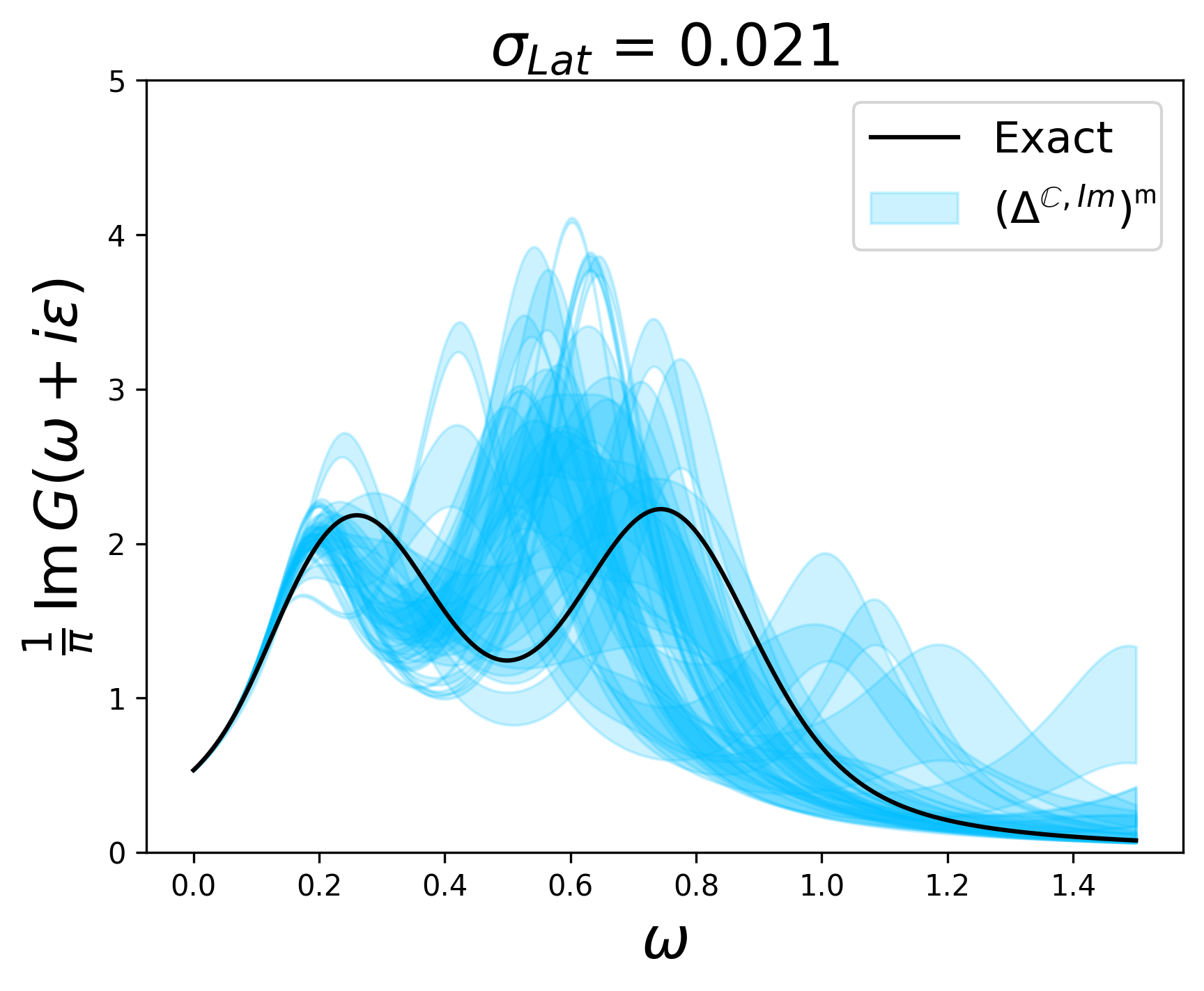}
    \includegraphics[width=0.35\linewidth]{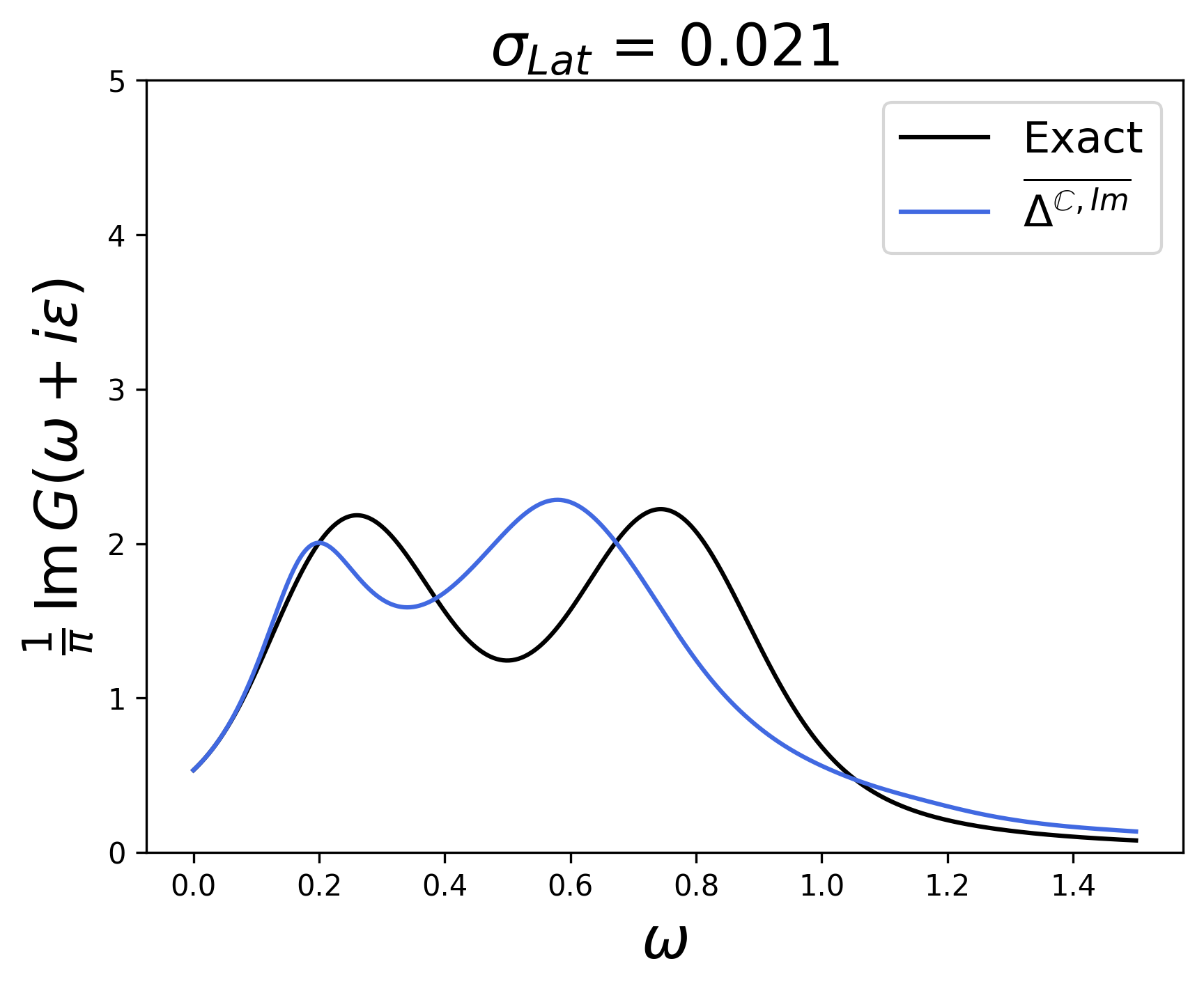}
    \includegraphics[width=0.35\linewidth]{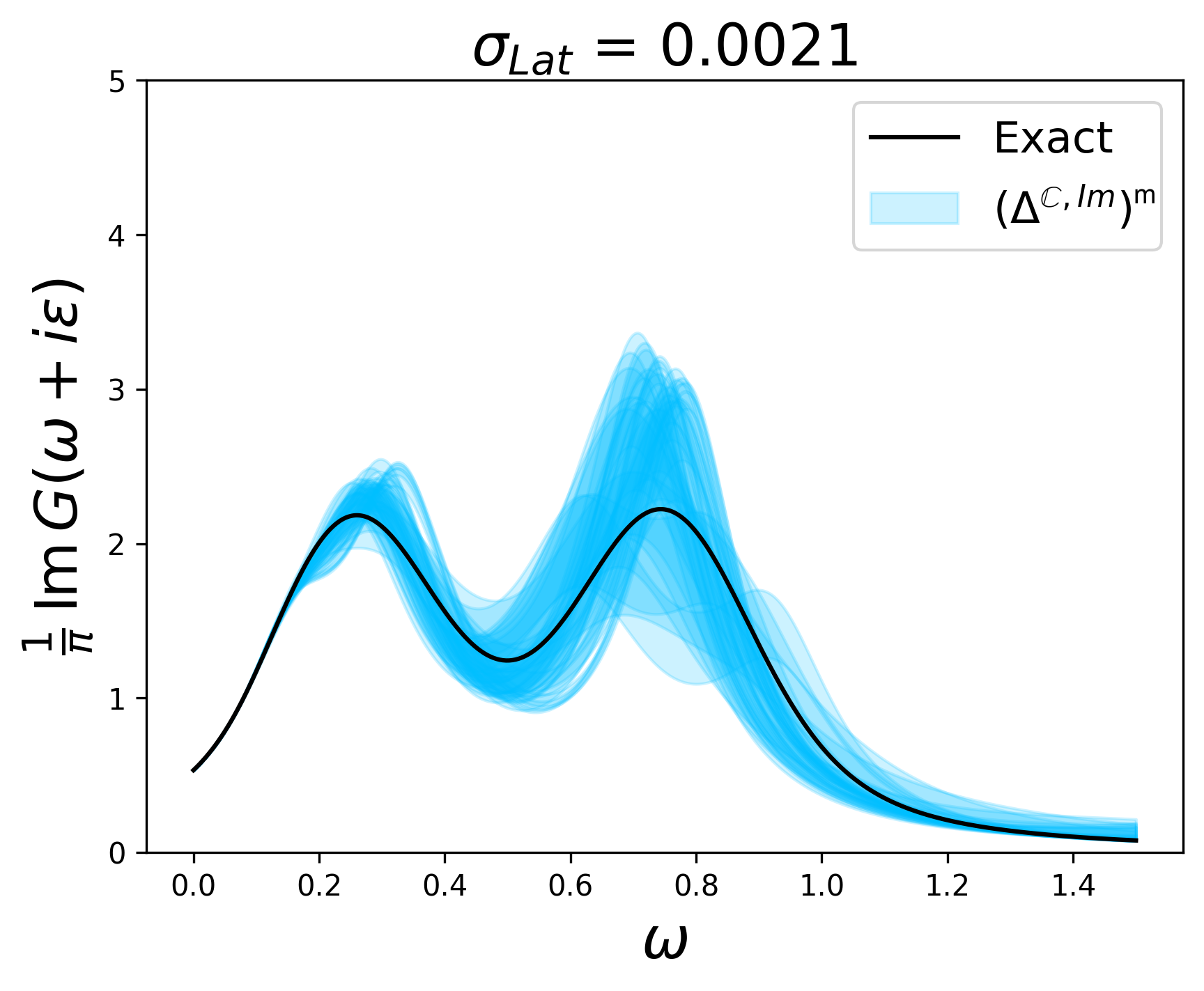}
    \includegraphics[width=0.35\linewidth]{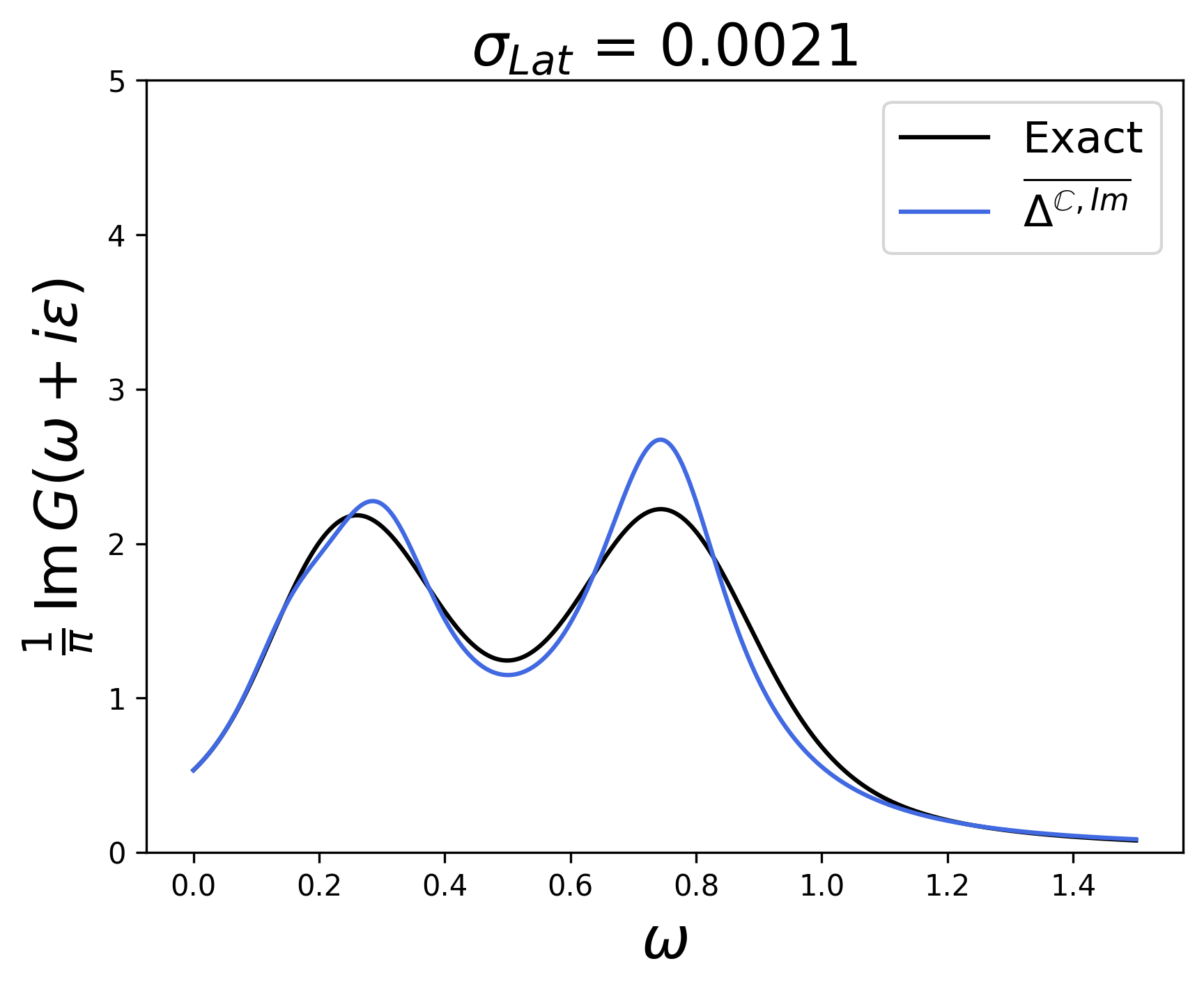}
    \includegraphics[width=0.35\linewidth]{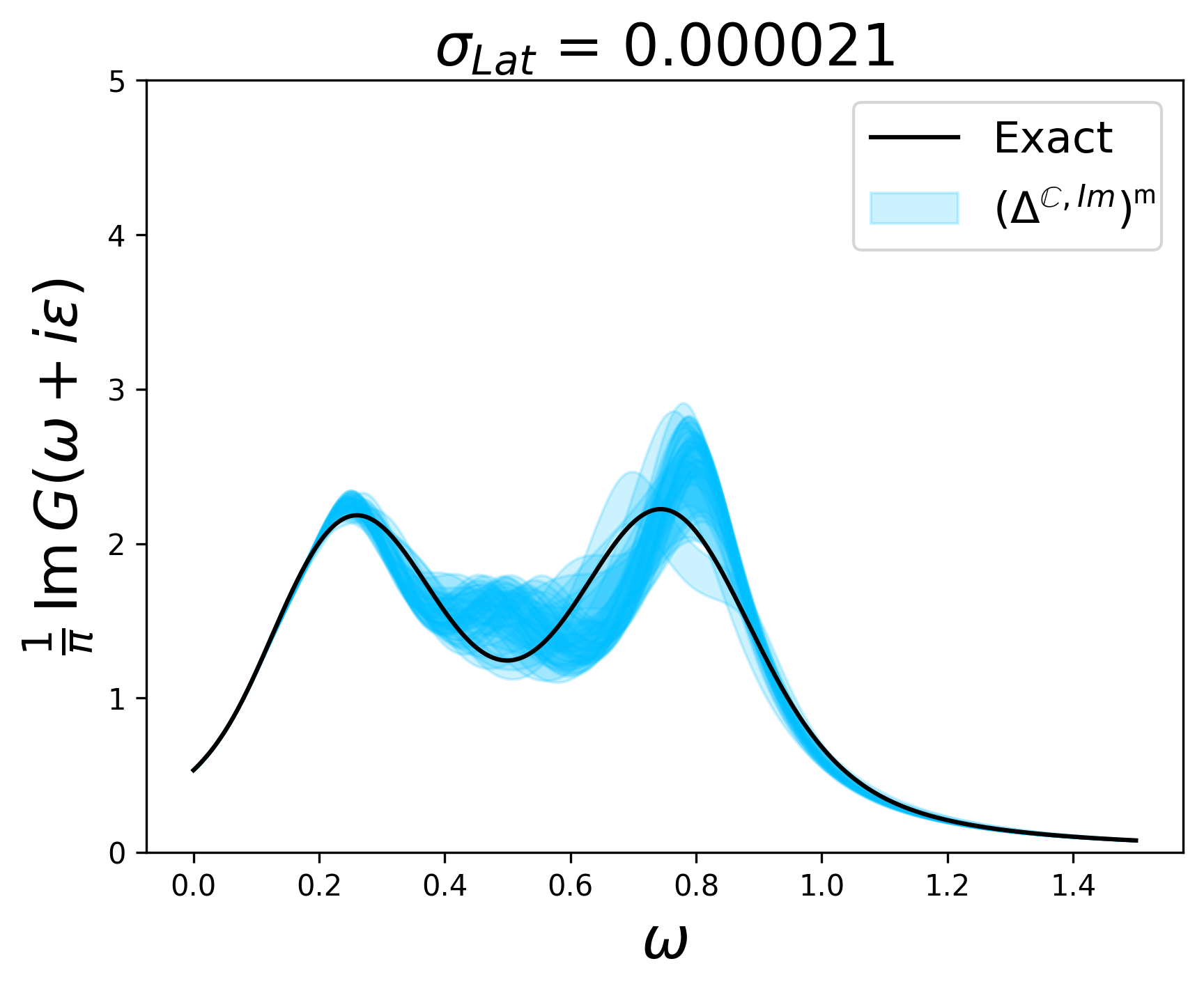}
    \includegraphics[width=0.35\linewidth]{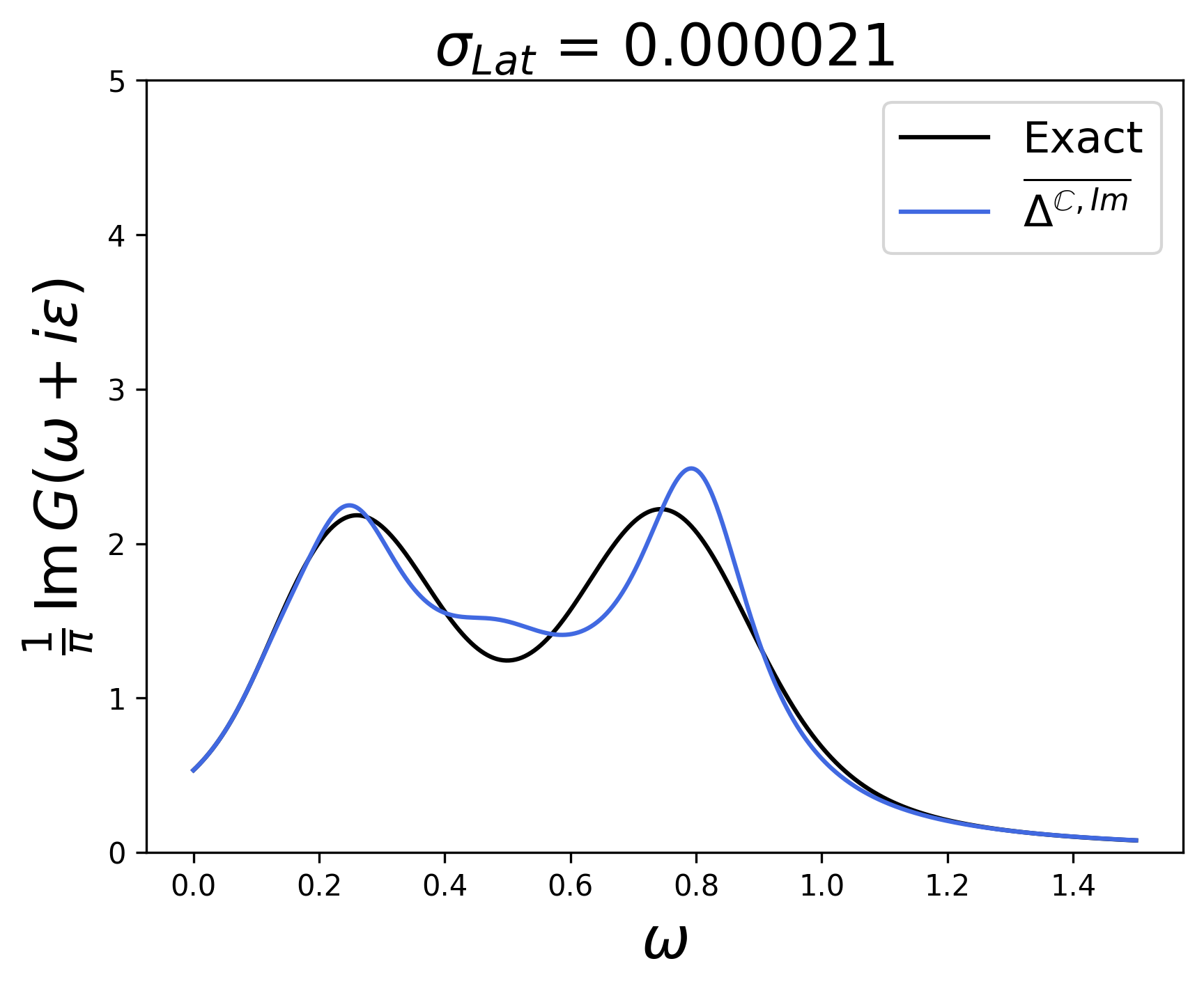}
    \caption{Plots showing the individual Wertevorrat regions for the integrand $\widetilde{G}^{\mathrm{Im},m}(z)/\pi$ as a function of $z$ (left) and the corresponding average curve produced by these samples (right). These results can be compared to the exact result shown as the black curve. Shown are three cases where the input error is determined by the error scale $\xi = 0.01$ (top row), 0.001 (middle row) and 0.00001 (bottom row). The initial locations $z_n$ were chosen as the same 10 equally spaced values along the interval $\{0.1i,2.0i\}$ and $\epsilon=0.1$ was used. As in Fig.~\ref{fig:Number_of_N} we plot 25 of the 3725 samples for the left-hand column while the average curve shown in the right column was generated from the full 3725 sample.}
    \label{fig:decreasing_error}
\end{figure}

\subsection{Dependence on size of input data errors}
\label{sec:Depend-errors}

It has been demonstrated above that as the number of samples $N$ increases the error assigned to the final interpolated result depends more on the fluctuations among the samples than the uncertainties of the interpolation process. Thus, it is important to understand the degree to which the fluctuation among the samples can be reduced by decreasing the error volume in which the sample data to be interpolated is chosen. Figure~\ref{fig:decreasing_error} confirms that decreasing the lattice error results in the interpolated integrand becoming a better approximation to the exact function.  

In Table~\ref{table:reducing_error} we examine this topic quantitatively comparing the three sizes of error, 1\%, 0.1\% and 0.001\%. We see that the total error on the final result decreases from 12\%, 7\% and 5\% for those three input errors. As can be seen from the two right-most columns of Table~\ref{table:reducing_error} this results from a rather imbalanced response of the sampling and Wertevorrat errors to this increasing input precision.  

As should be expected, the interpolation error, $\mathcal{E}^\mathrm{Im}_W$ is not much affected by the increasing precision of the input data. However, the fluctuations of the (`max'+`min)/2 giving the sampling error $\mathcal{E}^\mathrm{Im}_\mathrm{mean}$ decreases by a factor of 40 from this three order-of-magnitude drop in the input errors. Even the factor of 10 decrease in those input errors causes a more than fivefold decrease in the sampling error. The combined results from Section~\ref{sec:Depend-N} and the current Section suggest that significantly smaller final errors can be achieved by simultaneously reducing the errors on the input lattice data and increasing the number of lattice points from which the interpolation is being performed. 

\begin{table}[ht!]
    \centering
    \footnotesize
    \begin{tabular}{|c|c|c|c|}
    \hline
    $\xi$ &$\left\langle\mathcal{I}^\mathrm{Im}_\mathrm{avg}\right\rangle$ & $ \mathcal{E}_\mathrm{mean}^\mathrm{Im}$  & $\mathcal{E}_\mathrm{W}^\mathrm{Im}$ \\
    \hline
0.01\phantom{00}    & 1.682 $\pm 0.204$ & 0.044 $\pm 0.004$     & 0.199 $\pm 0.013$ \\
0.001 \phantom{0}   & 1.703 $\pm 0.112$ &  0.008 $\pm 0.001$     & 0.112 $\pm 0.011$\\
0.00001 & 1.707 $\pm 0.078$ & 0.001 $\pm 0.0000794$ & 0.077 $\pm 0.005$\\
    \hline
    $\mathcal{I}^\mathrm{Im}_\mathrm{Exact}$ & 1.705 &- & - \\
    \hline
\end{tabular}
\caption{Table of the interpolated results for $\left\langle\mathcal{I}^\mathrm{Im}_\mathrm{avg}\right\rangle$ and their errors as the error volume associated with the lattice result from which the 3725 samples were chosen was decreased by three orders of magnitude: corresponding to the fractional errors shown in the left-most column. The values $N=10$, the initial data locations in the $\{0.1i,2.0i\}$ and a displacement $\epsilon = 0.1$ from the real axis were used in each case. The errors on $\mathcal{E}_\mathrm{mean}^\mathrm{Im}$ and $\mathcal{E}_\mathrm{W}^\mathrm{Im}$ were calculated using jackknife resampling on the 50 boundary points used to obtain the original 3725 samples. }
\label{table:reducing_error}
\end{table}

\begin{figure} [hbt!]
    \centering
    \includegraphics[width=0.41\linewidth]{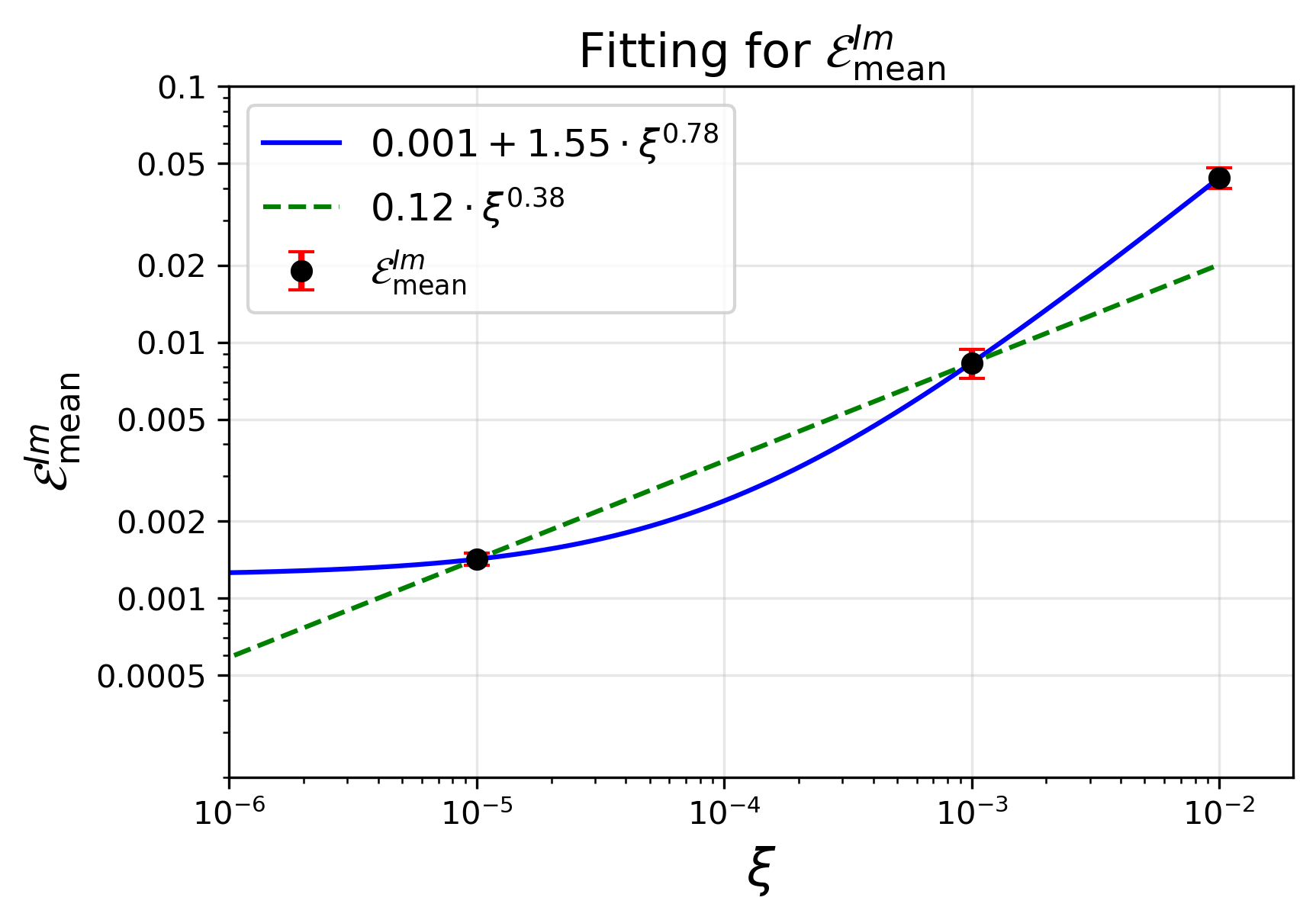}
    \includegraphics[width=0.4\linewidth]{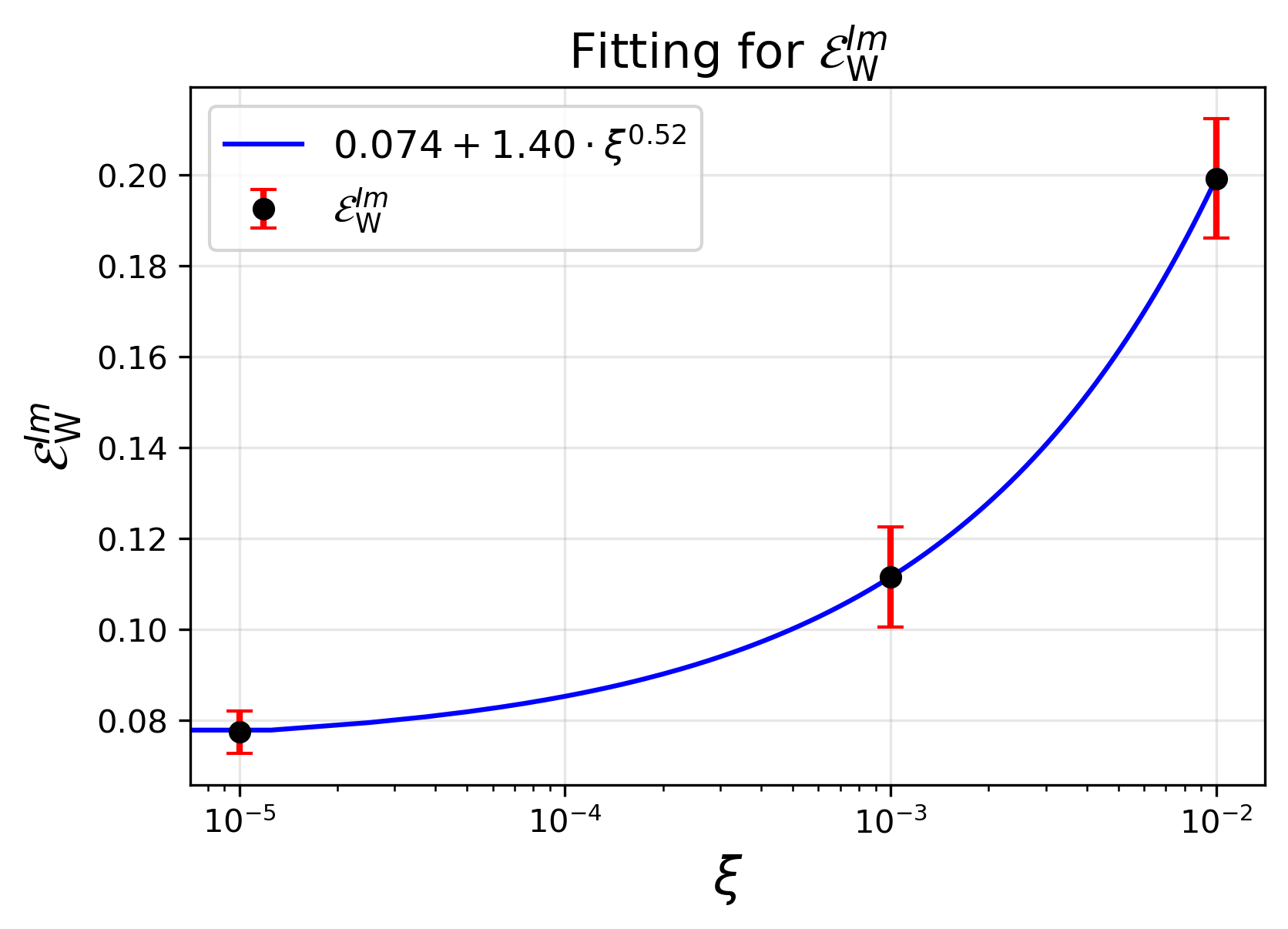}
    \caption{Fitted curves compared with the data points for $\mathcal{E}^\mathrm{Im}_\mathrm{mean}(\xi)$ (left) and $\mathcal{E}^\mathrm{Im}_\mathrm{W}(\xi)$ (right).  Both the $x$ and $y$ axies in the $\mathcal{E}^\mathrm{Im}_\mathrm{mean}(\xi)$ plot are shown on a logarithmic scale while only the $x$-axis is shown on a logarithmic scale for the $\mathcal{E}^\mathrm{Im}_\mathrm{W}(\xi)$ plot. The results for the errors $\mathcal{E}^\mathrm{Im}_\mathrm{mean}(\xi)$ and $\mathcal{E}^\mathrm{Im}_\mathrm{W}(\xi)$, with their errors are given in Table~\ref{table:reducing_error}. The fitted functional form is indicated in the legend, and the errors on the fit parameters are listed in Table ~\ref{tab:fitting_paramters}.}
    \label{fig:fitting_plots}
\end{figure}

In an attempt to learn more from Table~\ref{table:reducing_error} we have performed power-law fits to the $\xi$ dependence of the two components of the error on our interpolated result, $\mathcal{E}_\mathrm{mean}^\mathrm{Im}$ and $\mathcal{E}_\mathrm{W}^\mathrm{Im}$, beginning with the interpolation error $\mathcal{E}_\mathrm{W}^\mathrm{Im}$ where we use
\begin{equation}
    \mathcal{E}^\mathrm{Im}_W(\xi) = 0.074+1.40\cdot\xi^{0.52}.
    \label{eq:fit-Epsilon_W}
\end{equation}
Here we show the three fit parameters determined from the three results for $\mathcal{E}_\mathrm{W}^\mathrm{Im}$ that appear in Table~\ref{table:reducing_error}.
The errors for the three fit parameters are shown in Table ~\ref{tab:fitting_paramters}. We should expect the constant in Eq.~\eqref{eq:fit-Epsilon_W} to agree with the difference $\frac{1}{2}\left(\mathcal{I}^\mathrm{Im, exact}_\mathrm{max} - \mathcal{I}^\mathrm{Im, exact}_\mathrm{min}\right)$ for Nevanlinna-Pick interpolation carried out from the exact input data provided by our simple example.  This result is given by
\begin{equation}
   \frac{1}{2}\left(\mathcal{I}^\mathrm{Im, exact}_\mathrm{max} - \mathcal{I}^\mathrm{Im, exact}_\mathrm{min}\right) = 0.066, 
\end{equation}
a result in good agreement with the value $0.074\pm0.007$ given in Eq.~\eqref{eq:fit-Epsilon_W} and Table~\ref{table:reducing_error}.  While the exponent $\nu = 0.524 \pm 0.178$ of the error scale $\xi$ may be universal, the factor $b$ in the fit depends on the data being examined.  For the case at hand the exact result, at which our error volume is centered, lies near the edge of the Pick-consistent region and therefore has a smaller than average Wertevorrat.  Hence the coefficient $b$ is positive and the interpolation error shrinks as the error scale $\xi$ approaches zero.~\footnote{While paradoxical, the case of negative $b$ is possible and the resulting increasing error as the lattice results become more accurate but approach a point with a larger-than-average Wertevorrat, the correct behavior.}  The fit shown in Eq.~\eqref{eq:fit-Epsilon_W} is plotted in Fig.~\ref{fig:fitting_plots}.

We performed a similar three-parameter fit to the $\xi$ behavior of $\mathcal{E}^\mathrm{Im}_\mathrm{mean}(\xi)$:
\begin{equation}
    \mathcal{E}^\mathrm{Im}_\mathrm{mean}(\xi) = 0.001+1.55\cdot\xi^{0.78},
    \label{eq:fit-Epsilon_mean}
\end{equation}
again fitting the three data points shown in Table~\ref{table:reducing_error}.
The errors on the fit parameters are given in Table ~\ref{tab:fitting_paramters}. Table~\ref{table:reducing_error} shows that the interpolation result for $\mathcal{I}^\mathrm{Im}$ accurately approaches the exact value with the result when $\xi=0.00001$, agreeing within errors with the exact value.  However, the non-zero constant in the three-parameter fit given in Eq.~\eqref{eq:fit-Epsilon_mean}, $a=0.001 \pm 0.0001$ suggests that the corresponding error may not go continuously to zero.  Of course, this may be a failure of the form of the attempted three-parameter fit, at least for an error scale factor as large as $\xi=0.01$.  We therefore performed a second power-law fit to the $\xi$ dependence of the error $\mathcal{E}^\mathrm{Im}_\mathrm{mean}(\xi)$ of the form:
\begin{equation}
    \mathcal{E}^\mathrm{Im}_\mathrm{mean}(\xi) = 0.12\cdot\xi^{0.38},
    \label{eq:fit-Epsilon_mean_2}
\end{equation}
fitting only the results for the two smallest values of $\xi$, {\it i.e.} using data from the third and fourth rows of Table~\ref{table:reducing_error}.  
This second model for the $\xi$ dependence of the error on the mean, (`max'+'min')/2, of the Wertevorrat bounds now goes to zero as $\xi\to 0$. However, as shown in Fig.~\ref{fig:fitting_plots}, this fit form lies 50\% below the actual value of $\mathcal{E}^\mathrm{Im}_\mathrm{X}$ at $\xi=0.01$ .

\begin{table}[hbt!]
    \centering
\begin{tabular}{|c|c|c|c|} 
\hline
quantity  & model & parameter & value \\
\hline
\multirow{3}{4em}{$\mathcal{E}^\mathrm{Im}_\mathrm{W}(\xi)$} & \multirow{3}{4em} {$a+b\cdot\xi^{\nu}$} & a & $0.074 \pm0.007$ \\ 
& & b & $1.398 \pm 1.228$  \\ 
& &  $\nu$ & $0.524\pm 0.178$ \\ 
\hline
\multirow{3}{4em}{$\mathcal{E}^\mathrm{Im}_\mathrm{mean}(\xi)$} & \multirow{3}{4em}{$a+b\cdot\xi^{\nu}$} & a & $0.001 \pm 0.0001$ \\
&& b & $1.551 \pm 0.708$ \\ 
&& $\nu$ & $0.780 \pm 0.086$ \\ 
\hline
\multirow{2}{4em}{$\mathcal{E}^\mathrm{Im}_\mathrm{mean}(\xi)$} & \multirow{2}{2em}{$a\cdot\xi^{\nu}$} & a & $0.118 \pm 0.0001$ \\ && $\nu$ & $0.383 \pm 0.031$ \\ 
\hline
\end{tabular}
    \caption{Table showing the three models of $\xi$ dependence for parameters results for $\mathcal{E}^\mathrm{Im}_W(\xi)$ and $\mathcal{E}^\mathrm{Im}_\mathrm{mean}(\xi)$.  The direct fitting refers to the results obtained by doing the 3-point or 2-point fit using the average values in Table ~\ref{table:reducing_error}. The two rightmost columns show the average parameter values and associated errors obtained via a super-jackknife analysis on the boundary-sample jackknife distributions of $\mathcal{E}^\mathrm{Im}_X(\xi)$ for $X= W$ or mean.}
    \label{tab:fitting_paramters}
\end{table}

\subsection{Dependence on the distance $\epsilon$ between the contour $C_2$ and the real axis}
\label{sec:Depend-epsilon}

As expected from Nevanlinna-Pick interpolation~\cite{Bergamaschi:2023xzx} the interpolation errors grow as the distance $\epsilon$ between the integration contour $C_2$ and the real axis decreases. We have studied this effect further for the case of inexact data to determine how the propagated error changes. In Table~\ref{table:increased_epsilon} we examine five choices for $\epsilon$ varying between 0.06 and 0.14. As can be seen, the total error increases by a factor of nearly 5 as $\epsilon$ is decreased from 0.14 to 0.06. The bulk of this increase comes from the change in the interpolation error $\mathcal{E}^\mathrm{Im}_W$ which increases by a factor twice as large as the factor describing the increase in the sampling error $\mathcal{E}^\mathrm{Im}_\mathrm{mean}$, a behavior that might have been expected.

In a realistic calculation, the strategy for choosing $\epsilon$ may be complicated. Because of the logarithmic interpolation singularity expected as the integration contour approaches the real axis, it is possible that a different method will be needed to close the integration contour. The range over which such a second method is employed and even the shape of the contour on which Nevanlinna-Pick interpolation is employed will depend on details of the particular physical application.

\begin{table}[hbt!]
    \centering
    \footnotesize
    \begin{tabular}{|c|c|c|c|c|}
    \hline
     $\epsilon$ & $\left\langle\mathcal{I}^\mathrm{Im}_\mathrm{avg}\right\rangle$ & $\mathcal{E}_\mathrm{mean}^\mathrm{Im}$ & $\mathcal{E}_\mathrm{W}^\mathrm{Im}$ & $\mathcal{I}^\mathrm{Im}_\mathrm{Exact}$  \\
    \hline
    
     0.06 & 1.833 $\pm$ 0.476 & 0.082 & 0.469 &1.793\\
     0.08 & 1.765 $\pm$ 0.353 & 0.059 & 0.348 &1.748\\
     0.10 & 1.682 $\pm$ 0.204 & 0.044 & 0.199 &1.705\\
     0.12 & 1.638 $\pm$ 0.142 & 0.042 & 0.135 &1.664\\
     0.14 & 1.600 $\pm$ 0.099 & 0.030 & 0.095 &1.624\\

    \hline
\end{tabular}
\caption{Table of results with errors for varying distances $\epsilon$ between the integration contour and the real axis. M=3725 Pick-consistent points for all calculations. The table shows results when the initial $N=10$ locations $z_n$ were chosen as 10 equally spaced values along the interval $\{0.1i,2.0i\}$. The error scale was $\xi = 0.01$ with a corresponding lattice error of 0.021.  }
\label{table:increased_epsilon}
\end{table}

\section{Conclusion}
\label{sec:conclusion}

In this paper we develop the application of Nevanlinna-Pick interpolation to lattice QCD proposed in Ref.~\cite{Bergamaschi:2023xzx} to incorporate the intrinsic uncertainty in the lattice QCD results that are being interpolated. We propose in detail how lattice QCD errors can be propagated through the interpolation process, yielding meaningful errors on the interpolated results.  
This method is applied to a simple example and the dependence of the results on the number of input lattice data and their errors is studied.

We begin with the lattice results for a Laplace-transformed correlation function and their systematic and statistical errors at a series of $N$ energies. These $N$ Euclidan-space lattice results define the center of an $N$-dimensional complex ``error volume'' whose extent is determined by the errors on that central value. We then choose a large number (typically 3725) of possible lattice data samples within this error volume and use Nevanlinna-Pick interpolation to determine the bounds (Wertevorrat) within which the resulting Green's function in the complex plane must lie for each of these samples. We then evaluate a specific integral of such a family of Green's functions for our simple example, an integral chosen to be a quantity of potential physical importance. The variation of this integral over the large number data samples is used to determine an error. The distribution of results from these samples appears localized without long tails giving meaning to our assigned errors and playing a similar role to the Gaussian distribution of the usual lattice QCD averages.  For the tens of cases examined the known exact result was consistent with our interpolated result within the calculated errors.

An important challenge that was overcome arises from the fact that the analytic properties of the physical Green's function being studied is extraordinary restrictive so that only an extremely small sub-volume within the lattice error volume obeys the necessary ``Pick criterion''. We overcome this difficulty by distributing points at random within the lattice error volume and using extended precision arithmetic to evolve those initial points in a process of gradient ascent toward a place where the least positive eigenvalue of what is known as the $N\times N$ Pick matrix becomes positive, the Pick criterion referred to above.

The results of this procedure are encouraging. With thirty lattice samples and input errors of 1\%, final errors of 2\% were found -- errors potentially competitive with the errors determined by other methods for interesting inclusive processes. However, there are many essential problems and some opportunities that have not been addressed.
\begin{enumerate}
    \item The simple example studied was constructed from a Gaussian spectral density. In a realistic problem this density would grow with increasing energy and require careful treatment of a difficult asymptotic behavior.
    \item In this exploratory study we have not identified specific physical quantities to which these interpolation methods can be competitively applied.
    \item Nevanlinna-Pick interpolation typically fails for quantities integrated on complex contours which intersect the real axis.  Likely all physically interesting examples involve such an intersection and may then represent a portion of the calculation that must be done using other methods.
    \item The naive lattice error volume sampled in this study is a simple hypercube with no correlations between the lattice data values.  Incorporating an appropriate correlation matrix may lead to a reduction in the resulting errors since additional lattice input data has been provided.
    \item Since the Green's function being determined by interpolation is defined in the entire upper half plane, it may be advantageous to include additional interpolation data from points at large complex energies that can be determined accurately from QCD perturbation theory.
\end{enumerate}
Considering the above difficulties and opportunities, the next step should be to apply this approach to compute a particular quantity of physical interest.  In addition to its physical interest, such an application would determine if the assignment of errors developed here for an example spectral density, which has the same statistical validity as the errors determined in a conventional lattice QCD calculation of particle masses or matrix elements, can also used in a realistic case.  The study of a physical case would merit larger-statistics studies of the distribution of interpolated errors and the rate at which those errors decrease as additional lattice data with reduced errors are analyzed.

\acknowledgments
We thank our colleagues of the RBC and UKQCD collaborations for insightful discussions and ideas, in particular Robert Mawhinney, Mattia Bruno and Luchang Jin.  We thank Will Jay for valuable improvements to the manuscript.  We are grateful for the computational resources of Brookhaven National Laboratory, which is supported by the Office of Science of the U.S. This material is based on work supported by the National Science Foundation Graduate Research Fellowship Program under Grant No. DGE-2036197. Any opinions, findings, conclusions or recommendations expressed in this material are those of the author(s) and do not necessarily reflect the views of the National Science Foundation. This work was supported in part by U.S. DOE grant No. DE-SC0011941.


%

\end{document}